\documentclass{article}

\usepackage{arxiv}

\usepackage[utf8]{inputenc} 
\usepackage[T1]{fontenc}    
\usepackage{hyperref}       
\usepackage{url}            
\usepackage{booktabs}       
\usepackage{amsfonts}       
\usepackage{nicefrac}       
\usepackage{microtype}      
\usepackage{lipsum}		
\usepackage{graphicx}
\usepackage{natbib}
\usepackage{doi}
\usepackage{caption}
\usepackage{subcaption}
\usepackage{algorithm}
\usepackage{algorithmic}
\usepackage{amsmath}

\title{Development of Context-Sensitive Formulas to Obtain Constant Luminance Perception for a Foreground Object in Front of Backgrounds of Varying Luminance}

\date{February 27, 2024}	

\author{ \href{https://orcid.org/0000-0003-3618-4166}{\includegraphics[scale=0.06]{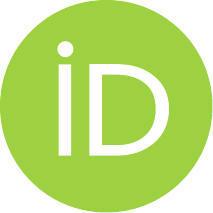}\hspace{1mm}Ergun Akleman} \\
	Visual Computing \& Computational Media, SPVFA\\
 Computer Science and Engineering, COE\\
 Texas A\&M University, College Station, TX, 77831\\
	\texttt{ergun@tamu.edu} \\
 	\And
 \href{https://orcid.org/0000-0002-9726-1340}{\includegraphics[scale=0.06]{orcid.pdf}\hspace{1mm}Bekir Tevfik Akgun} \\
Faculty of Computer and Information Sciences\\
	Yeditepe University, Istanbul, Turkey\\
	\texttt{bekirtevfik.akgun@yeditepe.edu.tr} \\
 	\And 
  \href{https://orcid.org/0000-0001-7695-196X}{\includegraphics[scale=0.06]{orcid.pdf}\hspace{1mm}Adil Alpkocak} \\
Department of Computer Engineering \\
İzmir Bakırçay University, Menemen, Turkey\\
	\texttt{adil.alpkocak@bakircay.edu.tr} 
}

\renewcommand{\shorttitle}{Constant Color Perception}

\hypersetup{
pdftitle={A template for the arxiv style},
pdfsubject={q-bio.NC, q-bio.QM},
pdfauthor={David S.~Hippocampus, Elias D.~Striatum},
pdfkeywords={First keyword, Second keyword, More},
}

\begin{document}
\maketitle

\begin{abstract}
In this article, we present a framework for developing context-sensitive luminance correction formulas that can produce constant luminance perception for foreground objects. Our formulas make the foreground object slightly translucent to mix with the blurred version of the background. This mix can be used to quickly produce any desired illusion of luminance in foreground objects based on the luminance of the background. The translucency formula has only one parameter; the relative size of the foreground object, which is a number between $0$ and $1$. We have identified the general structure of the translucency formulas as a power function of the form $y=s^f(s)$ where $s \in [0,1]$ is the relative size of the foreground object, $f(s)$ is a polynomial, and $y$ is the opacity of the foreground object. We have implemented a web-based interactive program in Shadertoy. Using this program, we determined the coefficients of the polynomials $f(s)$. To intuitively control the coefficients of the polynomial functions, we have used a B\'{e}zier form. Our final translucency formula uses a quadratic polynomial and requires only three coefficients. We also identified a simpler affine formula, which requires only two coefficients. We made our program publicly available in Shadertoy so that anyone can access and improve it. In this article, we also explain how to intuitively change the polynomial part of the formula. Using our explanation, users change the polynomial part of the formula to obtain their own perceptively constant luminance. This can be used as a crowd-sourcing experiment for further improvement of the formula. 
\end{abstract}

\begin{figure}[hbtp]
\centering
     \begin{subfigure}[t]{0.495\textwidth}
\includegraphics[width=0.99\textwidth]{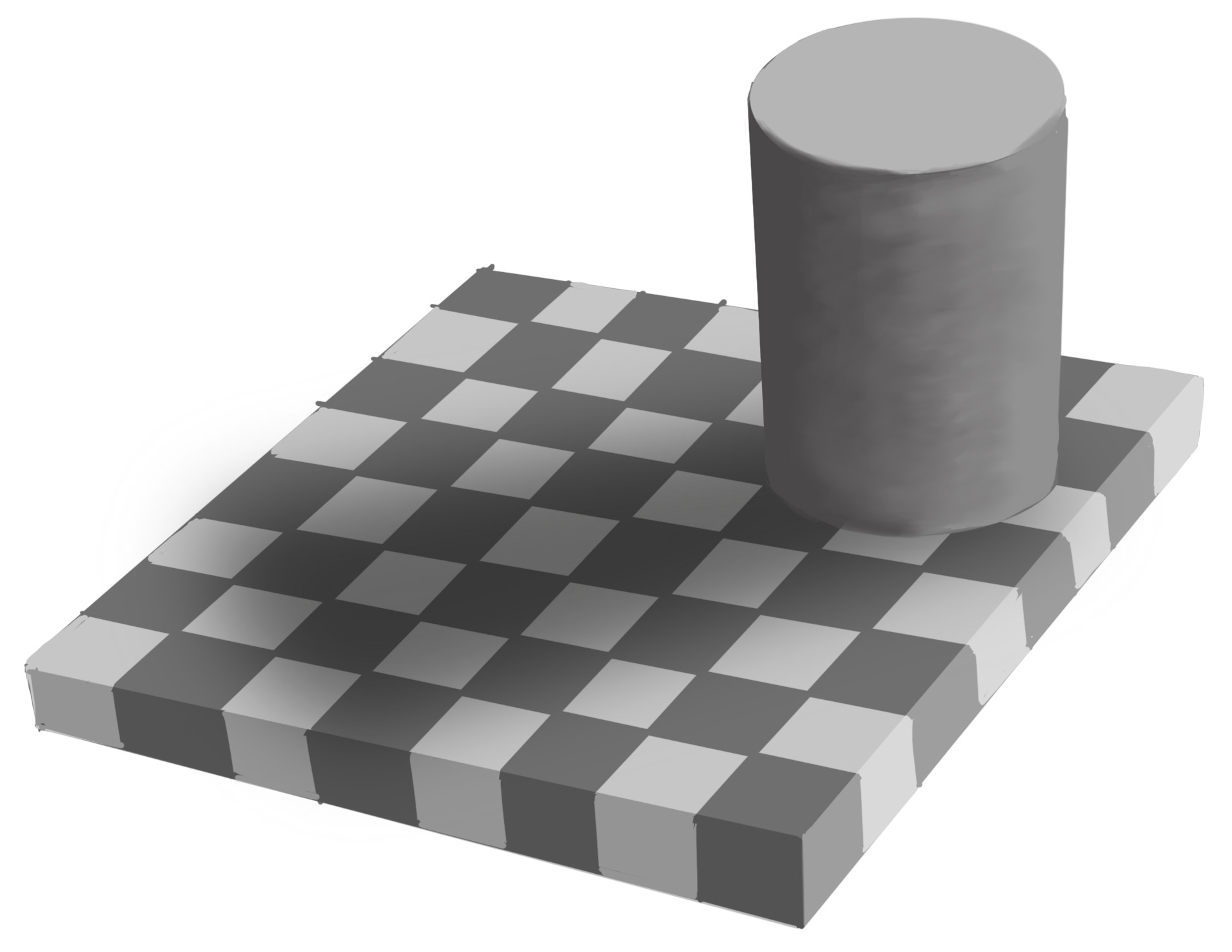}
    \caption{A chessboard with a shadow casting cylinder. }
    \label{chessboard/10}
 \end{subfigure}
 \hfill
      \begin{subfigure}[t]{0.495\textwidth}
\includegraphics[width=0.99\textwidth]{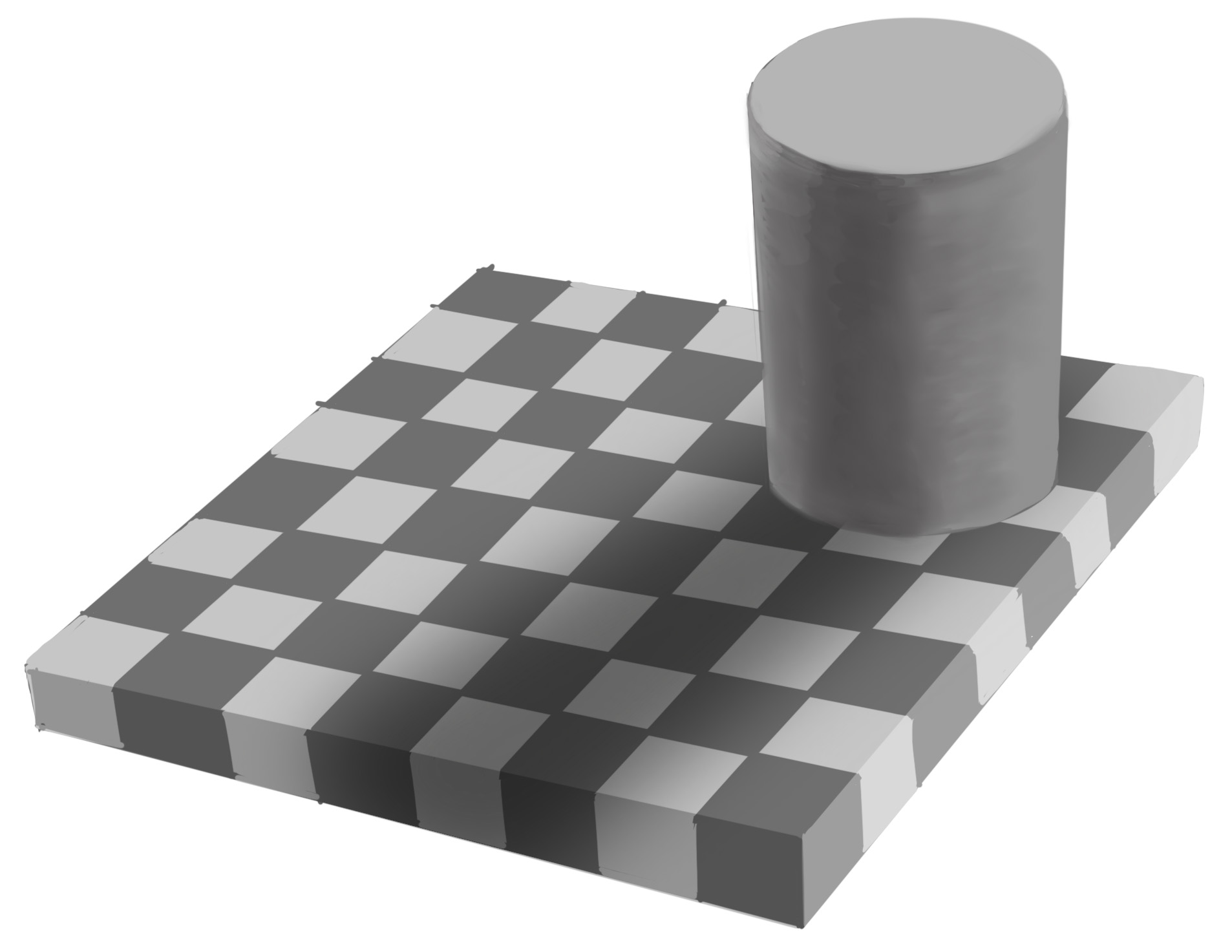}
    \caption{A chessboard with the same shadow-casting cylinder after slightly moving the light source. }
    \label{chessboard/11}
 \end{subfigure}
 \hfill
    \caption{This is a perspective recreation of Adelson's famous chessboard puzzle using moving shadow \cite{adelson200024}. We have created a new version to demonstrate that it is easy to move the light source. This is not a rendered image. It is created by one of the authors using Procreate and Photoshop. }
\label{fig_chessboard1}
\end{figure}

\section{Introduction and Motivation}

\begin{figure}[hbtp]
\centering
     \begin{subfigure}[t]{0.495\textwidth}
\includegraphics[width=0.99\textwidth]{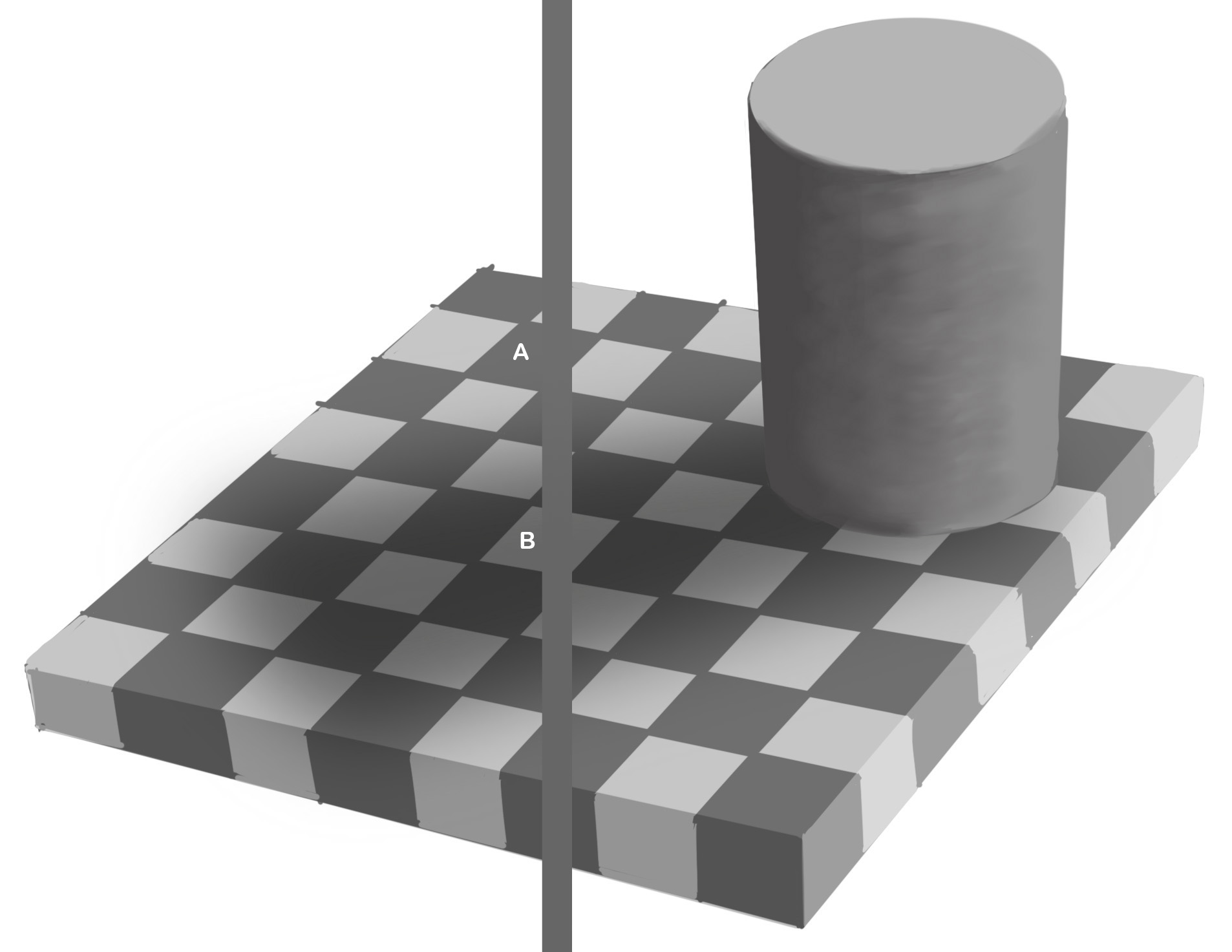}
    \caption{Similar luminance squares in the chessboard with a shadow-casting cylinder. }
    \label{chessboard/10b}
 \end{subfigure}
 \hfill
      \begin{subfigure}[t]{0.495\textwidth}
\includegraphics[width=0.99\textwidth]{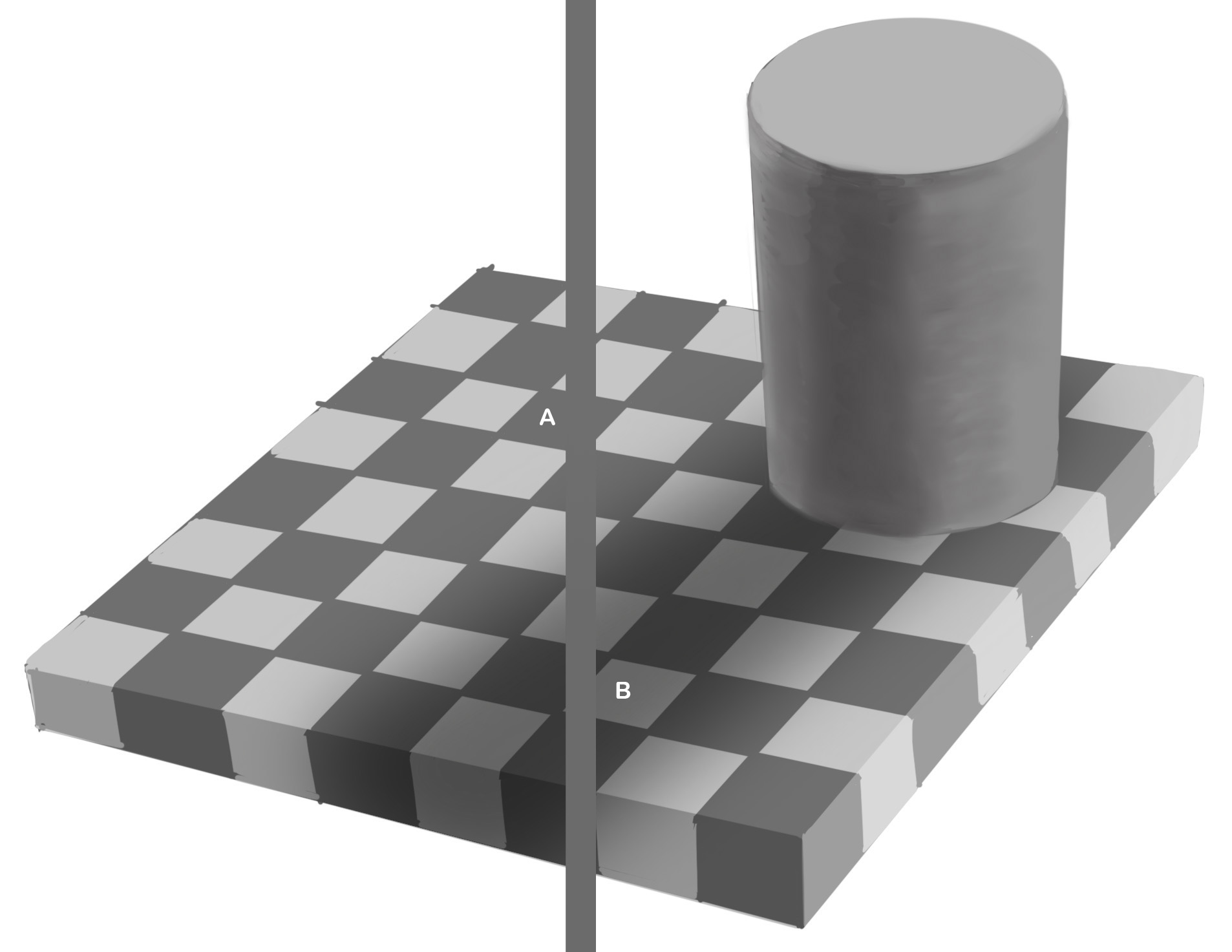}
    \caption{Similar luminance squares in the chessboard with the same shadow-casting cylinder after slightly moving the light source. }
    \label{chessboard/11b}
 \end{subfigure}
 \hfill
    \caption{Note that the squares A and B are almost the same with B being slightly darker. It is even possible to make B much darker in both cases and B squares will still look brighter.}
\label{fig_chessboard1b}
\end{figure}

\begin{figure}[hbtp]
\centering
     \begin{subfigure}[t]{0.495\textwidth}
\includegraphics[width=0.99\textwidth]{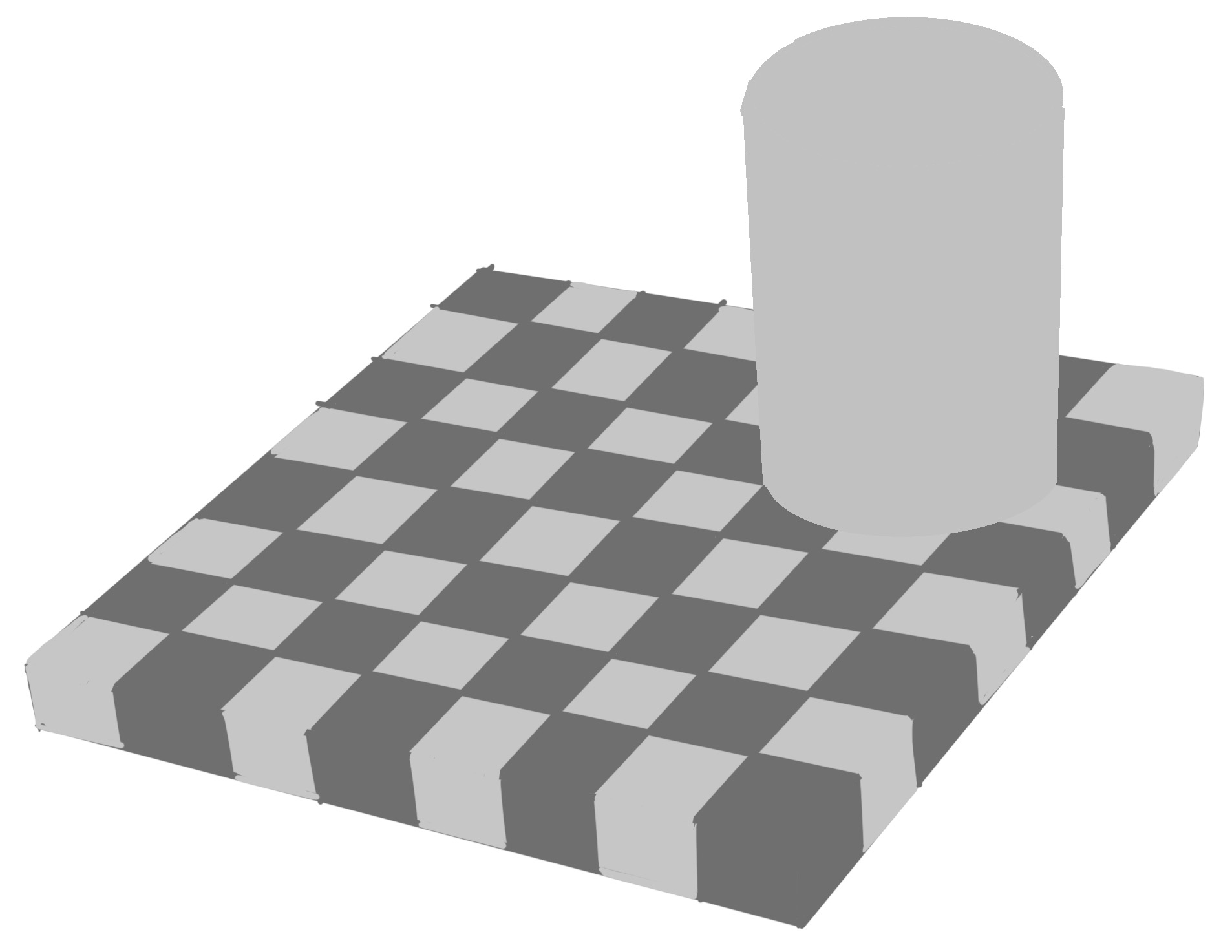}
    \caption{Reflectence image: Diffuse reflection term for all the materials. }
    \label{chessboard/materials}
 \end{subfigure}
 \hfill
      \begin{subfigure}[t]{0.495\textwidth}
\includegraphics[width=0.99\textwidth]{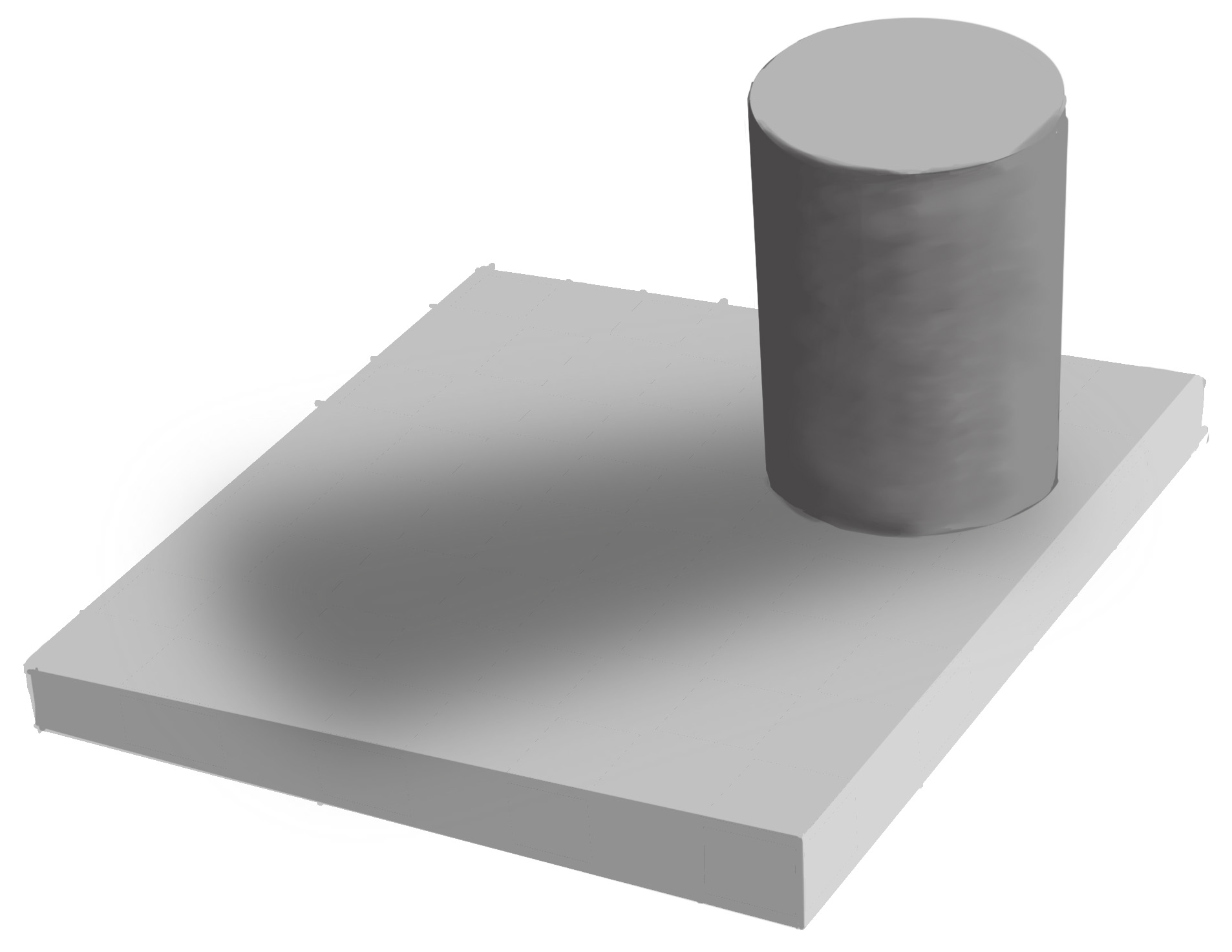}
    \caption{Illuminance Image: Illumination for Figure~\ref{chessboard/10} where all objects have the same materials. }
    \label{chessboard/illumination}
 \end{subfigure}
 \hfill
    \caption{Intrinsic image analysis \cite{barrow1978recovering}: Our conjecture is that animal visual systems are evolved to differentiate diffuse reflection terms to differentiate different objects.}
\label{fig_chessboardinfo}
\end{figure}

The illusion of the appearance of the color presents a large number of interesting problems \cite{foster2011color,fairchild2013color}. A famous type of illusion is the perception of luminosity \cite{adelson200024,gilchrist1999an}. A recreation of a well-known example of such illusions is shown in Figures~\ref{fig_chessboard1} and~\ref{fig_chessboard1b} \cite{force2020cheker}. In this article, we present an approach that can be used to correct the perceived illusion of luminosity in foreground objects based on the luminosity of the background. Our approach is motivated by intrinsic image analysis \cite{adelson1996perception, barrow1978recovering}, in which perceptual tasks are framed as computing reflectance and shading images that represent the underlying physical properties of a scene, which were also called intrinsic images (see Figure~\ref{fig_chessboardinfo}). This idea comes from the observation that colors are produced by a complex interaction of material properties of foreground objects and illumination of the environment. These complex interactions can be expressed by a bidirectional scattering distribution function (BSDF), which consists of a bidirectional reflectance distribution function (BRDF) and a bidirectional transmittance distribution function (BTDF). 

Intrinsic image analysis is extremely useful in explaining the reasons behind many interesting perceptual phenomena \cite{adelson1996perception}.  However, the main problem with intrinsic image analysis is that it is physically based. All these physics-based functions and the illumination of the environment are very complicated, which took many centuries \cite{darrigol2012history} and paradigm shifts for scientists to understand \cite{kuhn1962structure} \footnote{For example, Plato and Euclid supported emission theory, believing that visual perception is accomplished by rays emitted by the eyes.... This was not settled until the 11th century when Alhazen (Ibn al-Haytham) demonstrated that the emission theory cannot be valid \cite{smith1998ptolemy}. Newton rejected the wave theory \cite{cantor2006physical}. }. We cannot expect such complicated processes, which took centuries for the brightest scientists to explain, can easily be handled by primitive animal brains or early humans. There is a need for simpler yet somewhat meaningful explanations that can allow us to develop algorithms and formulas to explain how primitive visual systems could be processing the 3D world, which can result in some interesting perceptual phenomena.

In this article, we present a simple and somewhat meaningful explanation that is not physically based. This explanation allows us to develop a context-sensitive formula to obtain constant luminance perception. We observe that the overall luminosity of the color produced by the complex physical interactions can be estimated from a few parameters. We hypothesized that luminance perception in visual systems must have evolved to recognize the diffuse reflection term of the bidirectional scattering distribution function by considering only qualitatively essential parameters to easily identify similarities and differences in the material properties of foreground objects. 

\begin{figure}[hbtp]
\centering
     \begin{subfigure}[t]{0.32\textwidth}
\includegraphics[width=0.99\textwidth]{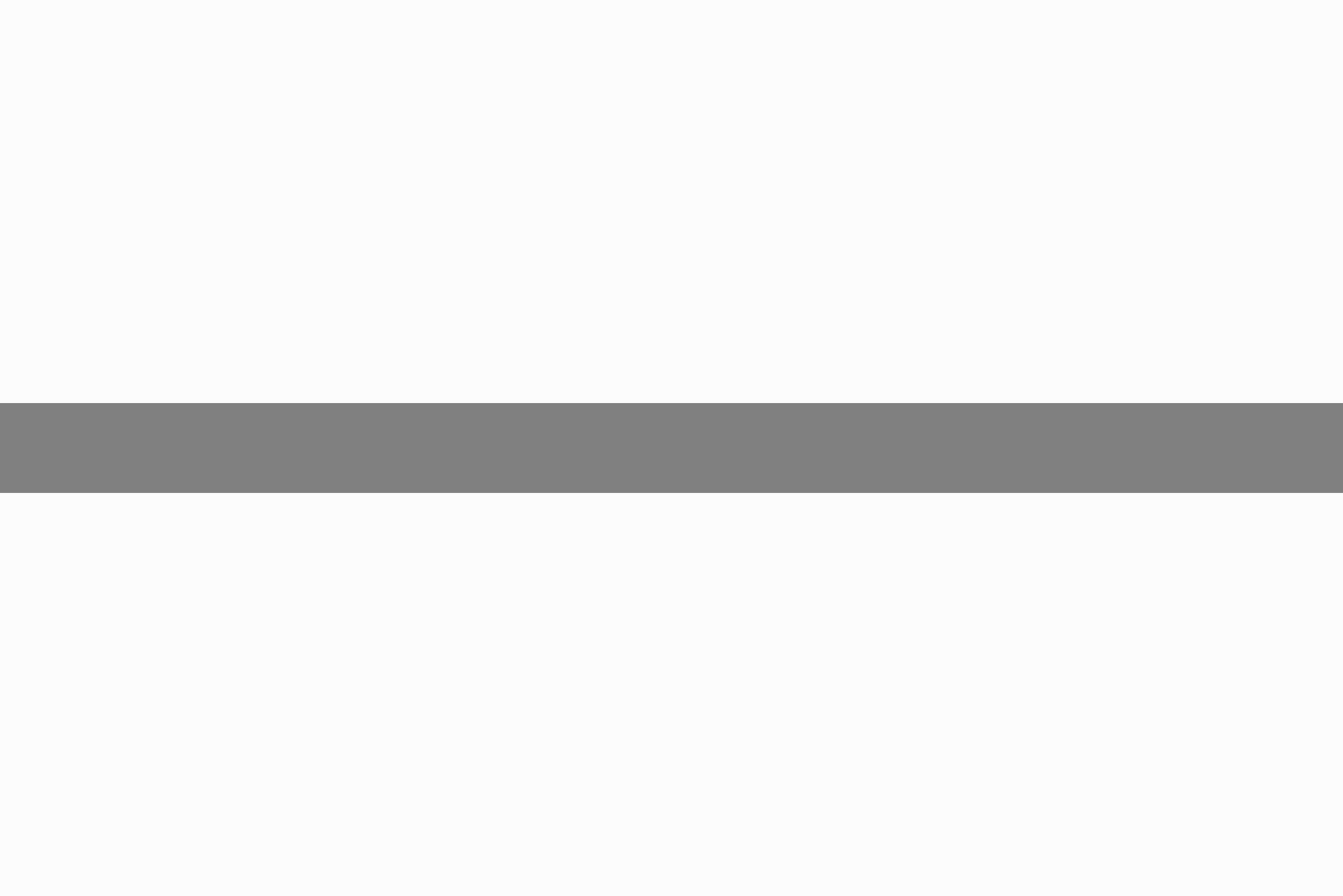}
    \caption{A constant color rectangle covering 10\% of the area in front of a white background. }
    \label{WhiteBGConstantColor/010}
 \end{subfigure}
 \hfill
     \begin{subfigure}[t]{0.32\textwidth}
\includegraphics[width=0.99\textwidth]{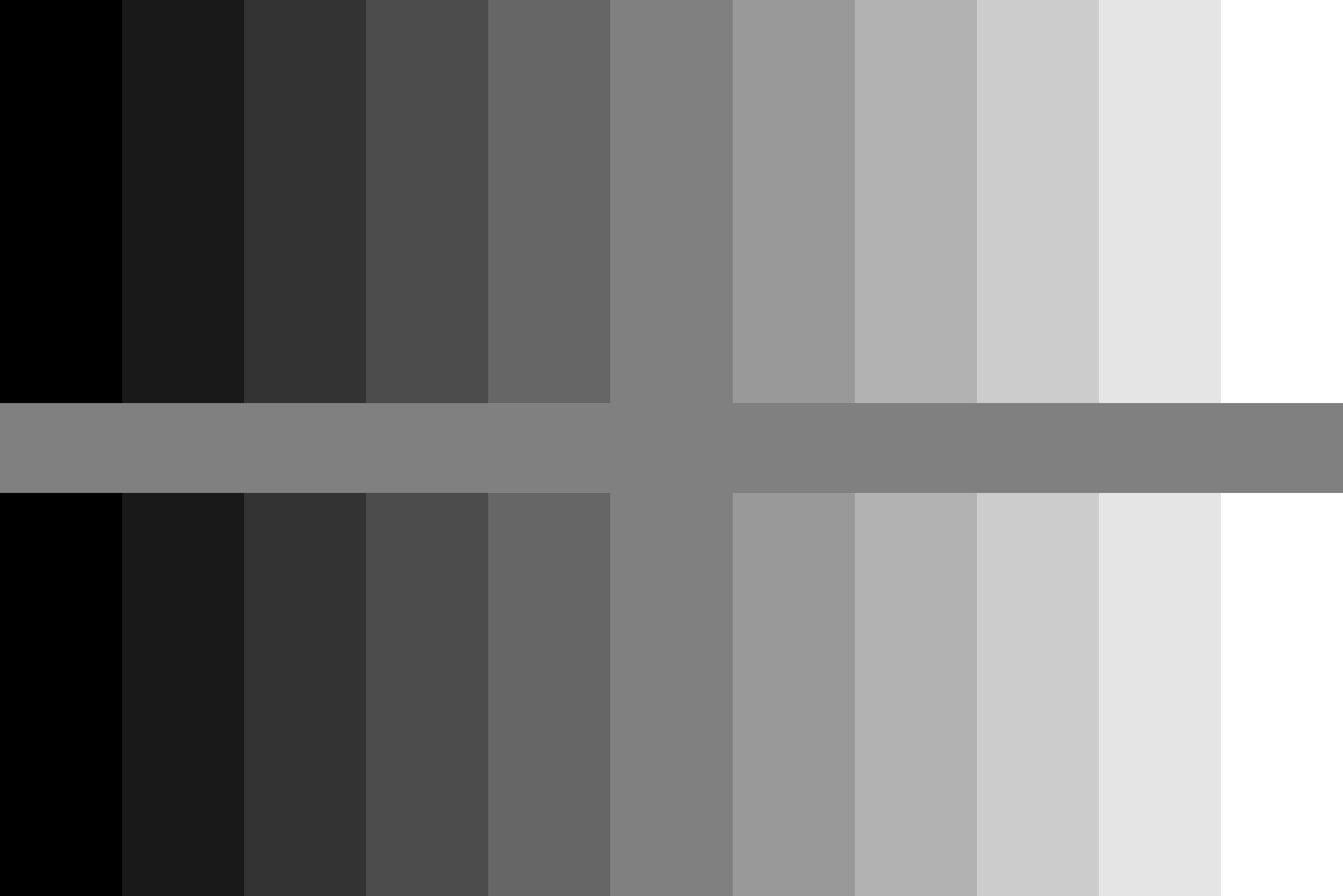}
    \caption{A constant color rectangle covering 10\% of the area in front of a lightness-scale background image that consists of 10 bands. }
    \label{fig_10BandBGConstantColor/010}
 \end{subfigure}
 \hfill
      \begin{subfigure}[t]{0.32\textwidth}
\includegraphics[width=0.99\textwidth]{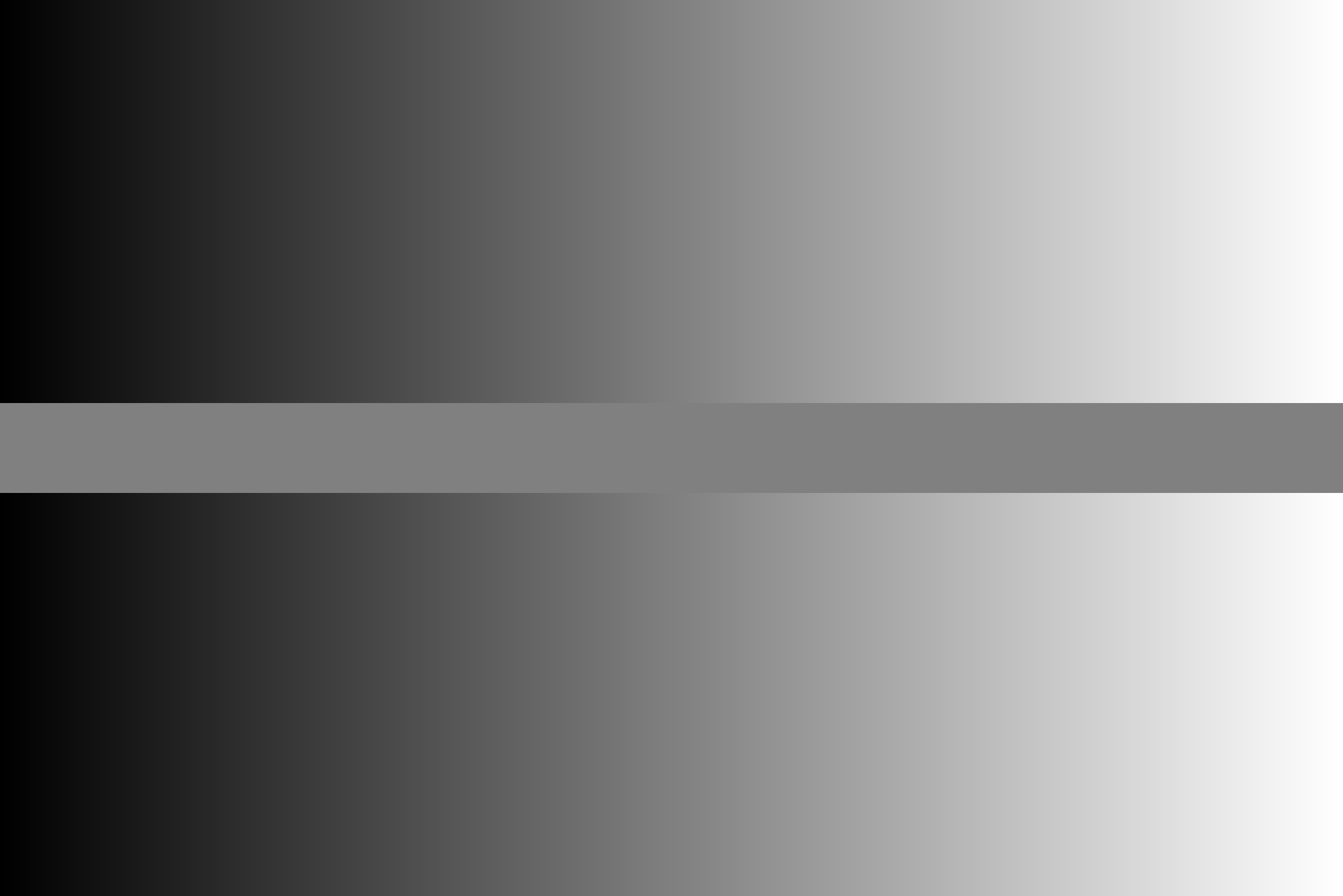}
    \caption{A constant color rectangle covering 10\% of the area in front of a continuous lightness-scale background image. }
    \label{fig_ContinousBGConstantColor/010}
 \end{subfigure}
 \hfill
     \begin{subfigure}[t]{0.32\textwidth}
\includegraphics[width=0.99\textwidth]{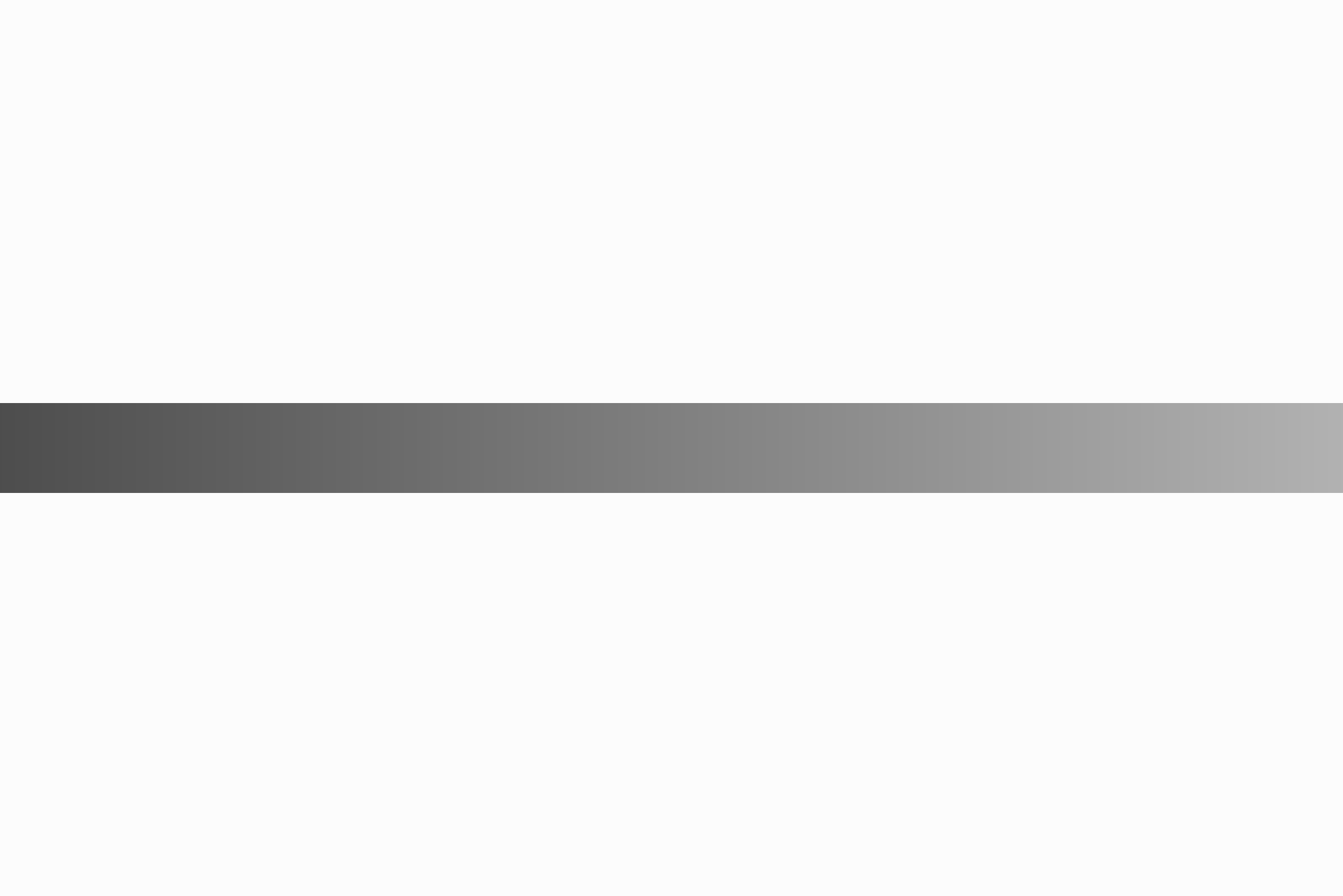}
    \caption{A variable color rectangle covering 10\% of the area in front of a white background. }
    \label{fig_WhiteBGConstantPerception/010}
 \end{subfigure}
 \hfill
     \begin{subfigure}[t]{0.32\textwidth}
\includegraphics[width=0.99\textwidth]{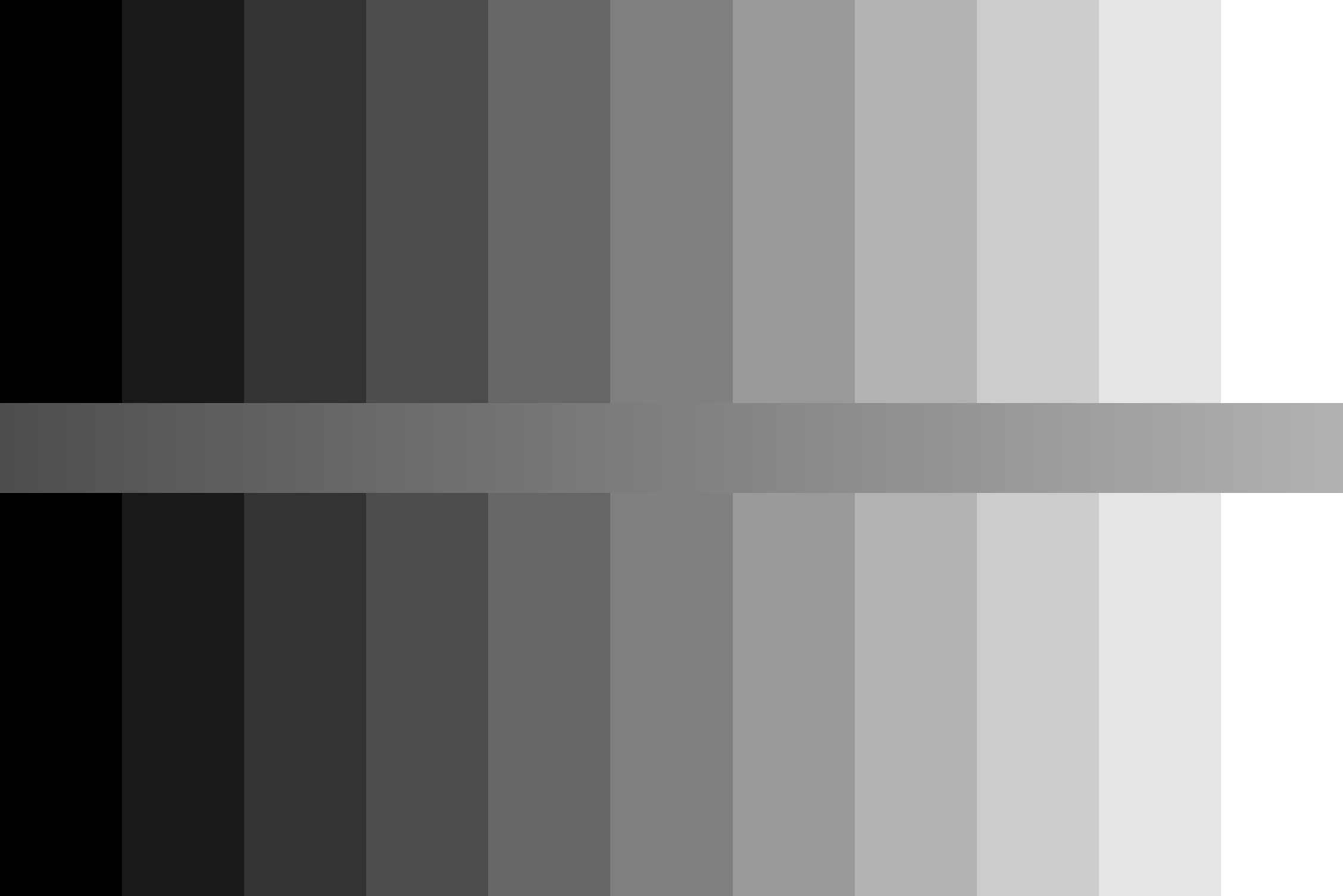}
    \caption{The same rectangle in \ref{fig_WhiteBGConstantPerception/010} in front of a lightness-scale background image that consists of 10 bands.}
    \label{fig_10BandBGConstantPerception/010}
 \end{subfigure}
 \hfill
      \begin{subfigure}[t]{0.32\textwidth}
\includegraphics[width=0.99\textwidth]{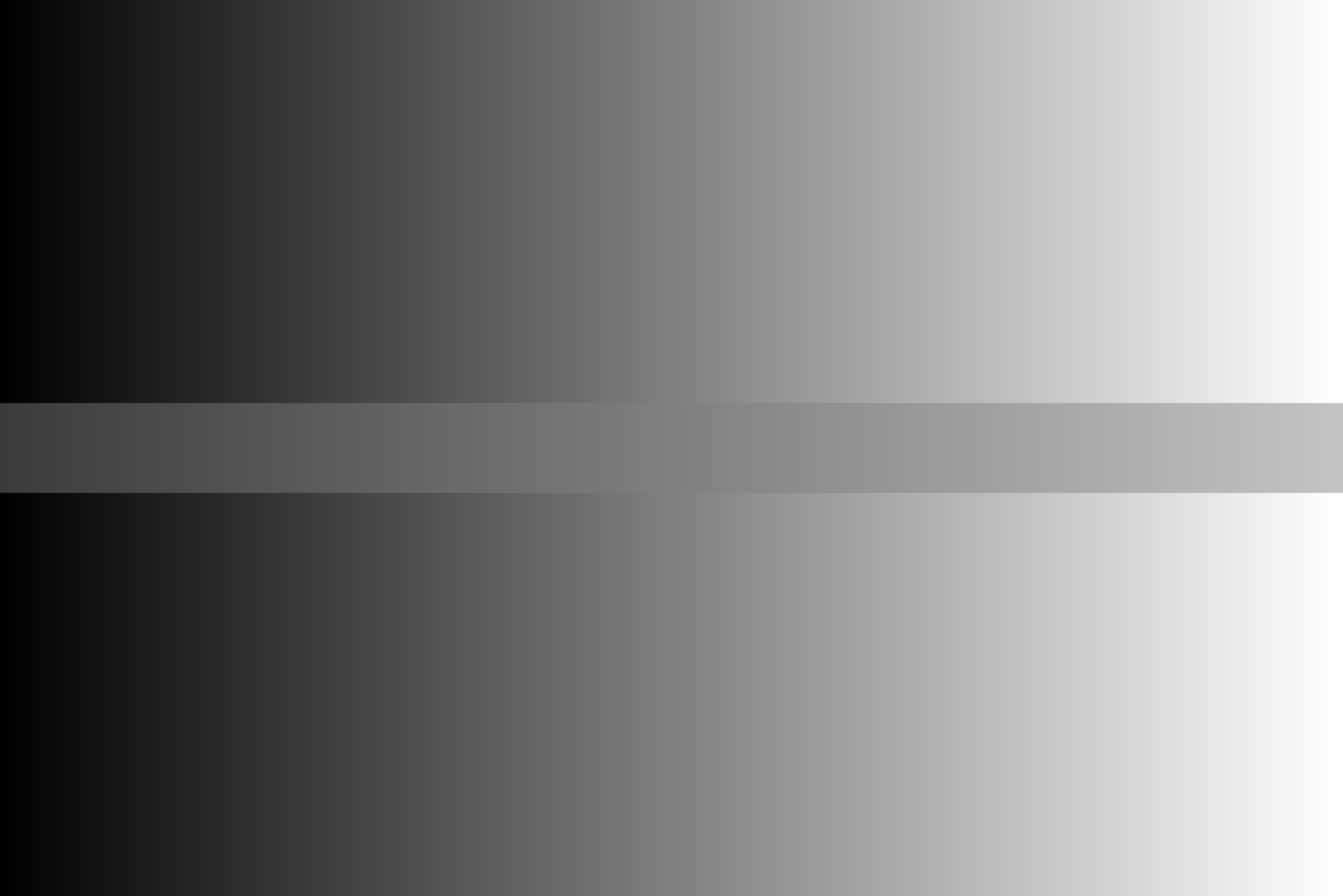}
    \caption{The same rectangle in \ref{fig_WhiteBGConstantPerception/010} in front of a continuous lightness-scale background. }
    \label{fig_ContinousBGBGConstantPerception/010}
 \end{subfigure}
 \hfill 
    \caption{Comparison of our method with constant color. Note that the constant-luminescence rectangle appears to have variable luminosity in front of lightness-scale backgrounds. On the other hand, the rectangle with variable-luminescence rectangle band creates an illusion of constant-luminescence in front of lightness-scale backgrounds.}
\label{fig_010}
\end{figure}

To test our hypothesis, we developed one simple test case by placing a rectangular foreground in front of a standard lightness scale background, as shown in Figure~\ref{fig_010}. We implement this test case in Shadertoy as an interactive program. We can change the size and luminance of the rectangle with the positions of $x$ and $y$ of the mouse. We also control the number of discrete levels in the lightness-scale background image. We can also choose any random image as a background.

Based on our hypothesis, we also conjecture that there must exist a context-sensitive luminance correction formula that can produce a constant luminance effect by making the foreground object slightly transparent to mix with the blurred version of the background. Using this interactive program in Shadertoy, we have adjusted the coefficients of this translucency function and quickly obtained the desired constant-luminance perception for all colors and sizes of the rectangles. 

The results for a specific luminance and size are shown in Figure~\ref{fig_010}. In this paper, we provide almost all sizes for one specific luminance value, so we can see the effect directly in the article. We have made the code publicly available so that anyone can access our program to test our results by changing all the variables. Users can also change the coefficients of our translucency formula to see if they can further improve our results. We also provide how to intuitively change the polynomial part of the formula, so that users can also change the degree of the polynomial part of the formula to see if higher degree polynomials are necessary. 

\subsection{Context \& Motivation}

Josef Albers in 1949 stated that "Every perception of color is an illusion. [...] We do not see the colors as they really are. In our perception, they alter [each other]."\cite{albers1949homage}. To understand how colors alter each other, Albers produced more than two thousand images for his Homage to the Square series between 1950 and 1976 \cite{kooning1950albers}. This is exactly the problem for studying perception-related problems. The parameter space can be quite large, and therefore it is difficult to design user studies that can effectively cover such large design spaces. As a result, most studies deal with some small set of parameters, and they can only help to prove the existence of some specific perceptual effects. 

We think instead that we need an approach similar to that of Albers. He viewed his method as unscientific, saying that ``science aims to solve the problems of life, [whereas] art depends on unsolved problems \cite{kooning1950albers}.'' 
On the other hand, we observe that Albers' approach is essentially scientific, resembling the discovery process in some scientific fields such as chemistry or physics. In such fields, we first perform a large number of experiments to collect data. The experts then make several bold claims in terms of mathematical theory to explain some behavior in the data produced\footnote{Some of these mathematical theories may even help to explain some behavior, which is not yet known, and may eventually create a paradigm shift in the field \cite{kuhn1962structure}.} 

In this paper, we present an art-based scientific approach to formulate the perception of luminosity. We hypothesized that luminance perception in (human and other) visual systems must have evolved to identify the brightness values of the diffuse reflection properties of foreground objects. Our hypothesis is simple and closely related to intrinsic image analysis. Actual luminance values are useless for differentiating objects. Animals who cannot identify materials cannot differentiate objects and cannot recognize prey and predators hiding in a shadow region. Such animals cannot survive to pass on their defective genes to the next generations. As a result, surviving animals are more likely to recognize the most basic material property, which is the diffuse reflection term, instead of the actual luminance values. 

\subsection{Basis and Rationale}
\label{BasisAndRationale}

Note that the diffuse reflection term is always a percentage; therefore, it is a number between $0$ and $1$. These percentages can be represented by very low dynamic ranges with a limited number of levels. For example, an example of a very low dynamic range can be given in four levels, such as $\mbox{Black} \approx 0/3$, $\mbox{Dark} \approx 1/3$, $\mbox{Light} \approx 2/3$, and $\mbox{White} \approx 3/3$. 
Note that all this particular primitive animal needs to do is differentiate these four potential diffuse reflection terms. The better the visual system, the number of levels it uses can increase; but the basic concept does not change. There will always be a need to differentiate a limited number of different diffuse reflection terms\footnote{In the examples in this paper, we use 256 different levels based on standard low-dynamic range (LOD) gray-level colors.}. 

We observe that this differentiation is essentially a qualitative process \cite{forbus1984,wang2014qualitative}; therefore, it does not have to be very complicated or precise. Animals only need to perceive the luminosity values in shadow regions as brighter than they appear, and vice versa. Therefore, the perception process can be extremely simple to handle even for primitive brains. There is only a need for a simple, yet logical operation that can estimate the diffuse reflection terms of the actual material properties. 

A precise calculation of the actual diffuse reflection of a particular requires a complex illumination process that includes multiple diffuse reflections that can produce radiosity effects such as soft shadows, color bleeding, and ambient occlusions \cite{goral1984modeling}. On the other hand, it is possible to ignore all these complicated effects and consider the brightness part of the diffuse reflection of a given point as just a monotonically increasing function of the average illumination of that point \cite{gouraud1998continuous}. This average can include some radiosity \cite{goral1984modeling} and some subsurface scattering and transmittance effects \cite{hanrahan2023reflection}. Fortunately, estimating the average illumination is not difficult: we can simply calculate it as the average luminance in regions around that particular point. 

One of our key hypotheses is that the visual systems of animals evolved to estimate average values in any given region, which can be simply computed using a blur operation. If we assume that an animal brain can compute the average illumination around a given point, the same brain also needs to construct a monotonically increasing function to qualitatively estimate the diffuse reflection term of the actual material properties. Note that the construction of functions that increase monotonically is essential to identify qualitative similarities \cite{forbus1984,wang2014qualitative}. 

In this article, we focus on estimating the basic structure and coefficients of monotonically increasing functions that can be used to differentiate a limited number of diffuse reflection terms. However, there is a caveat in this framework. This is an inverse process, whose correctness should have emerged through an evolutionary process. We, therefore, cannot verify the correctness of the monotonically increasing function through a similar inverse process since we cannot know the ground truth for diffuse terms. On the other hand, we can construct another (i.e. forward) monotonically increasing function that can produce constant luminance perception. The forward function can later be used to compute the inverse function.

\subsection{Contributions}

Our main contribution in this article is to frame the problem as a forward problem and to use the forward problem to construct a monotonically increasing forward function. We hypothesized that the monotonically increasing forward function can be formulated using a barycentric formula that can be calculated as a weighted average of the average illumination and brightness value of the diffuse reflection property. This formula is not physically based and is as rudimentary as possible, reflecting the simplicity of primitive visual systems. The formula in Equation~\ref{Eq_barycentric} increases monotonically simply by stating that the observed luminance becomes brighter if there is more illumination and vice versa. 

Our second contribution is the structure of the weighted average function, which works like translucency. We realized that the relative size of the foreground object must be a parameter. If it is extremely small, it should practically disappear. This means that its weight has to be zero. If the foreground covers the whole field of view, its weight has to be one. This suggests that the weight function has to be in the form of a power function as $y=s^f(s)$ where $x \in [0,1]$ is the relative size of the foreground object, $f(s)$ can be any polynomial and $y$ is the weight function that can be viewed as the opacity of the foreground object. We conjecture that the evolutionary process can produce such a function. The polynomial part of the formula can be different for different animals, human groups, individual humans, etc. 

Our third contribution is the development of a method to determine the polynomial $f(s)$, part of the equation of $y$. For this purpose, we have implemented a Web-based interactive program in Shadertoy. Three of us determined the coefficients of the polynomial $f(s)$ together. We need to point out that all the authors of the article have art experience. The coefficients can be different for others; however, we expect that the structure of the overall weighted average function will be the same. 

\begin{figure}[hbtp!]
\centering
\includegraphics[width=0.49\textwidth]{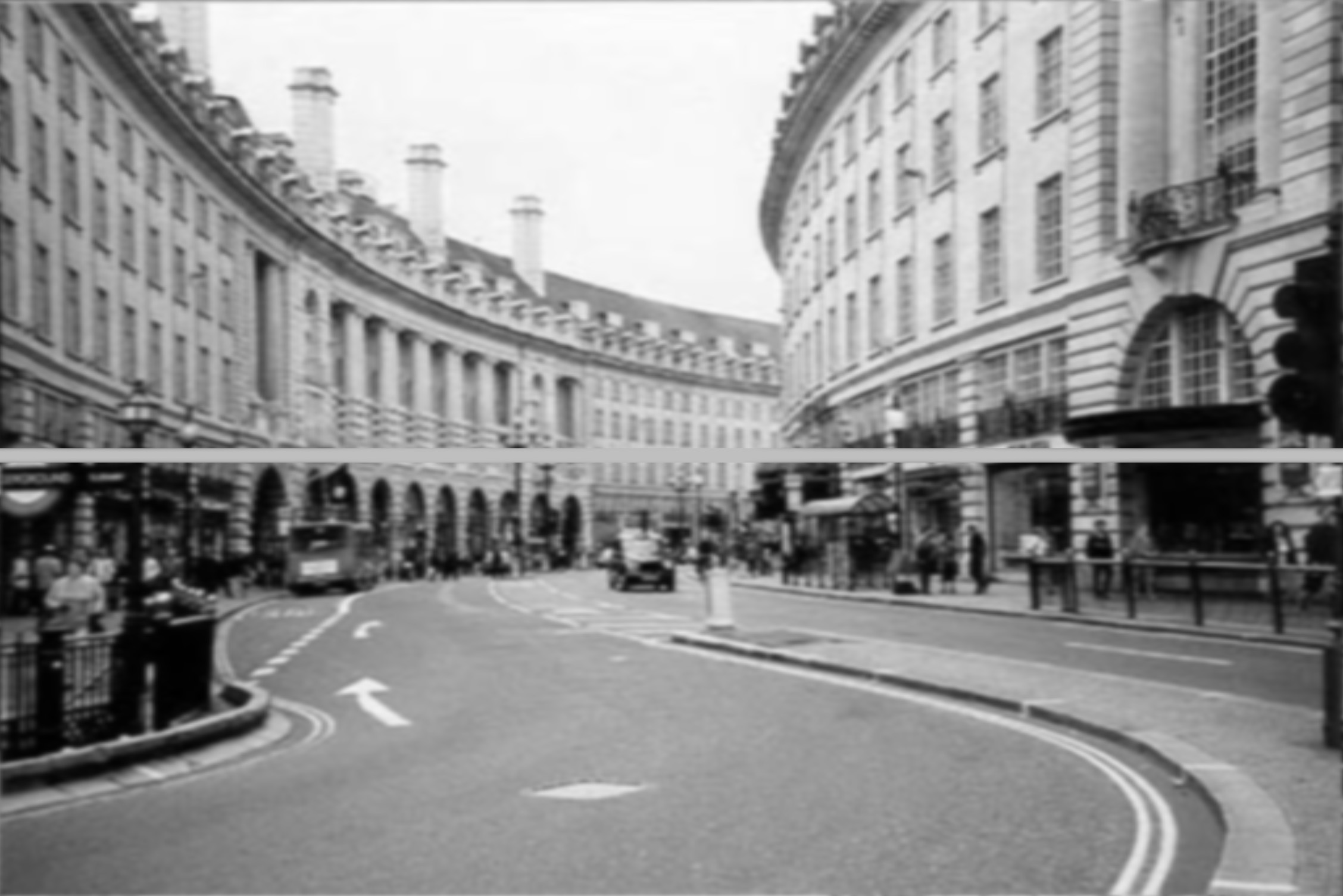}
\includegraphics[width=0.49\textwidth]{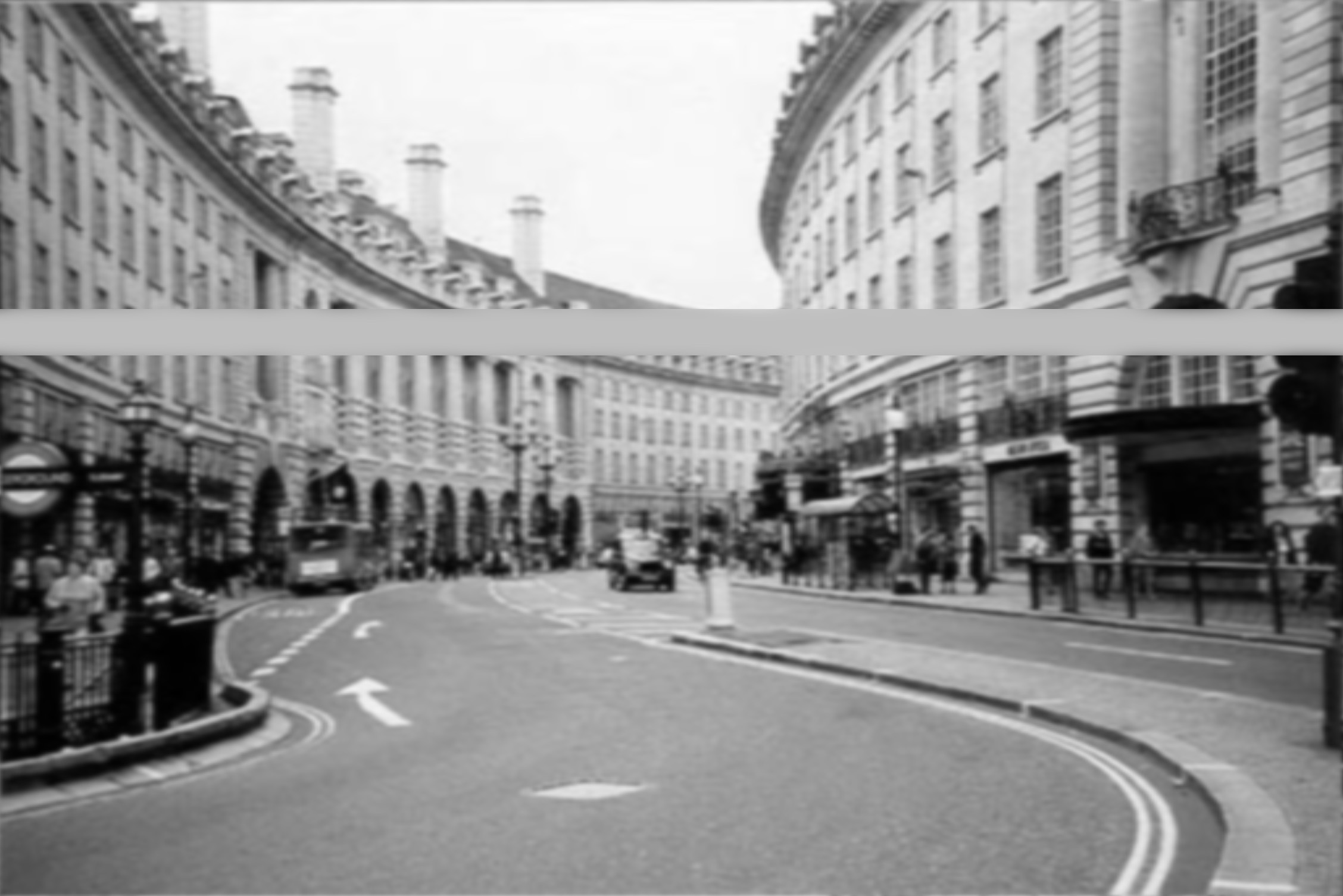}
\includegraphics[width=0.49\textwidth]{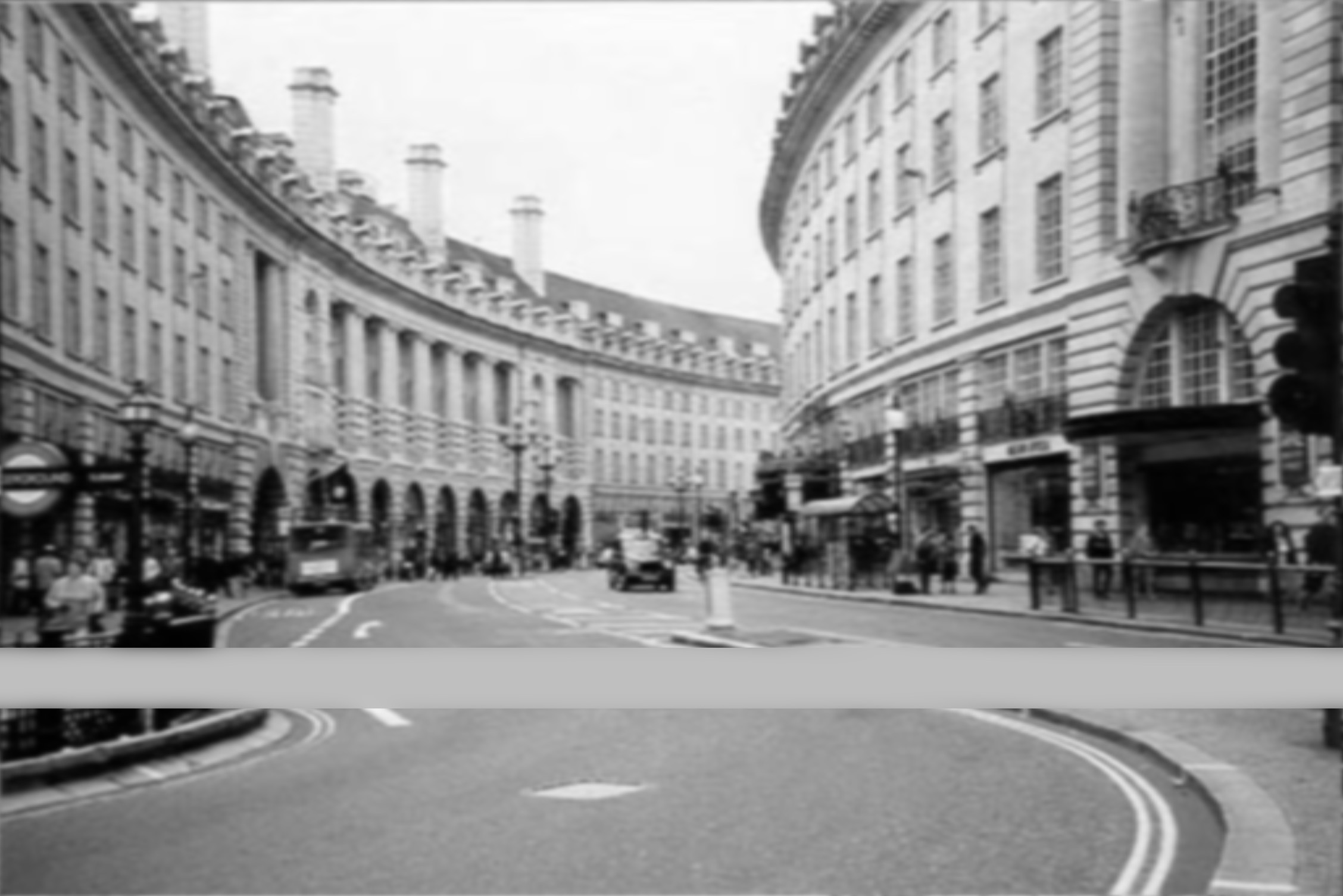}
\includegraphics[width=0.49\textwidth]{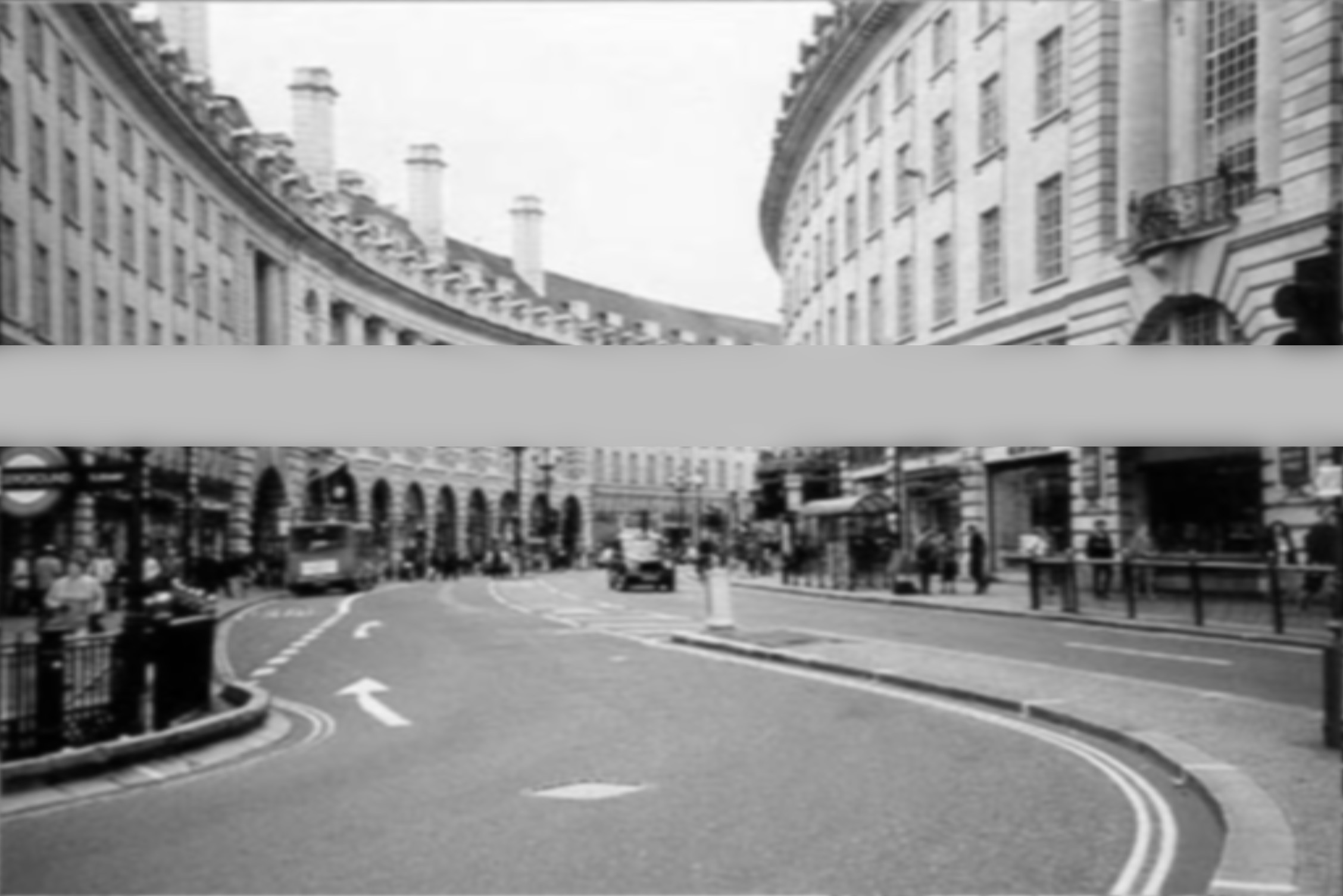}
\includegraphics[width=0.49\textwidth]{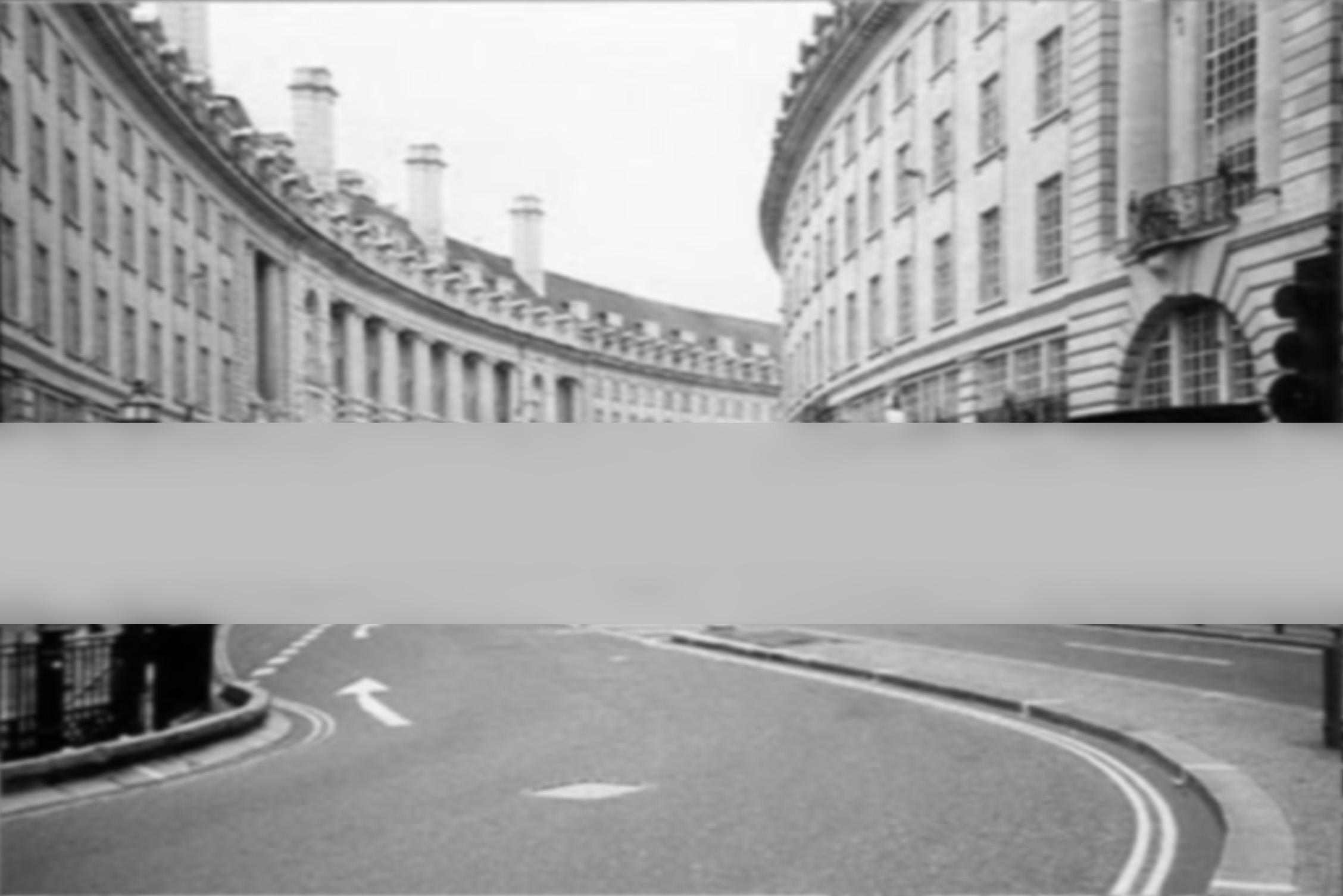}
\includegraphics[width=0.49\textwidth]{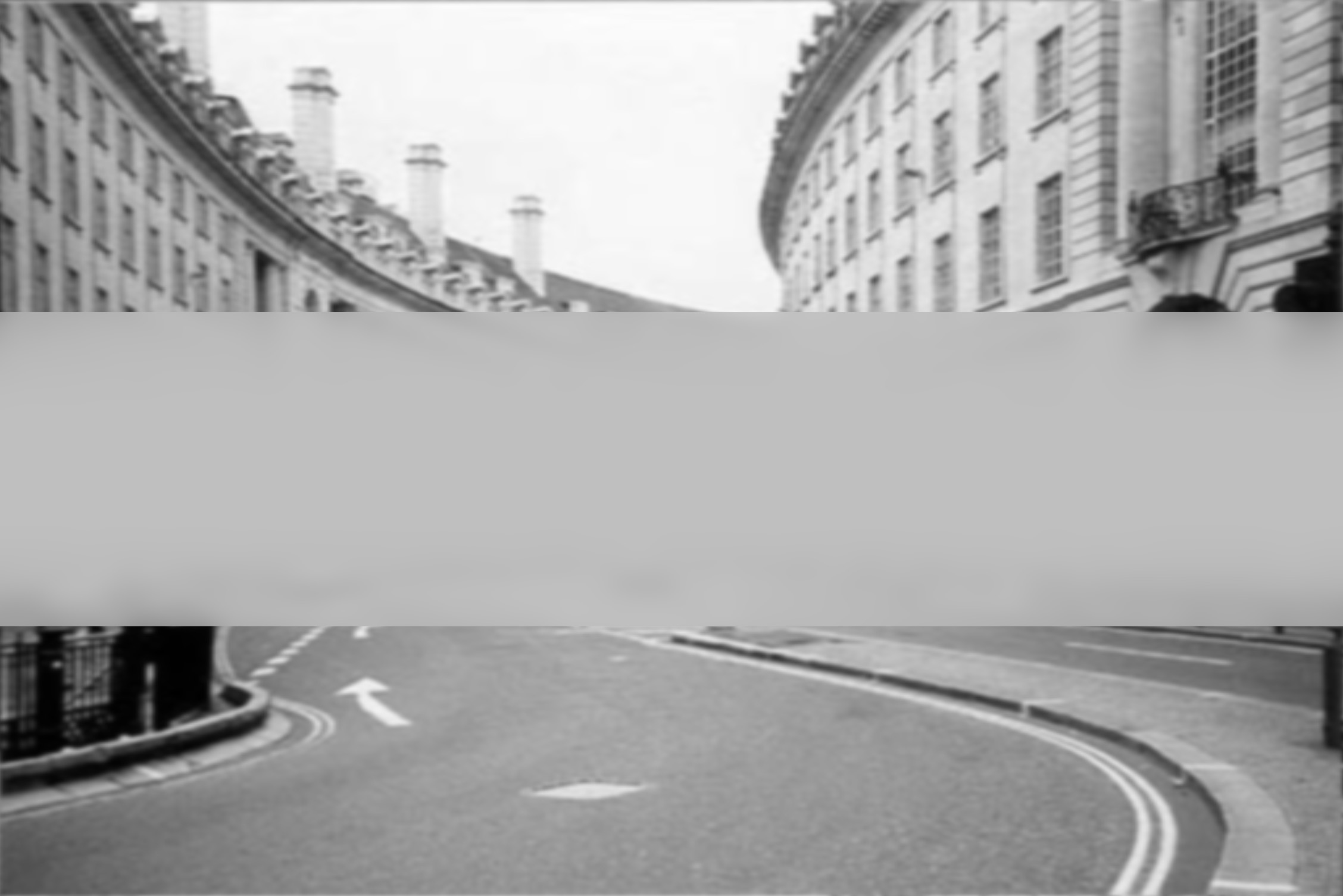}
    \caption{A foreground rectangle in front of a photograph. We made the center line of the rectangle completely opaque to avoid an appearance of translucency as if we were looking at a shower door. Only the boundaries are translucent with the Equation given in~\ref{Eq_barycentric_final_linear}. This creates an effect similar to subsurface scattering where the center is thicker and the boundaries are thinner. }
\label{fig_Vienna}
\end{figure}

Our fourth contribution is to enable others to determine the coefficients of the polynomial function by testing their perception of luminance. To provide intuitive control, we have used a B\'{e}zier form and explain how to change the coefficients intuitively. Using our program publicly available in Shadertoy, anyone can access the program and change the polynomial part of the formula to obtain their own perceptively constant luminance. This can be used as a crowd-sourcing experiment for further evaluation to identify the existence of individual and group differences based on culture, geography, gender, or any other classification. 

Our fifth and last contribution is the observation that for given coefficients, $y=s^f(s)$ resembles an affine equation when we exclude very small values of $s$. Based on this observation, we realized that it is possible to replace the equation of $y$ with a first-degree polynomial in the form of $y=as+b$ or $y=a_0(1-s)+a_1 s$. These formulas also provide constant luminance perception for the appropriate coefficients. The method works with photographs, but the center of the objects must be made opaque to avoid an effect of the shower door (see Figure~\ref{fig_Vienna}). Making the center opaque and the boundaries translucent creates an effect that is visually similar to seamless cloning \cite{perez2003poisson,agarwala2007efficient,farbman2009coordinates}. We think that combining with photographs requires more work by employing seamless cloning methods.

\section{Related Work}

Visual illusions have long been a topic of interest in the study of perception. A particularly interesting type of visual illusion is the perception of luminosity \footnote{We want to point out that some researchers prefer the term ``lightness perception \cite{gilchrist1999an}.'' We think it is more appropriate to call this perception of luminosity since luminosity is a physical entity. Its perception should be called lightness. }.  Gilchrist provides an excellent overview of research on luminosity perception and its history \cite{gilchrist2006seeing}. An important observation of Gilchrist is the inherent ambiguity in determining diffuse reflection terms since diffuse reflections are functions of illumination, which is also hard to determine \cite{gilchrist1977perceived, gilchrist1980does, gilchrist1999an}. The problem is even harder, since many other factors, such as surface normals, incoming light directions, and occlusions, can impact diffuse reflections. Therefore, intrinsic image analysis provides a strong conceptual framework, since it is based on relative luminance\cite{yarbus1967eye, whittle1969effect, shapley1984visual, bonato1994perception}. 

Our approach is also rooted in intrinsic image analysis.
Our conjecture in this article is that the structure of luminance perception evolved through an evolutionary process that began with the earliest visual systems in animals. Intrinsic image analysis appears to be based on the same underlying hypothesis. However, there is an important caveat. Intrinsic image analysis requires high-level operations such as the creation of intrinsic images \cite{chen2013simple, shen2013intrinsic, li2018learning} and ground truth \cite{grosse2009ground}. Handling such high-level operations remains a difficult problem in computer vision research, even with advanced computer vision techniques \cite{garces2022survey}. 

We believe that there must be no need for high-level operations such as the creation of intrinsic images, since the process is complicated for primitive brains. There must exist a very simple process that has naturally emerged from interactions between prey and predators or between animals and plants to find suitable food or avoid being killed \cite{regan2001fruits}. Survival requires the identification of animals and plants under all lighting conditions. This identification requires differentiation of the diffuse reflection terms of materials under different illumination conditions. In other words, we view the structure of luminance perception as a universal process that must extend even beyond human beings \footnote{Although, the idea could be correct for animals, one of the limitations is that this study can only be tested for human vision.}. 

On the other hand, we also recognize that there can still be geography-dependent or culturally-based perception differences in human (or animal) societies in terms of coefficients of the basic equation. These geographically or culturally dependent differences could also arise from parts of the formulas that are not essential for survival. We provide some examples of culturally based differences in perception between human societies. It can be a good idea to test if there are significant differences in the coefficients of our formula for culturally or geographically different groups of humans.

Therefore, we want to point out that our universality claim does not contradict some of the recent research that observed the role of cultural factors in shaping the way individuals perceive these illusions. The perception of \textit{certain} visual illusions that are not necessarily crucial for survival may not be universal, but is influenced by cultural factors. Understanding these differences is crucial for a comprehensive understanding of visual perception and has implications for fields such as psychology, design, and intercultural communication.

We provide an overview of the current understanding of cultural differences in visual perception through visual illusion perception, focusing on four main aspects: contextual influence, field dependence, experience with the environment, and depth perception.

\textit{Contextual Influence:} Western cultures, characterized by analytical thinking and individualism, tend to focus on individual elements in visual scenes. In contrast, East Asian cultures, known for their holistic thinking and collectivism, emphasize the context and relationships between elements. This difference in cognitive style affects the perception of illusions, such as illusions \cite{nisbett2005influence}.

\textit{Field Dependence} Holistic cognitive styles, common in East Asian cultures, are associated with greater field dependence. This means that individuals from these cultures are more influenced by the surrounding visual field, which impacts their susceptibility to certain illusions \cite{ji2000culture}.

\textit{Experience with the Environment} The environment in which a person is raised can influence their perception of illusions. For example, those raised in urban environments with many man-made structures may be more susceptible to geometric illusions than those raised in natural settings \cite{segall1966influence}.

\textit{Depth Perception}
Cultural differences in depth perception have been observed, with variations in susceptibility to illusions such as the Ponzo illusion. These differences are attributed to the variations of experiences with visual cues for depth \cite{deregowski1980illusions}.

\section{Theoretical Framework}
\label{Sec_TheoreticalFramework}

Let $L_P$ denote the perceived luminance of the given point, $L_O$ denote the actual observed luminance under average illumination, and $I_a$ denote the average illumination around that particular point; then we assume that the formula is of the following form. 
\begin{equation}L_O = y L_P + (1-y) I_a \label{Eq_barycentric}\end{equation}
where the weighted average $y$ can be considered as opacity. Note that this formula is not physically based and is as rudimentary as possible reflecting the simplicity of primitive visual systems. The formula in Equation~\ref{Eq_barycentric} increases monotonically simply by stating that the observed luminance becomes brighter if there is more illumination and vice versa. 

\begin{figure}[hbtp]
\centering
     \begin{subfigure}[t]{0.23\textwidth}
\fbox{\includegraphics[width=0.99\textwidth]{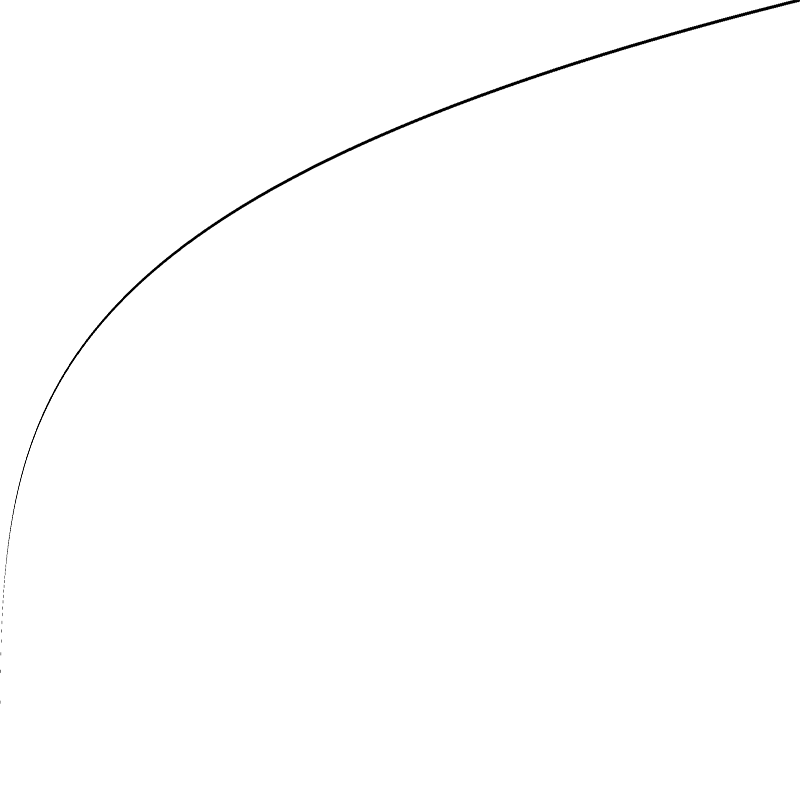}}
    \caption{$y=s^{0.25}$ }
    \label{functions/0.25}
 \end{subfigure}
 \hfill
     \begin{subfigure}[t]{0.23\textwidth}
\fbox{\includegraphics[width=0.99\textwidth]{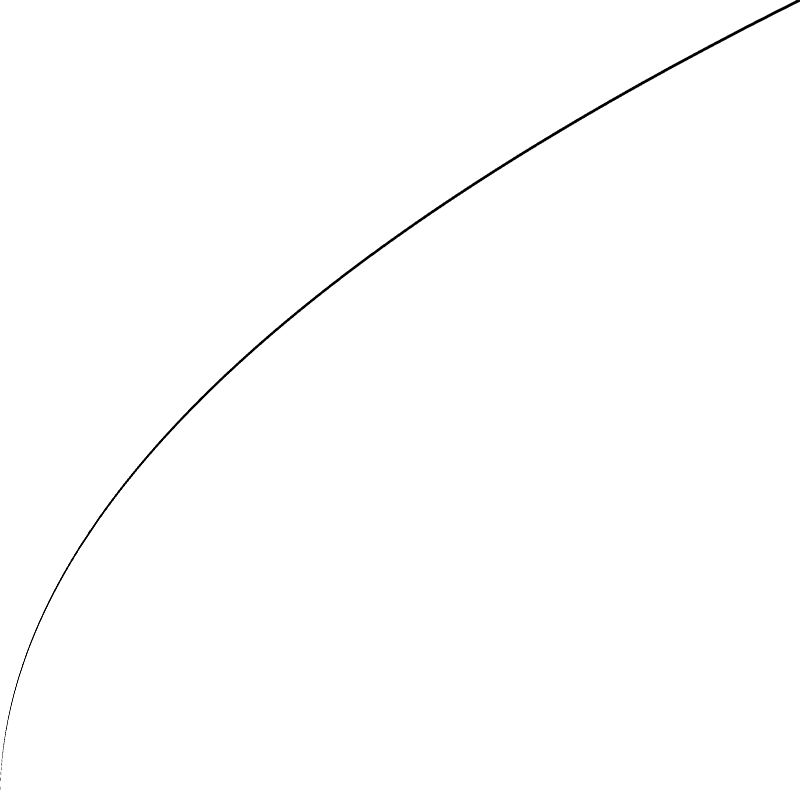}}
    \caption{$y=s^{0.5}$ }
    \label{functions/0.5}
 \end{subfigure}
 \hfill
      \begin{subfigure}[t]{0.23\textwidth}
\fbox{\includegraphics[width=0.99\textwidth]{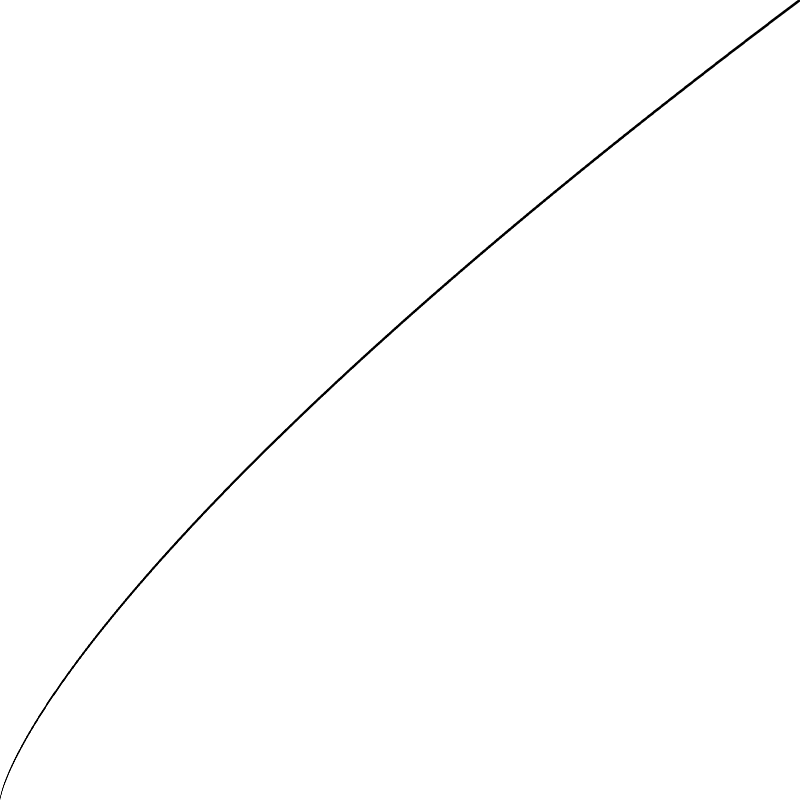}}
    \caption{$y=s^{0.75}$ }
    \label{functions/0.75}
 \end{subfigure}
 \hfill
     \begin{subfigure}[t]{0.23\textwidth}
\fbox{\includegraphics[width=0.99\textwidth]{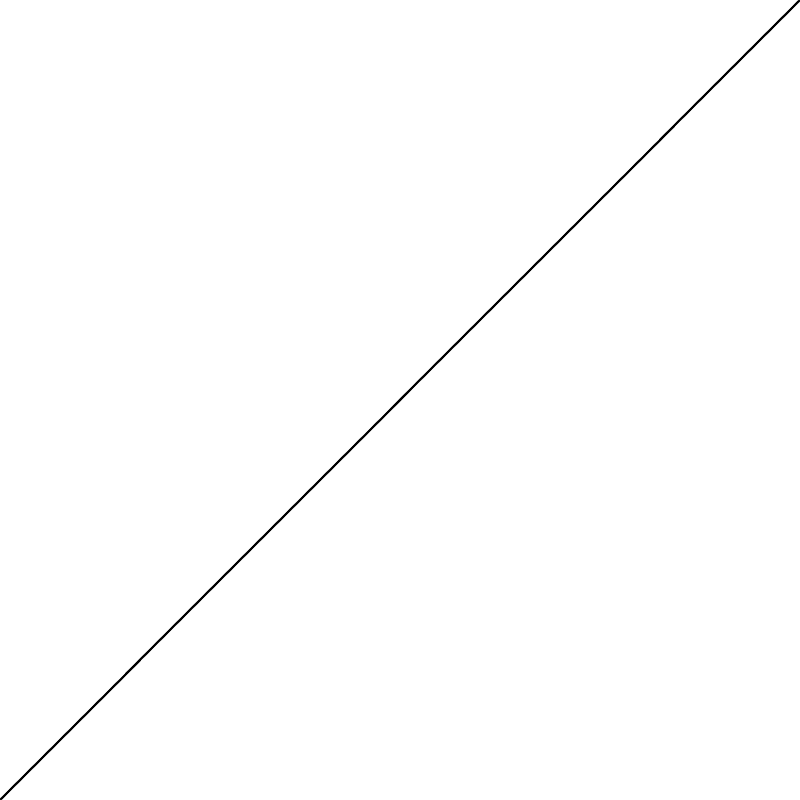}}
    \caption{$y=s^1$ }
    \label{functions/1}
 \end{subfigure}
 \hfill
     \begin{subfigure}[t]{0.23\textwidth}
\fbox{\includegraphics[width=0.99\textwidth]{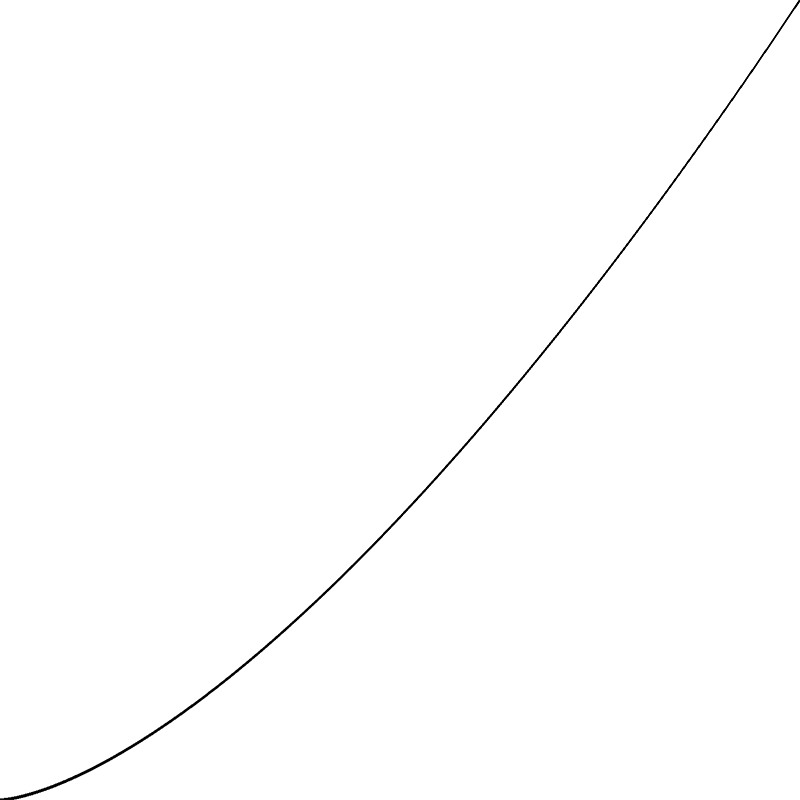}}
    \caption{$y=s^{1.5}$}
    \label{functions/1.5}
 \end{subfigure}
 \hfill
    \begin{subfigure}[t]{0.23\textwidth}
\fbox{\includegraphics[width=0.99\textwidth]{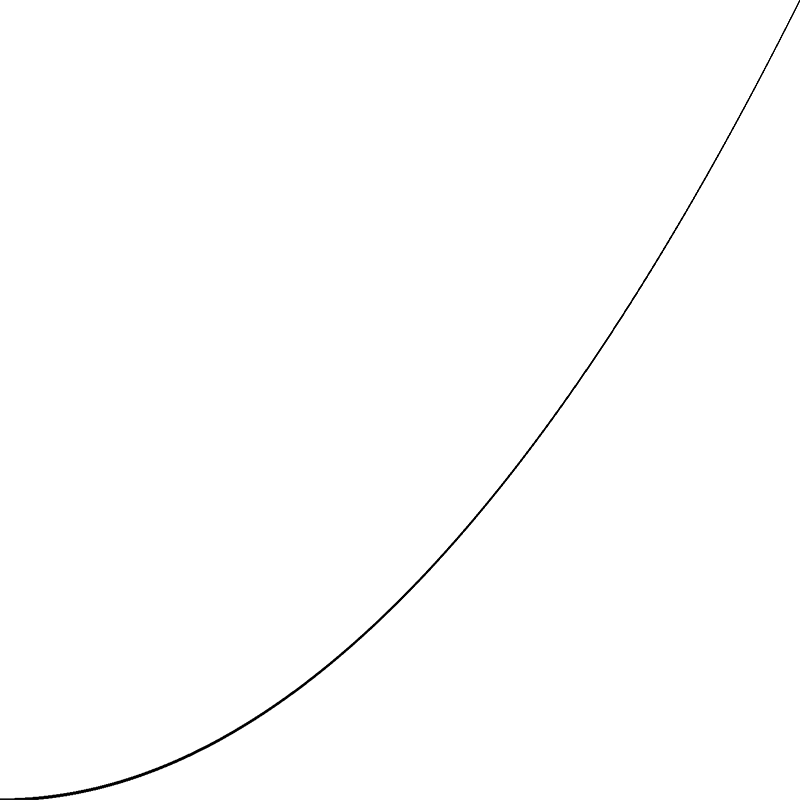}}
    \caption{$y=s^2$ }
    \label{functions/2}
 \end{subfigure}
 \hfill
     \begin{subfigure}[t]{0.23\textwidth}
\fbox{\includegraphics[width=0.99\textwidth]{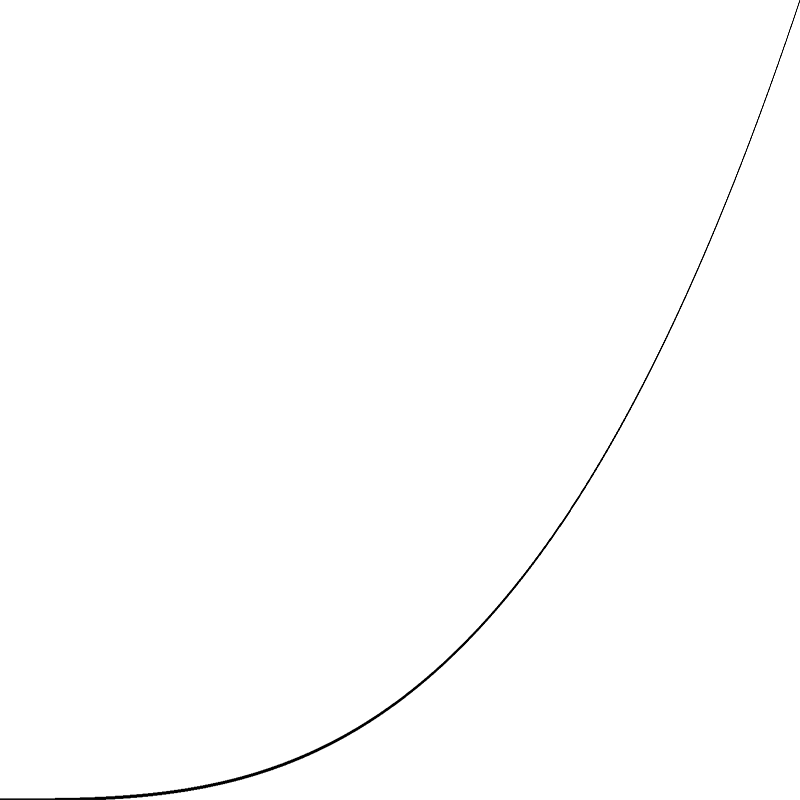}}
    \caption{$y=s^3$ }
    \label{functions/3}
 \end{subfigure}
 \hfill
      \begin{subfigure}[t]{0.23\textwidth}
\fbox{\includegraphics[width=0.99\textwidth]{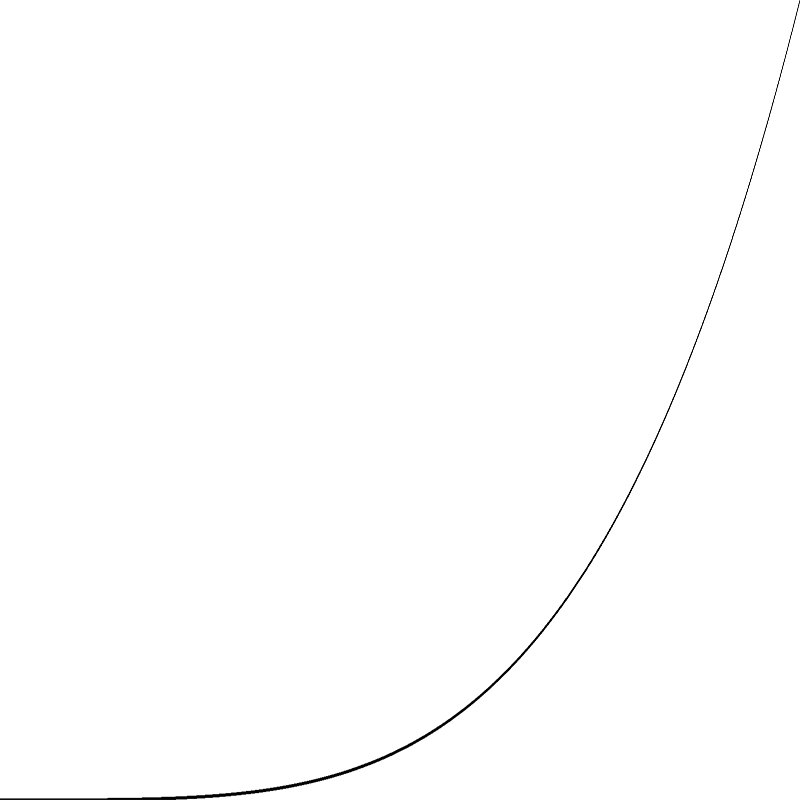}}
    \caption{$y=s^4$ }
    \label{functions/4}
 \end{subfigure}
 \hfill
    \caption{$y=s^{b_{00}}$ functions: zero degree polynomials as power term.}
\label{fig_functions_constant}
\end{figure}

\begin{figure}[hbtp]
\centering
     \begin{subfigure}[t]{0.23\textwidth}
\fbox{\includegraphics[width=0.99\textwidth]{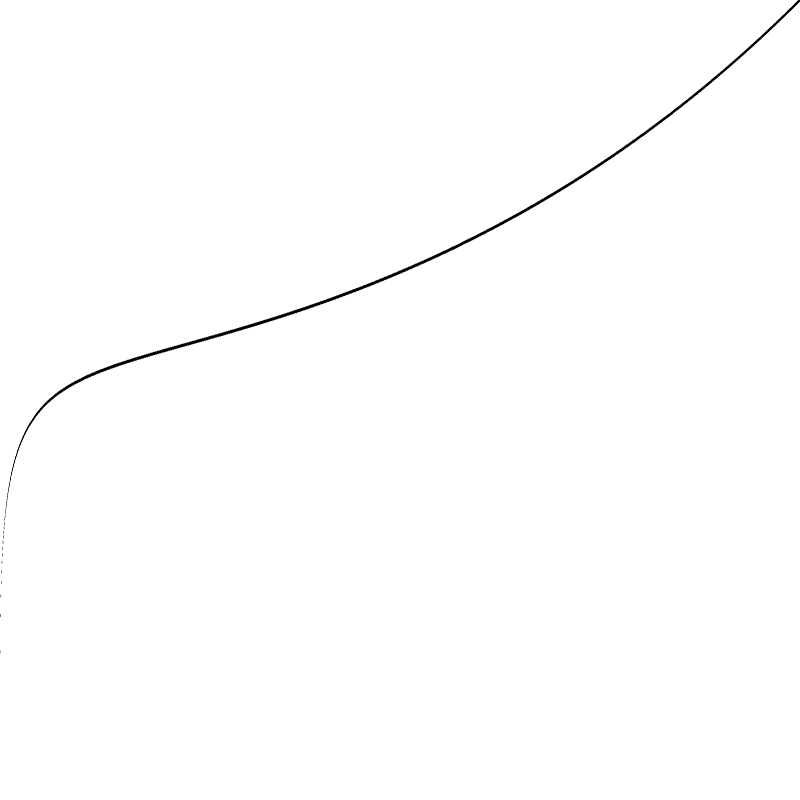}}
    \caption{$y=s^{0.2(1-s)+1.0s}$ }
    \label{functions/02b010b1}
 \end{subfigure}
 \hfill
     \begin{subfigure}[t]{0.23\textwidth}
\fbox{\includegraphics[width=0.99\textwidth]{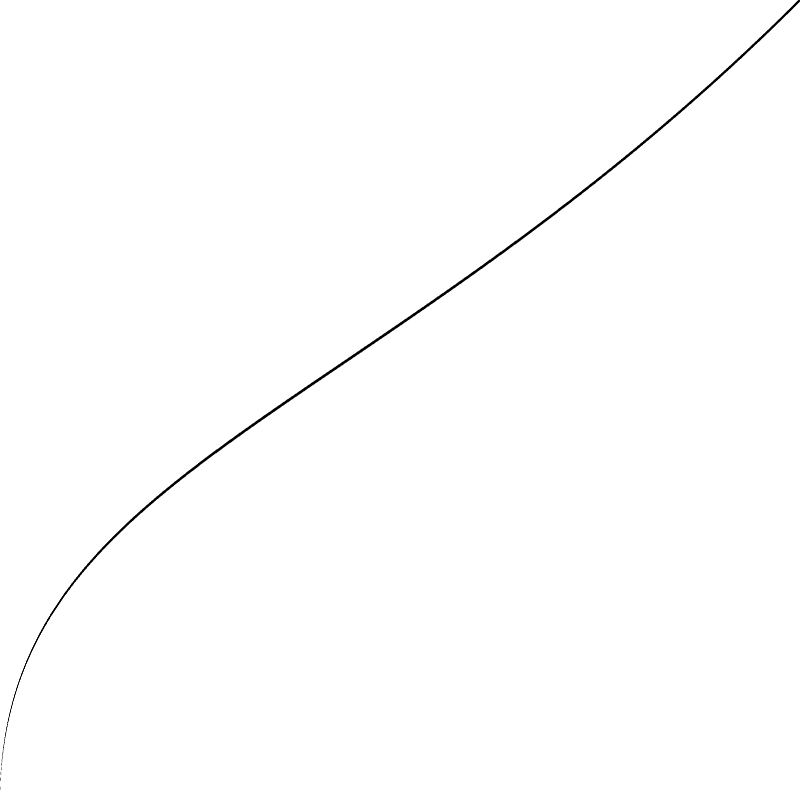}}
    \caption{$y=s^{0.5(1-s)+1.0s}$ }
    \label{functions/05b010b1}
 \end{subfigure}
 \hfill
      \begin{subfigure}[t]{0.23\textwidth}
\fbox{\includegraphics[width=0.99\textwidth]{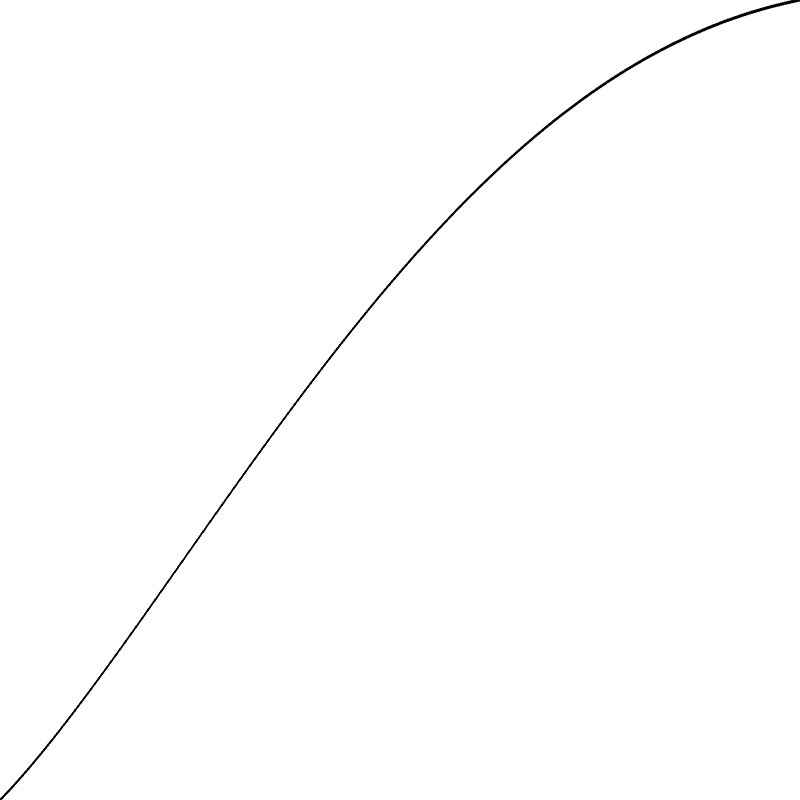}}
    \caption{$y=s^{1.0(1-s)+0.2s}$ }
    \label{functions/10b002b1}
 \end{subfigure}
 \hfill
     \begin{subfigure}[t]{0.23\textwidth}
\fbox{\includegraphics[width=0.99\textwidth]{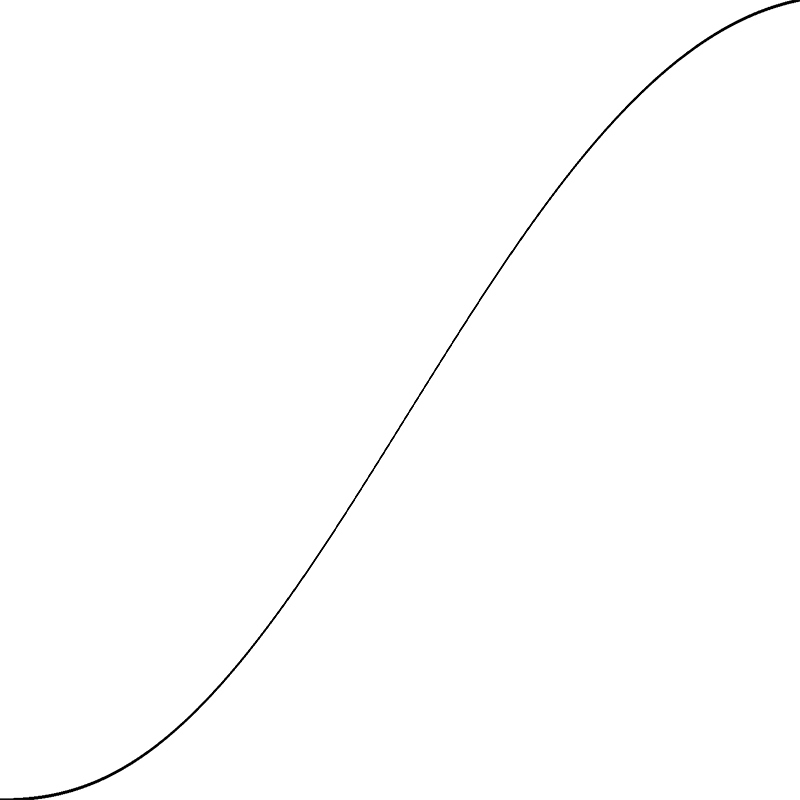}}
    \caption{$y=s^{2.0(1-s)+0.2s}$ }
    \label{functions/20b002b1}
 \end{subfigure}
 \hfill
    \caption{$y=s^{b_{10}(1-s) + b11 s}$ functions: first degree polynomials as power term.}
\label{fig_functions_linears}
\end{figure}

\begin{figure}[hbtp]
\centering
     \begin{subfigure}[t]{0.23\textwidth}
\fbox{\includegraphics[width=0.99\textwidth]{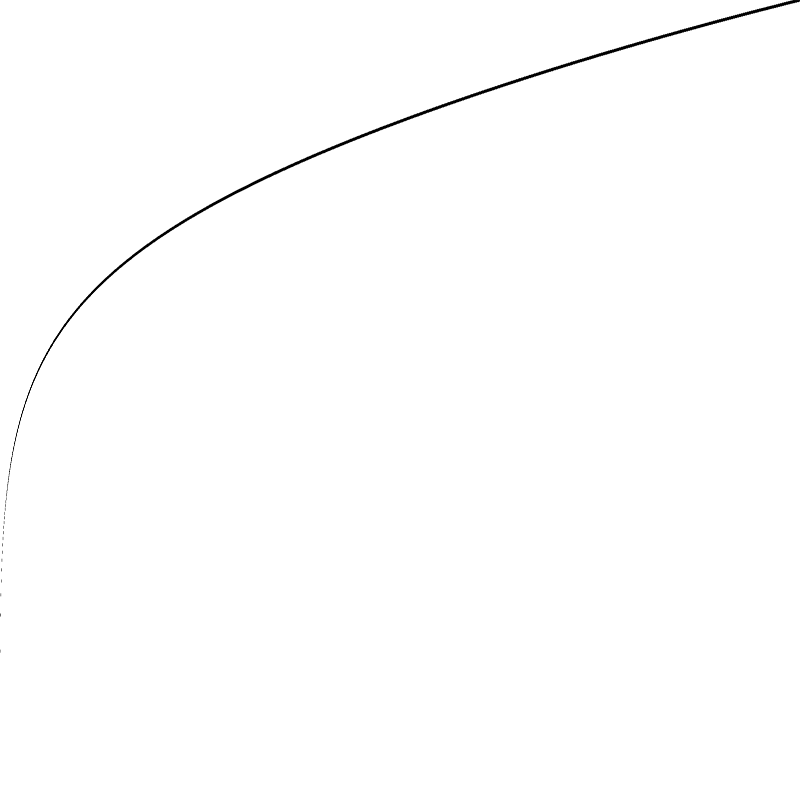}}
    \caption{$b_{20}=0.20$, $b_{21}=0.25$, and $b_{22}=1.0$ }
    \label{functions/020b0025b110b2}
 \end{subfigure}
 \hfill
     \begin{subfigure}[t]{0.23\textwidth}
\fbox{\includegraphics[width=0.99\textwidth]{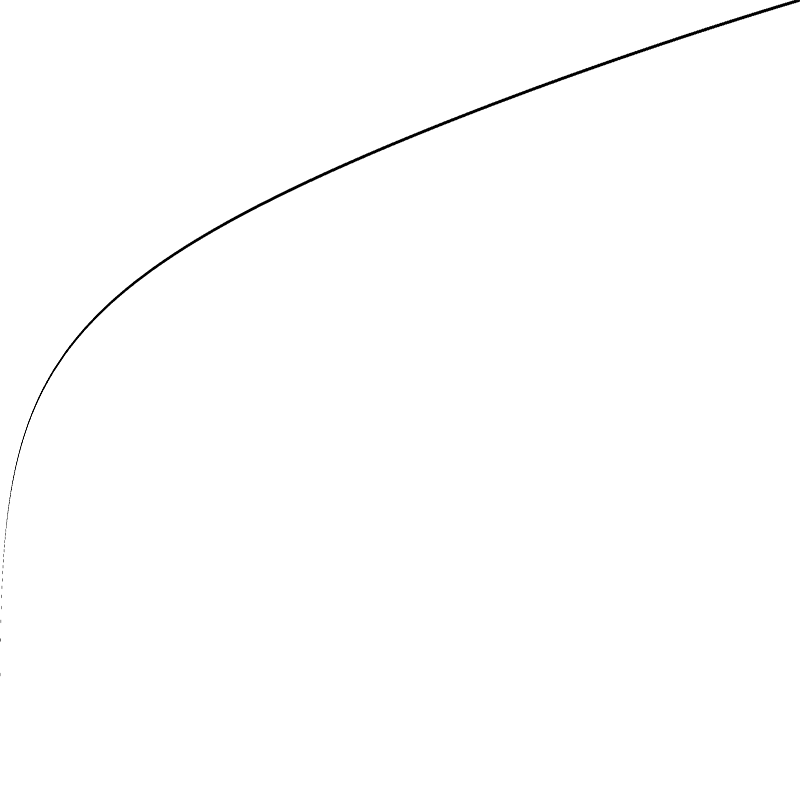}}
    \caption{$b_{20}=0.22$, $b_{21}=0.40$, and $b_{22}=1.0$ }
    \label{functions/022b0040b110b2}
 \end{subfigure}
 \hfill
      \begin{subfigure}[t]{0.23\textwidth}
\fbox{\includegraphics[width=0.99\textwidth]{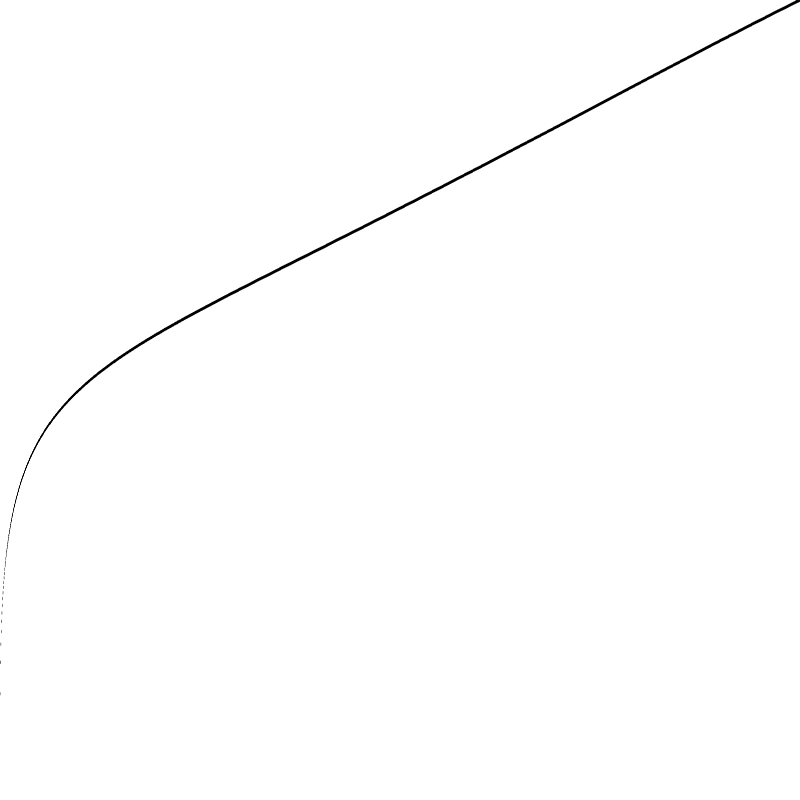}}
    \caption{$b_{20}=0.24$, $b_{21}=0.55$, and $b_{22}=1.0$ }
    \label{functions/024b0055b110b2}
 \end{subfigure}
 \hfill
      \begin{subfigure}[t]{0.23\textwidth}
\fbox{\includegraphics[width=0.99\textwidth]{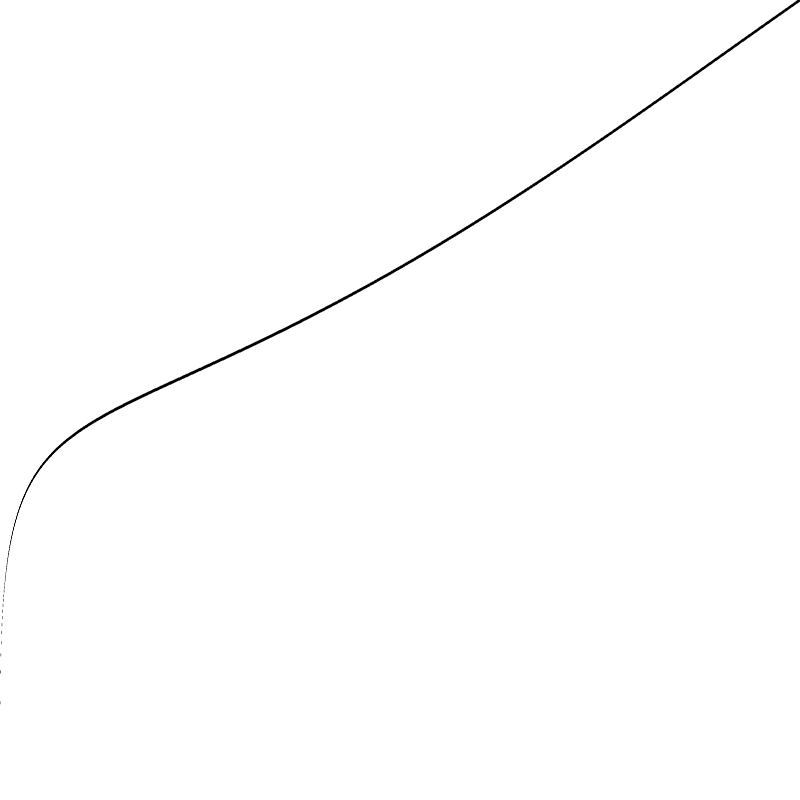}}
    \caption{$b_{20}=0.26$, $b_{21}=0.70$, and $b_{22}=1.0$ }
    \label{functions/026b007b110b2}
 \end{subfigure}
 \hfill
    \caption{$y=s^{b_{20}(1-s)^2 + 2b21 (1-s)s + b22 s^2}$ functions: second degree polynomials as power term.}
\label{fig_functions_quadrics}
\end{figure}

In this equation, $y$ should depend on the comparative size of the foreground object. The smaller the foreground object, the more affected it should be by the environment. In other words, $y$ should also be a monotonically increasing function of the relative size of the object in the foreground. Let $s \in [0,1]$ denote the relative size of the object in the foreground. Note that $s=0$ corresponds to an infinitely small foreground object. In this case, $y$ must be zero, since the foreground object must be invisible. In the case of $s=1$, the foreground object covers the entire viewing frustum, and the background will be completely invisible. In this case, $y$ must be one. In conclusion, we know that the monotonically increasing function $y$ must interpolate $(0,0)$ and $(1,1)$. This suggests that it should be a power function such as a gamma correction in the following form:
\begin{equation}
y = s^{f(s)} \label{Eq_gamma}
\end{equation}
where $f(s)$ is any always positive function as $f: [0,1] \rightarrow \Re^{+}$ with $\Re^{+}$ being the space of positive real numbers. To simplify the search for an appropriate solution, assume that $f(s)$ is a polynomial written in the B\'{e}zier form as 
$$f(s) = \sum_{i}^{n} \begin{pmatrix} n\\i\end{pmatrix} B_{n,i} (1-s)^{n-i} s^{i} $$
where $B_{n,i}$ are B\'{e}zier coefficients and $n$ is the degree of the polynomial \cite{bartels1995introduction,prautzsch2002bezier}. 
A polynomial of degree $0$ can also be considered in the B\'{e}zier form as
$$f (s) = B_{0,0}$$
and examples of functions $y=s^{f(s)}$  with zero-degree polynomials are shown in Figure~\ref{fig_functions_constant}. These are simply gamma correction functions. More interesting cases are obtained by using higher-degree polynomials as power terms. For example, first-degree polynomials in B\'{e}zier form are written as
$$ f (s) = B_{1,0} (1-s) + B_{1,1} s,$$
and examples of some functions of $y=s^{f(s)}$  are shown in Figure~\ref{fig_functions_linears}. 
A second-degree polynomial in B\'{e}zier form is written as 
$$f(s) = B_{2,0} (1-s)^2 + B_{2,1} 2 (1-s) s + B_{2,2} s^2,$$
and examples of $y=s^{f(s)}$ functions with degree-two polynomials are shown in Figure~\ref{fig_functions_quadrics}. 
We did not have to go to higher degrees, but for completeness we also provide third-degree polynomials in the B\'{e}zier form as 
$$f(s) = B_{3,0} (1-s)^3 + B_{3,1} 3 (1-s)^2 s + B_{3,2} 3 (1-s) s^2 + B_{3,3} s^3.$$
An important property of B\'{e}zier form is that the form satisfies the partition of the unity property. As a consequence, the resulting polynomial remains in a convex hull of the coefficient positions. In other words, if all coefficients are positive numbers, the function $f(s)$ is guaranteed to be positive. 

In conclusion, we conjecture that our vision system has evolved into equations~\ref{Eq_barycentric} and~\ref{Eq_gamma}. In other words, we believe that all animals can have a similar structure in their visual system since this appears to be the simplest structure that can emerge through an evolutionary process. Therefore, it can be embedded in genes. On the other hand, we expect that the function $f(s)$ can be different for different animals or individuals, as it does not have to be fixed. These differences can be attributed to the learning process. Therefore, we expect that the complexity of polynomials and their coefficients could have been different. 

We also need to point out that, as authors, we tested some polynomials and we all agree that the quadric polynomial is sufficient with B\'{e}zier coefficients $B_{2,0}=0.20, B_{2,1}=0.25, and B_{2,2}=1.00$. Note that the corresponding polynomial can be computed as 
$$0.20 (1-s)^2 + 0.5 s (1-s) + 1.0 s^2 =  0.7s^2 + 0.1s +0.2 $$
and the corresponding function
\begin{equation}y = s^{0.20 (1-s)^2 + 0.5 s (1-s) + 1.0 s^2} = s^{0.7s^2 + 0.1s +0.2} \label{Eq_y_final}\end{equation}
is shown in Figure~\ref{functions/02b010b1}. 
As a result, the full formula becomes the following: 
\begin{equation}L_O = s^{0.7s^2 + 0.1s +0.2} L_P + (1-s^{0.7s^2 + 0.1s +0.2}) I_a \label{Eq_barycentric_final}\end{equation}

To obtain the formula in Equation~\ref{Eq_barycentric_final}, we start first with degree-zero polynomials. We quickly establish that the best result is obtained by $y=s{0.25}$. Then we improved the formula using the degree-one polynomial. In this case, $y=s{0.2 (1-s) + 1.0 s}$ gave the best result, however, there was a problem in the sizes around $0.5$. To solve that, we increased the degree one more and kept the boundaries the same. This gave us $s^{0.20 (1-s)^2 + 0.5 s (1-s) + 1.0 s^2}$ a new function. We played with values as shown in Figures~\ref{functions/022b0040b110b2},~\ref{functions/024b0055b110b2}, and~\ref{functions/026b007b110b2}. Note that these functions are very close to each other and give an almost affine result for values of $s$ slightly larger than $0$. This suggests that the overall function can simply be a linear interpolation in the form of \begin{equation}y = 0.6 (1-s) + 1.0 s = 0.4 s + 0.6\label{Eq_y_linear_final}\end{equation} 
since the shapes of the two functions of $y$ are almost identical except for very small values of $s$. This gives us an even simpler formula as follows: 
\begin{equation}L_O = (0.4 - 0.4 s) L_P + (0.6 + 0.4 s) I_a \label{Eq_barycentric_final_linear}\end{equation}

This is an interesting result that shows how simple equations can naturally emerge. We have started a somewhat logical model using a barycentric approach. Once we identified coefficients, the shape of the curve given by the formula of Equation~\ref{Eq_barycentric_final} started to look even simpler if we ignore very small values of $s$. This is logical in the sense that very small foreground objects are either small, which may not be dangerous, or distant, which may not be immediately dangerous. However, as soon as their relative sizes become significant, the visual system needs to update its luminance values to decipher the actual diffuse reflection term. This can be simply approximated by an affine function such as the one given in Equation~\ref{Eq_barycentric_final_linear}. 

An important part of our approach is that we simply ignore how the visual system computes the relative size $s$ and the average illumination $I_a$. In our case, we know the ground truths and use them to obtain constant luminance perception. Furthermore, our test cases are simple; therefore, it can be easy for the visual system to estimate the relative size $s$ and the average illumination $I_a$. In the next subsection, we provide how we develop and use a program to obtain coefficients of the polynomial formulas (Equations such as~\ref{Eq_y_linear_final} and~\ref{Eq_barycentric_final}) to provide a guideline for others to change the coefficients. 

\section{Implementation and Results}
\label{Sec_ImplementationandResults}

\begin{figure}[hbtp]
\centering
     \begin{subfigure}[t]{0.32\textwidth}
\includegraphics[width=0.99\textwidth]{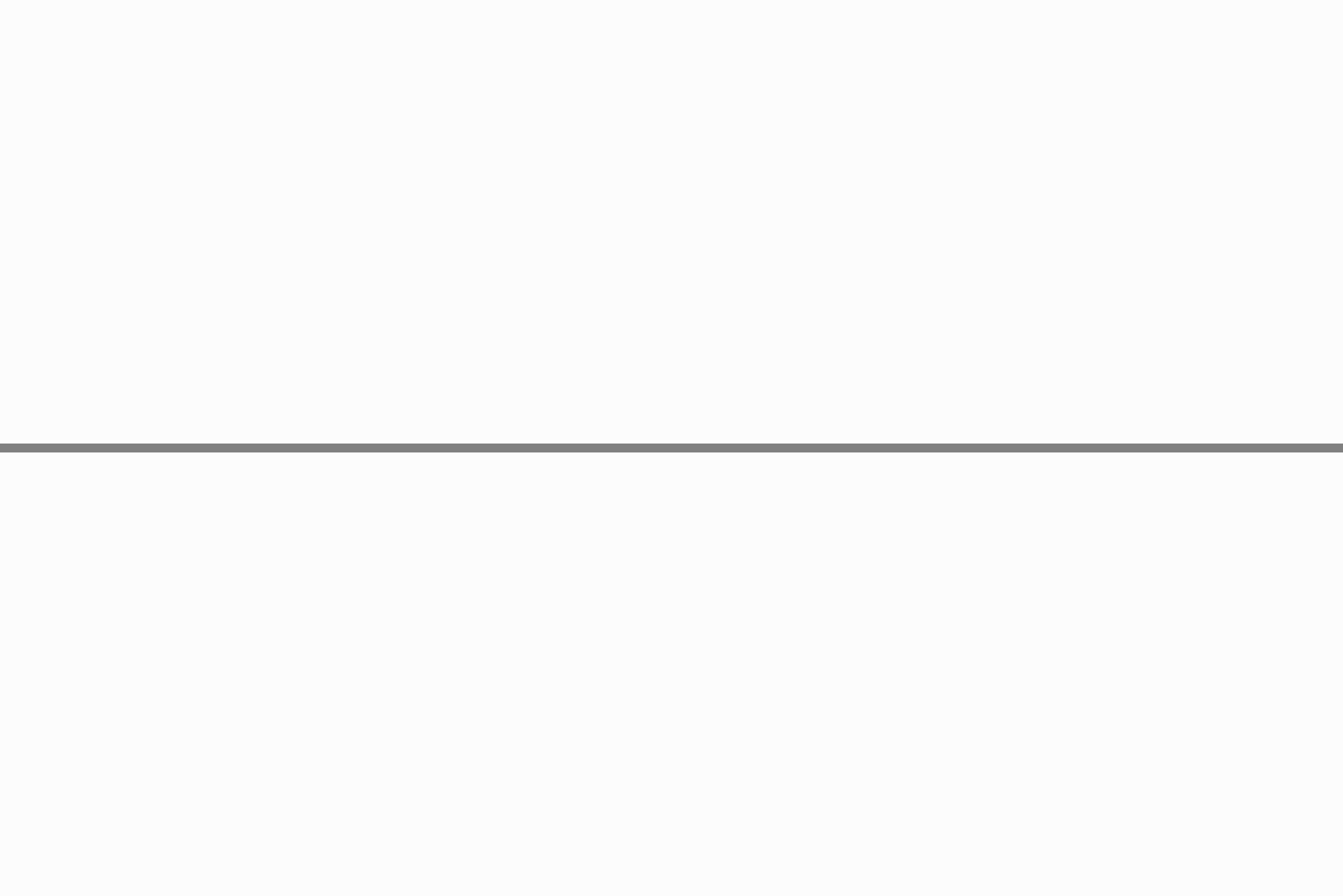}
    \caption{Constant color covering 01\% of the area in front of white background. }
    \label{fig_WhiteBGConstantColor/001}
 \end{subfigure}
 \hfill
     \begin{subfigure}[t]{0.32\textwidth}
\includegraphics[width=0.99\textwidth]{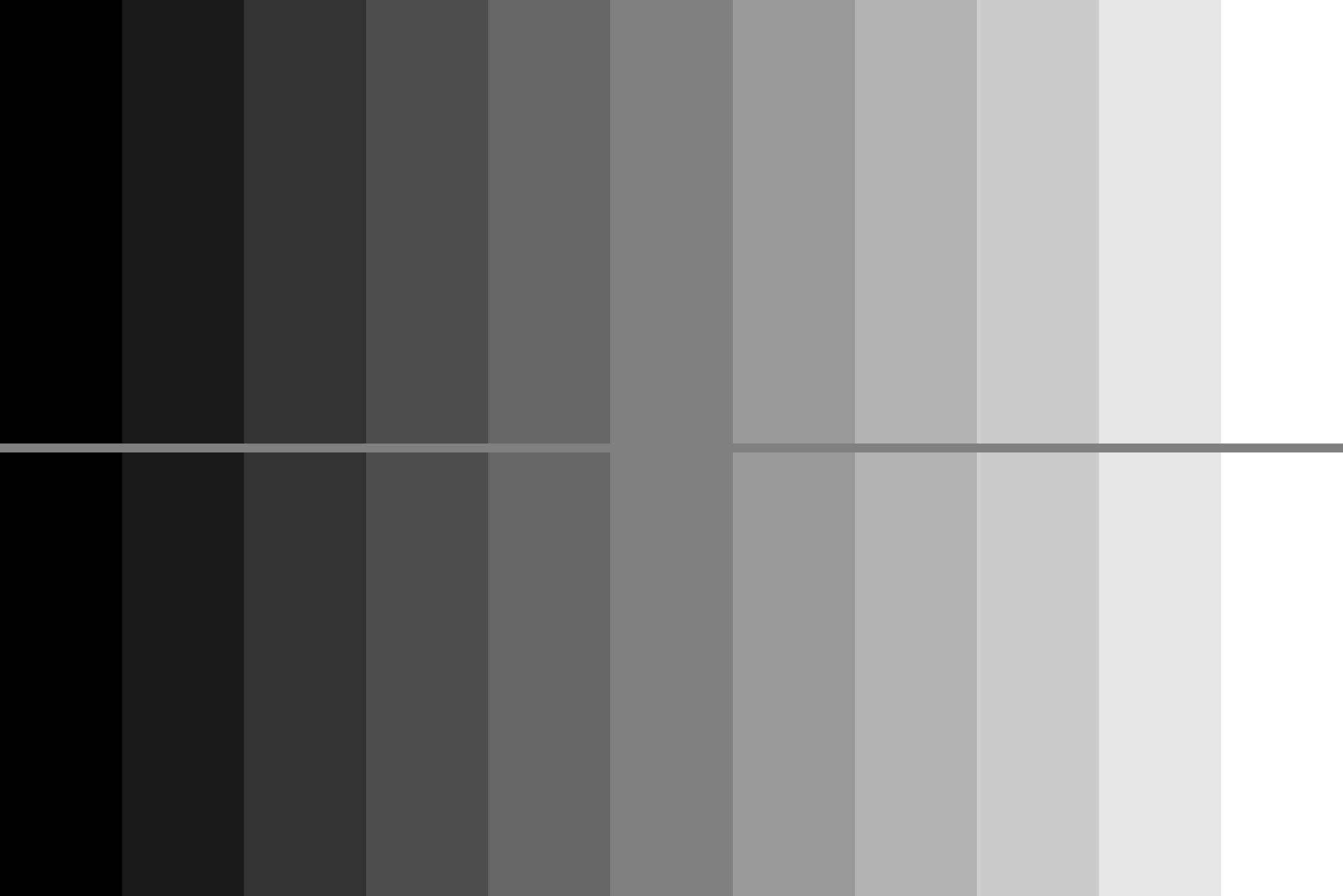}
    \caption{Constant color covering 01\% of the area in front of a graded color that consists of 01 bands. }
    \label{fig_10BandBGConstantColor/001}
 \end{subfigure}
 \hfill
      \begin{subfigure}[t]{0.32\textwidth}
\includegraphics[width=0.99\textwidth]{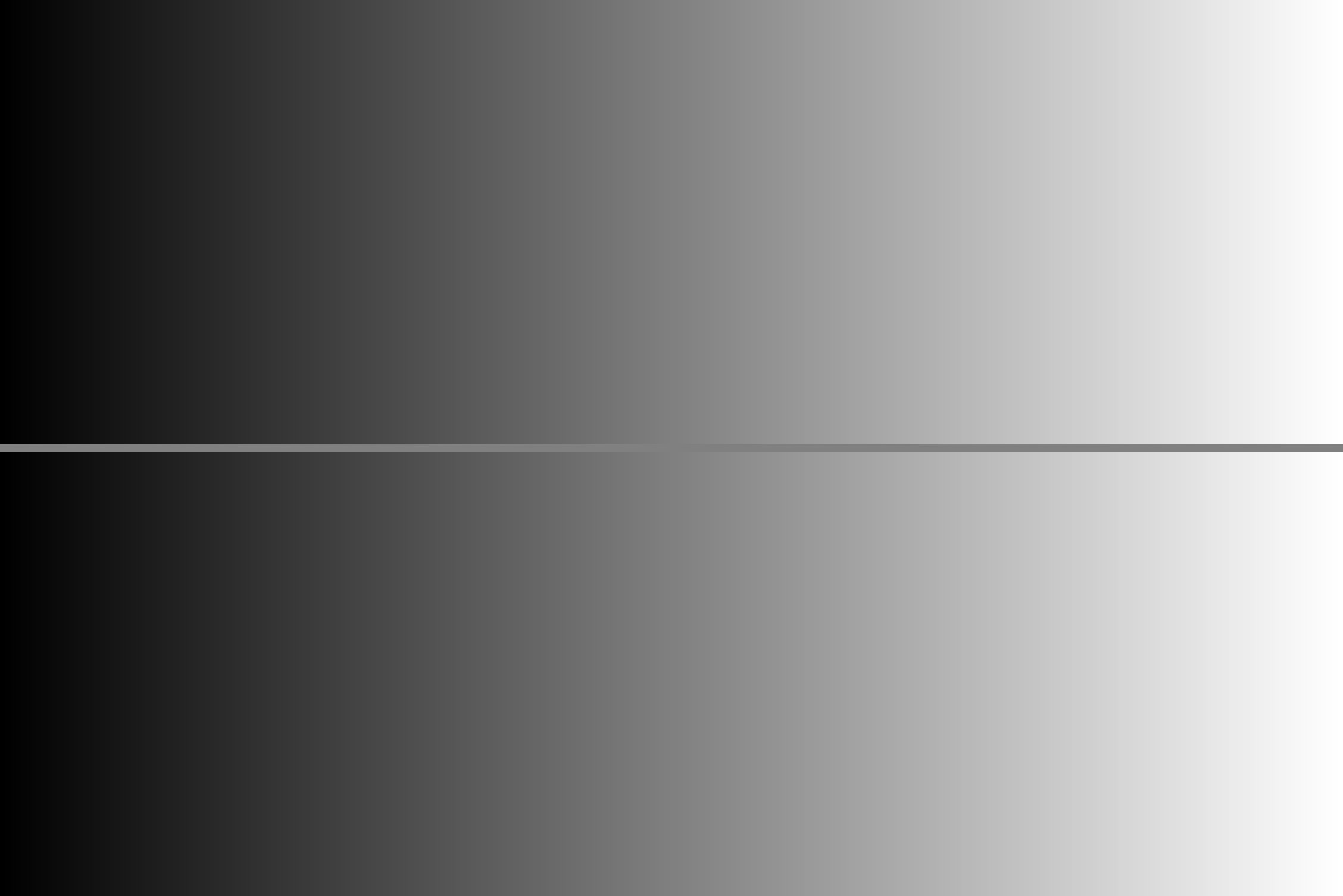}
    \caption{Constant color covering 01\% of the area in front of a continuously graded color. }
    \label{fig_ContinousBGConstantColor/001}
 \end{subfigure}
 \hfill
     \begin{subfigure}[t]{0.32\textwidth}
\includegraphics[width=0.99\textwidth]{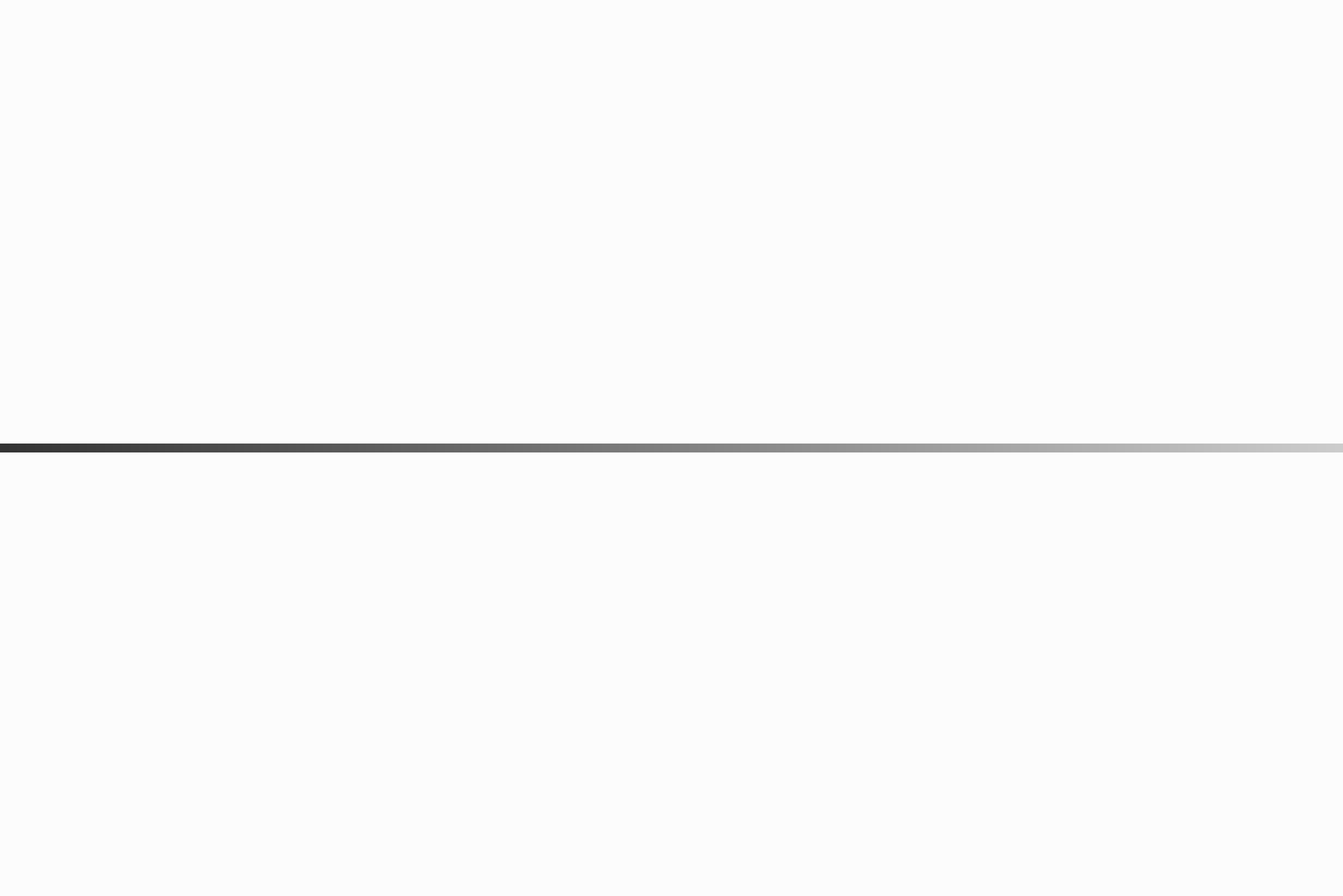}
    \caption{Our variable color band covers 01\% of the area in front of a white background. }
    \label{fig_WhiteBGConstantPerception/001}
 \end{subfigure}
 \hfill
     \begin{subfigure}[t]{0.32\textwidth}
\includegraphics[width=0.99\textwidth]{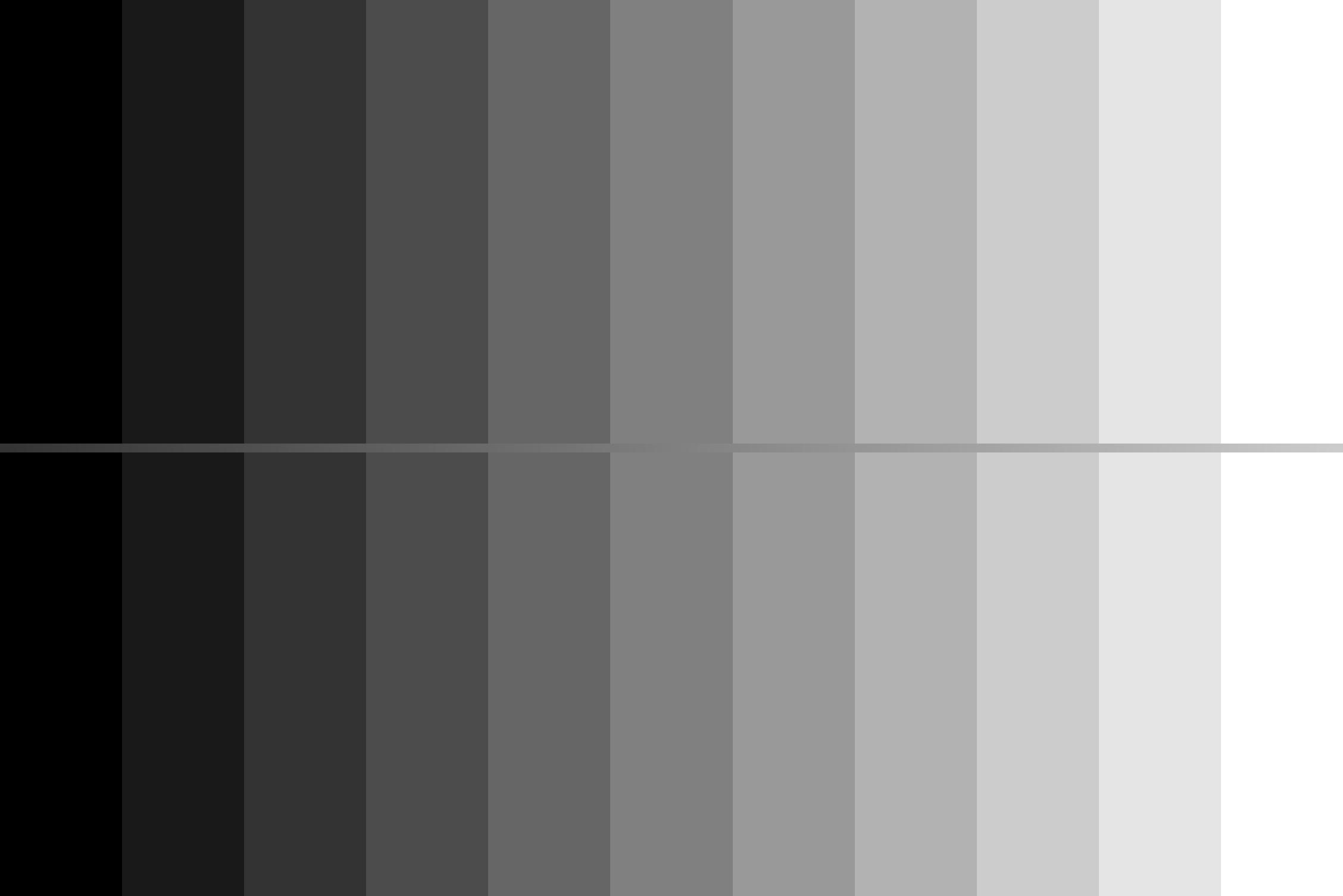}
    \caption{Our variable color band covers 01\% of the area in front of a graded color that consists of 01 bands. }
    \label{fig_10BandBGConstantPerception/001}
 \end{subfigure}
 \hfill
      \begin{subfigure}[t]{0.32\textwidth}
\includegraphics[width=0.99\textwidth]{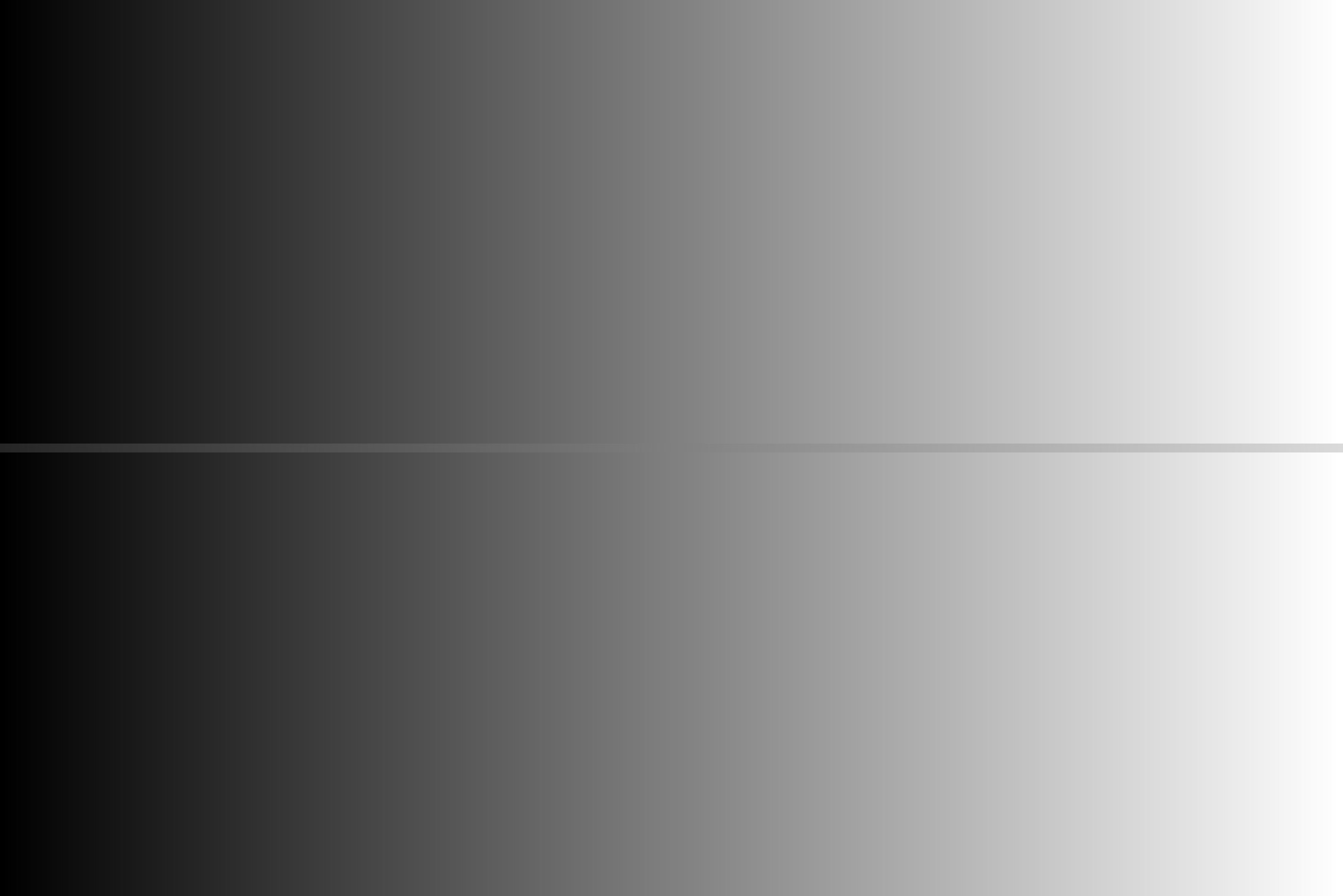}
    \caption{Our variable color band covers 01\% of the area in front of a continuously graded color. }
    \label{fig_ContinousBGBGConstantPerception/001}
 \end{subfigure}
 \hfill 
    \caption{Comparison of our method with constant color. Note that Our variable color band creates constant perception}
\label{fig_001}
\end{figure}

\begin{figure}[hbtp]
\centering
     \begin{subfigure}[t]{0.32\textwidth}
\includegraphics[width=0.99\textwidth]{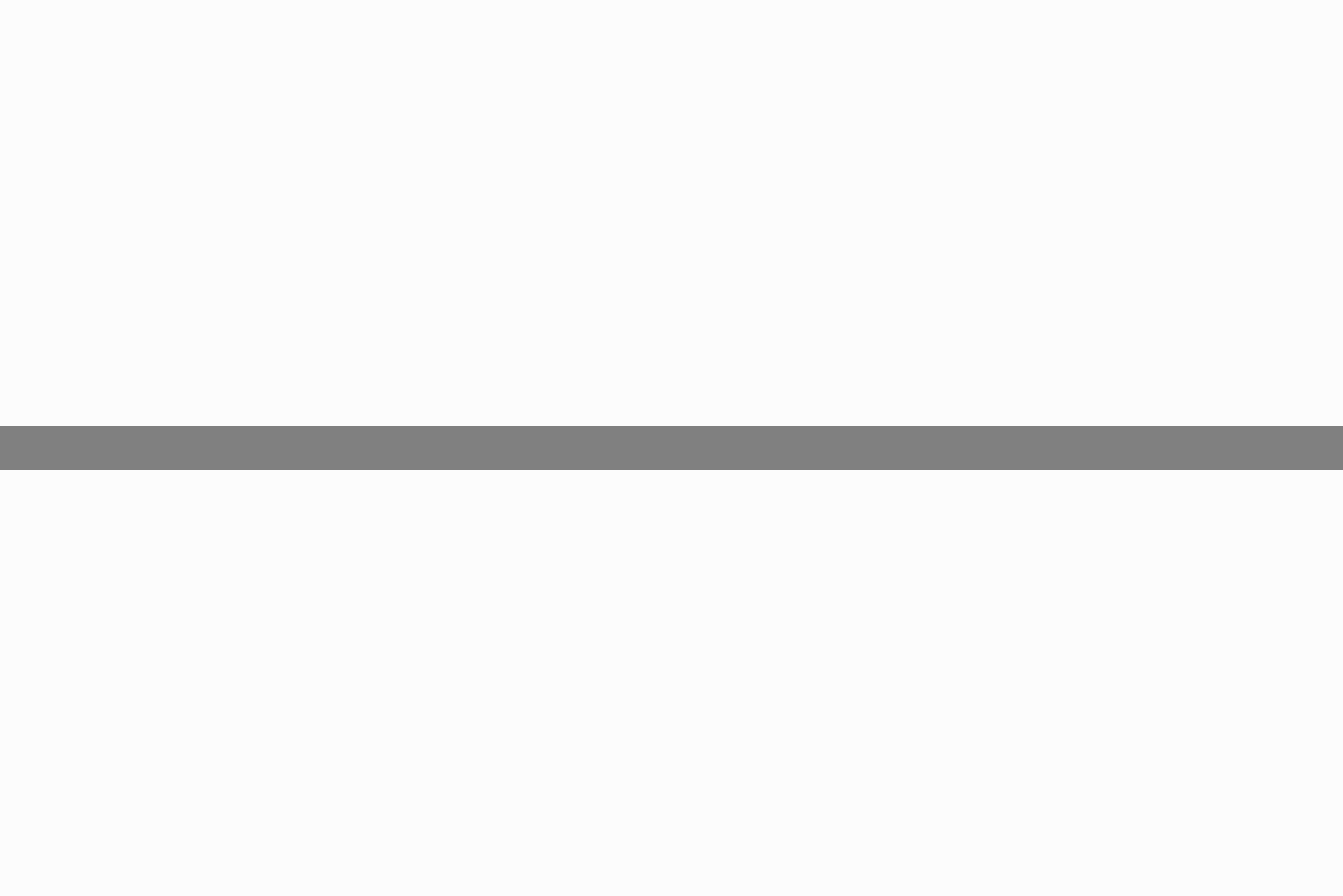}
    \caption{Constant color covering 05\% of the area in front of white background. }
    \label{fig_WhiteBGConstantColor/005}
 \end{subfigure}
 \hfill
     \begin{subfigure}[t]{0.32\textwidth}
\includegraphics[width=0.99\textwidth]{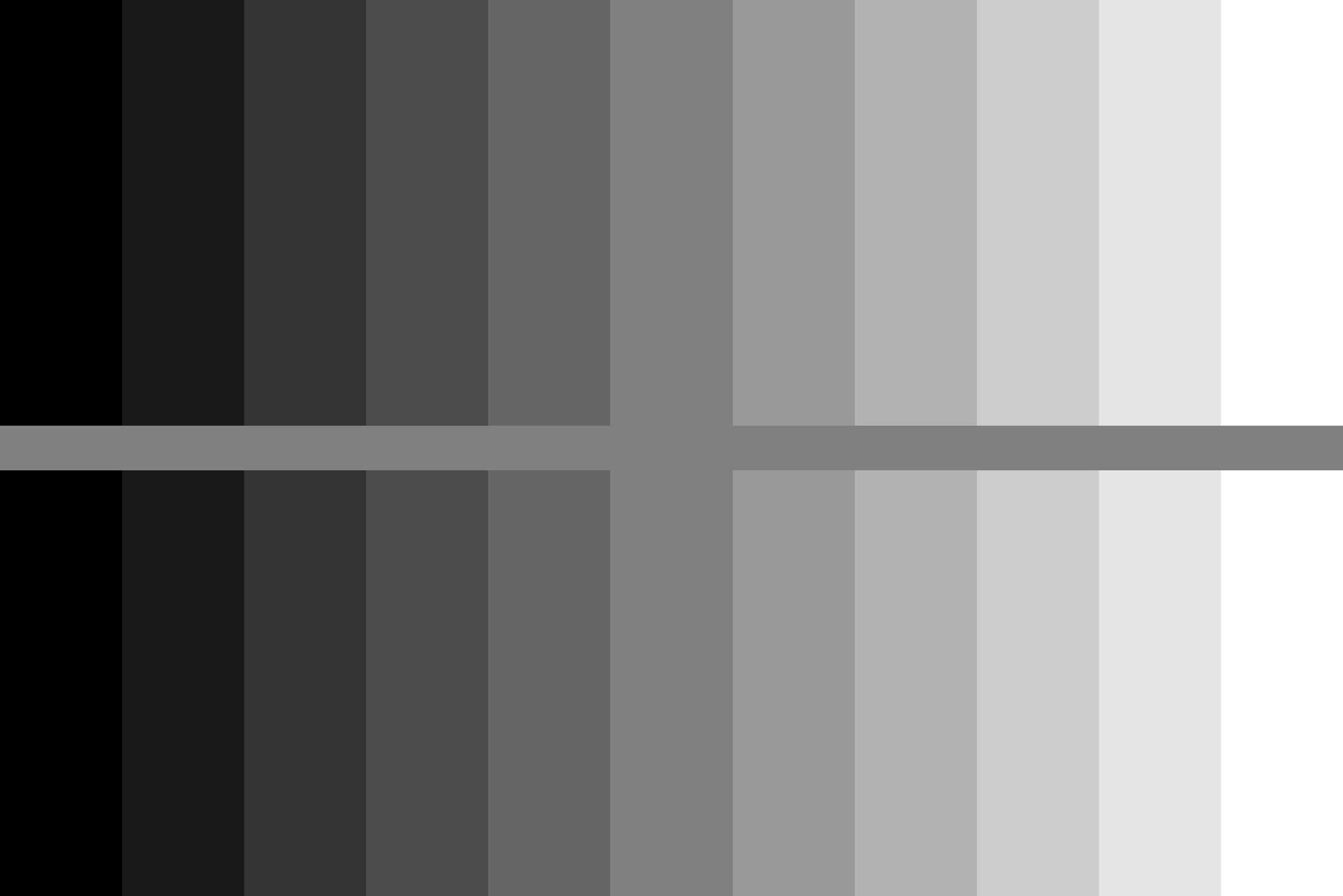}
    \caption{Constant color covering 05\% of the area in front of a graded color that consists of 05 bands. }
    \label{fig_10BandBGConstantColor/005}
 \end{subfigure}
 \hfill
      \begin{subfigure}[t]{0.32\textwidth}
\includegraphics[width=0.99\textwidth]{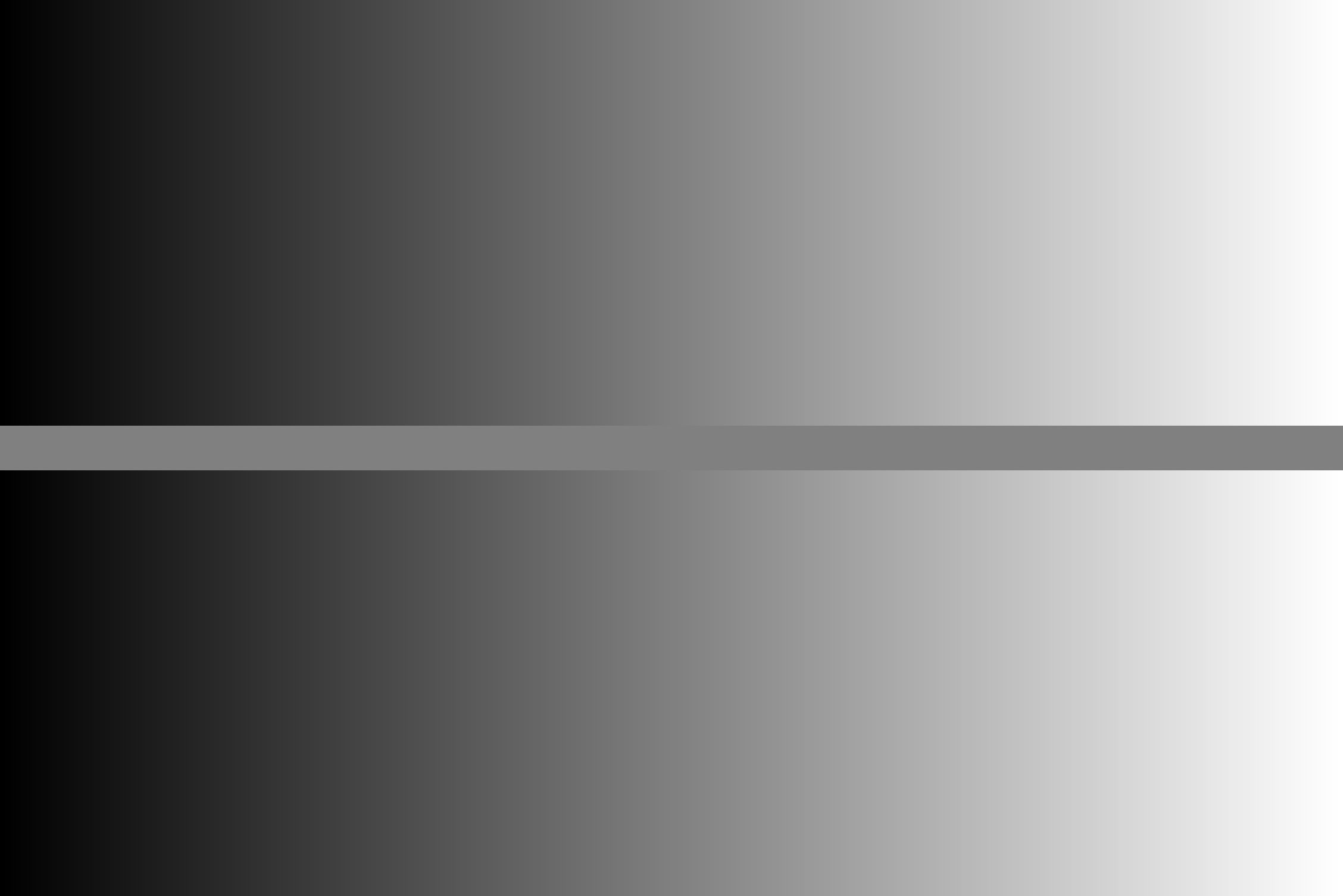}
    \caption{Constant color covering 05\% of the area in front of a continuously graded color. }
    \label{fig_ContinousBGConstantColor/005}
 \end{subfigure}
 \hfill
     \begin{subfigure}[t]{0.32\textwidth}
\includegraphics[width=0.99\textwidth]{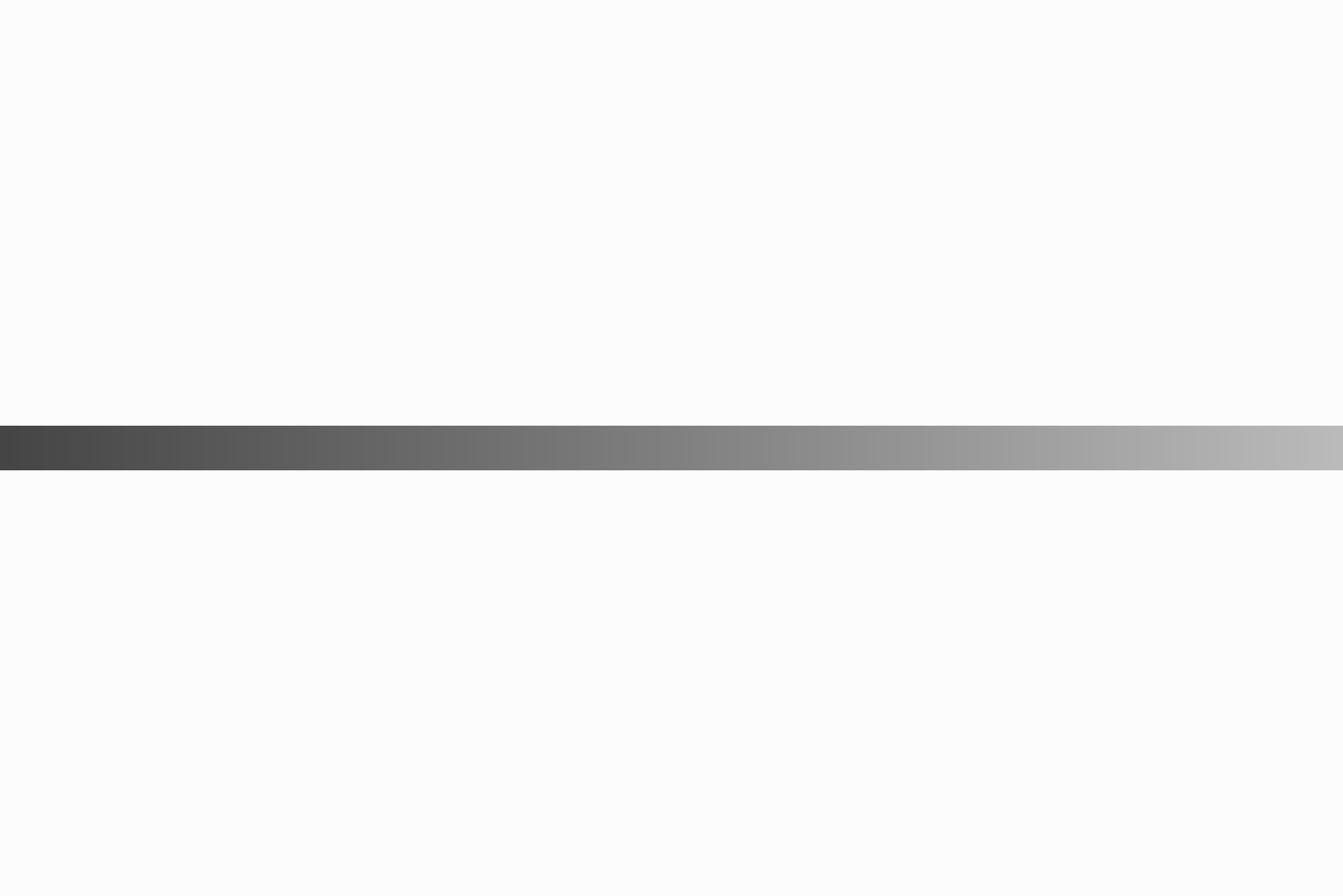}
    \caption{Our variable color band covers 05\% of the area in front of a white background. }
    \label{fig_WhiteBGConstantPerception/005}
 \end{subfigure}
 \hfill
     \begin{subfigure}[t]{0.32\textwidth}
\includegraphics[width=0.99\textwidth]{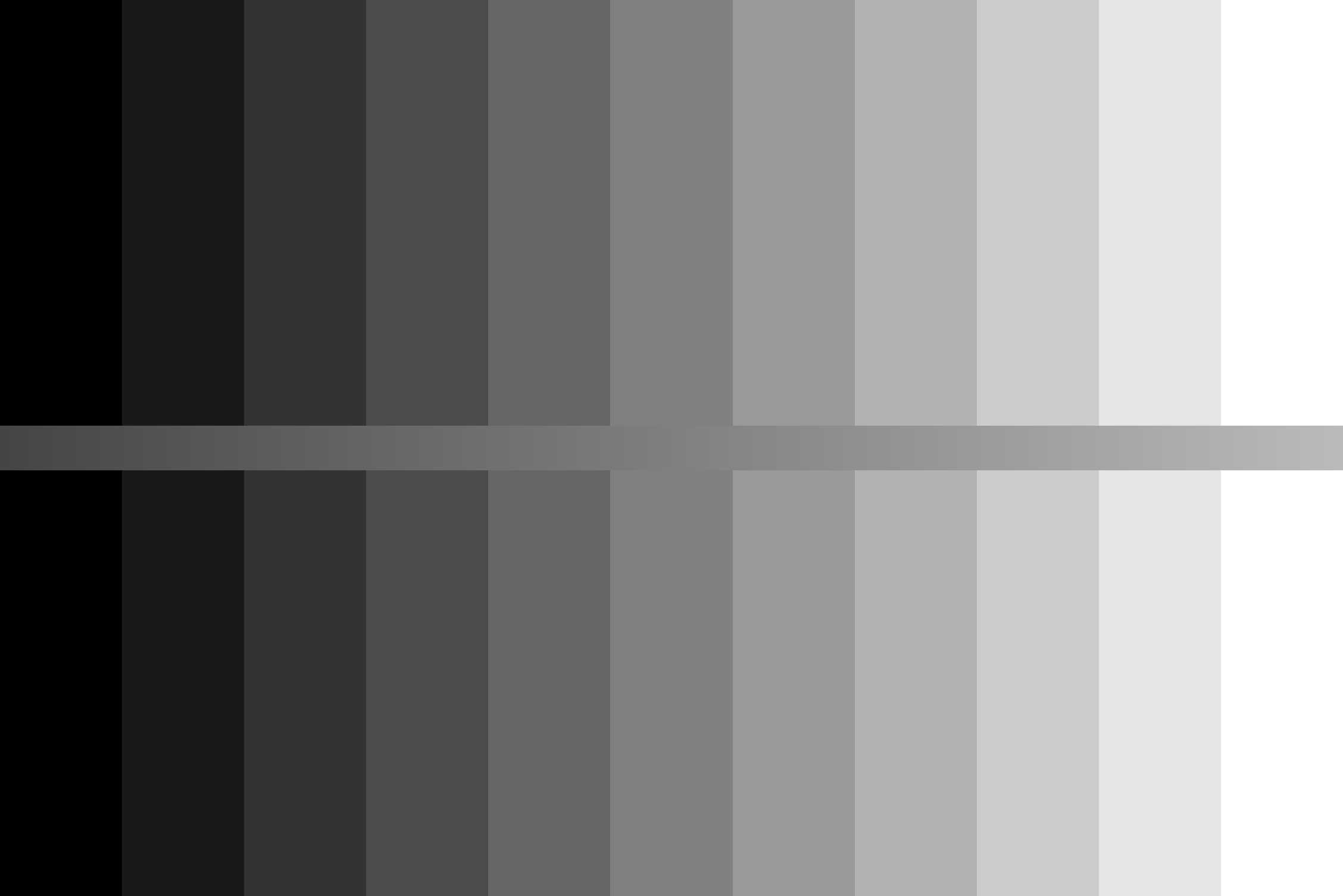}
    \caption{Our variable color band covers 05\% of the area in front of a graded color that consists of 05 bands. }
    \label{fig_10BandBGConstantPerception/005}
 \end{subfigure}
 \hfill
      \begin{subfigure}[t]{0.32\textwidth}
\includegraphics[width=0.99\textwidth]{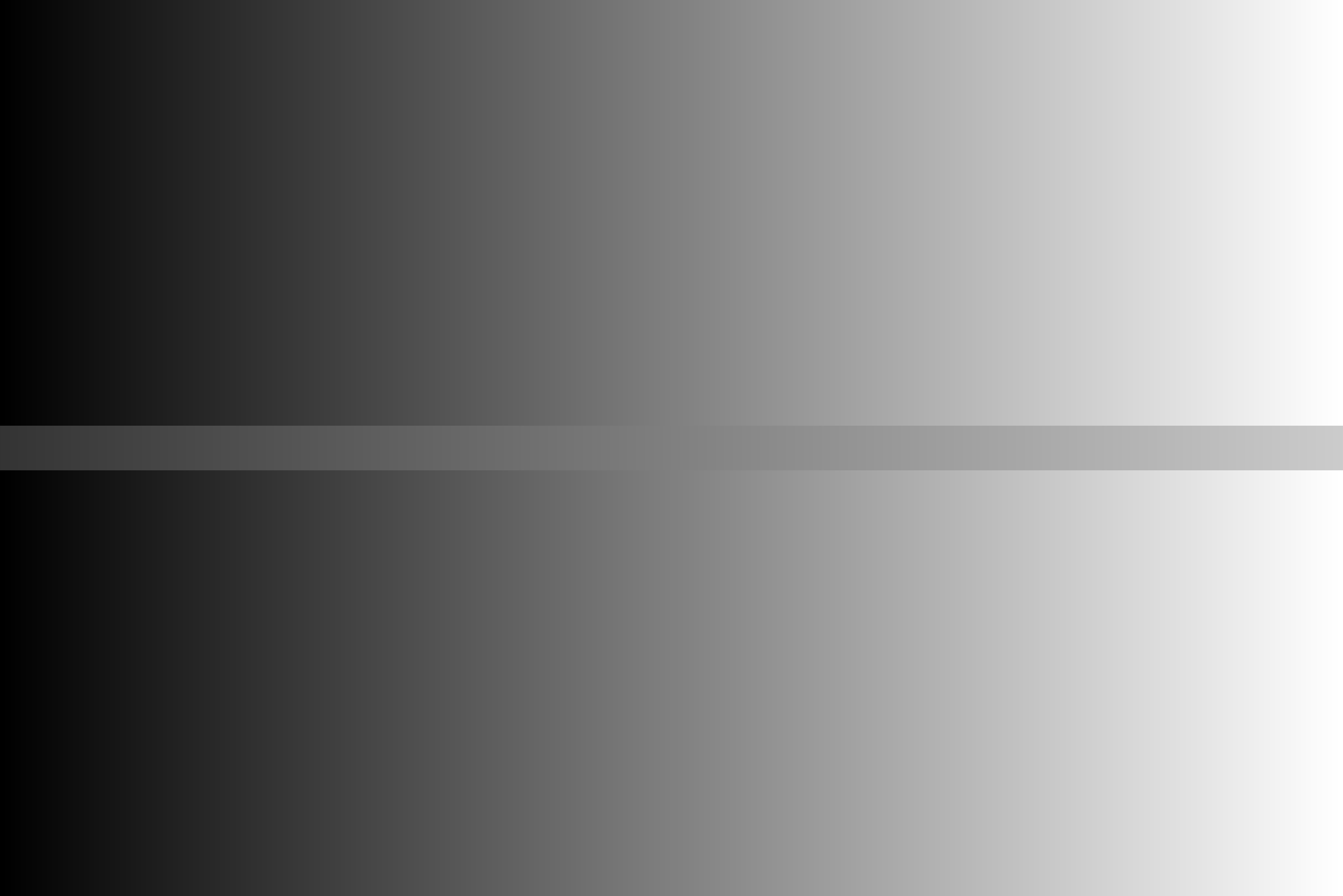}
    \caption{Our variable color band covers 05\% of the area in front of a continuously graded color. }
    \label{fig_ContinousBGBGConstantPerception/005}
 \end{subfigure}
 \hfill 
    \caption{Comparison of our method with constant color. Note that Our variable color band creates constant perception}
\label{fig_005}
\end{figure}

\begin{figure}[hbtp]
\centering
     \begin{subfigure}[t]{0.32\textwidth}
\includegraphics[width=0.99\textwidth]{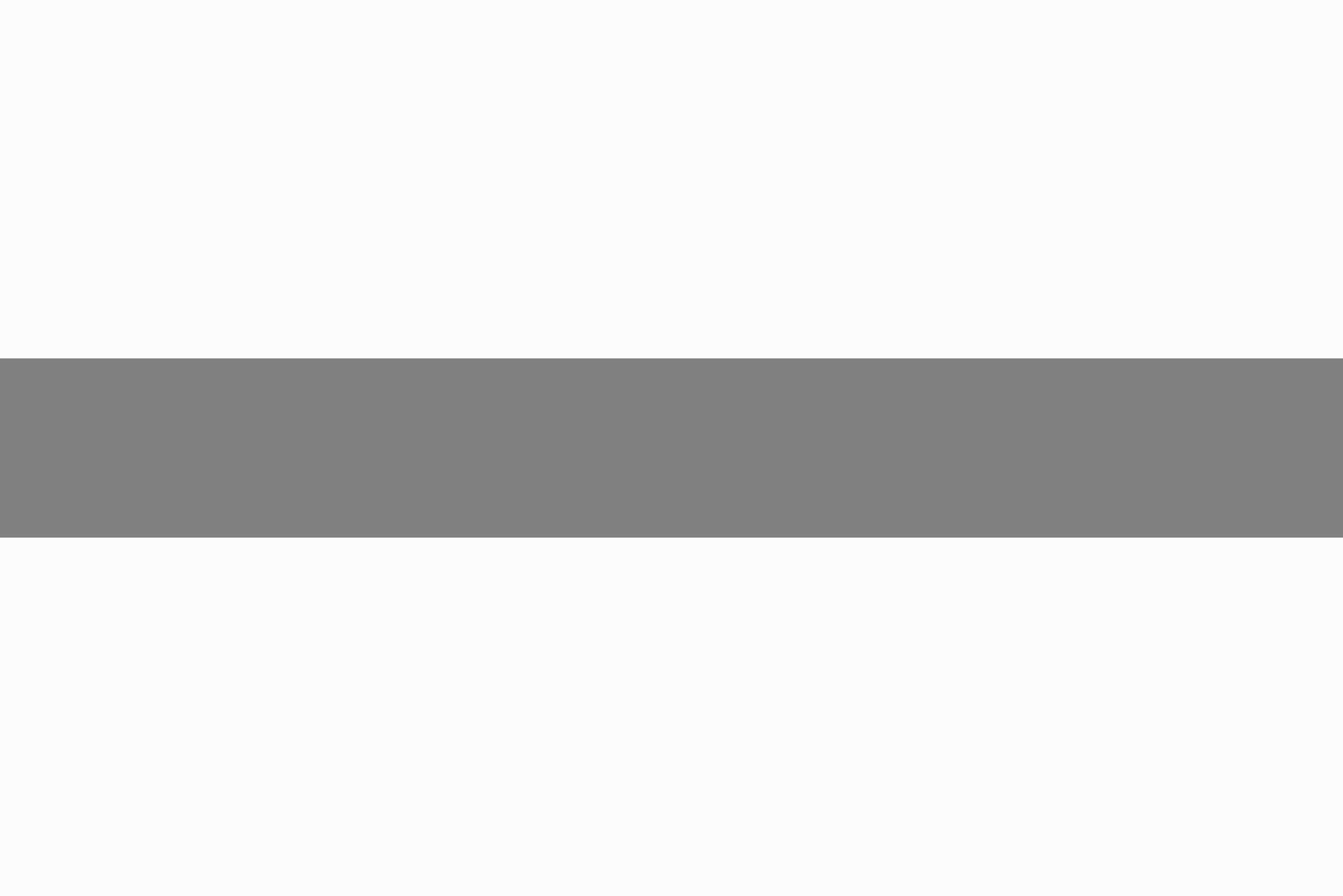}
    \caption{Constant color covering 20\% of the area in front of white background. }
    \label{fig_WhiteBGConstantColor/020}
 \end{subfigure}
 \hfill
     \begin{subfigure}[t]{0.32\textwidth}
\includegraphics[width=0.99\textwidth]{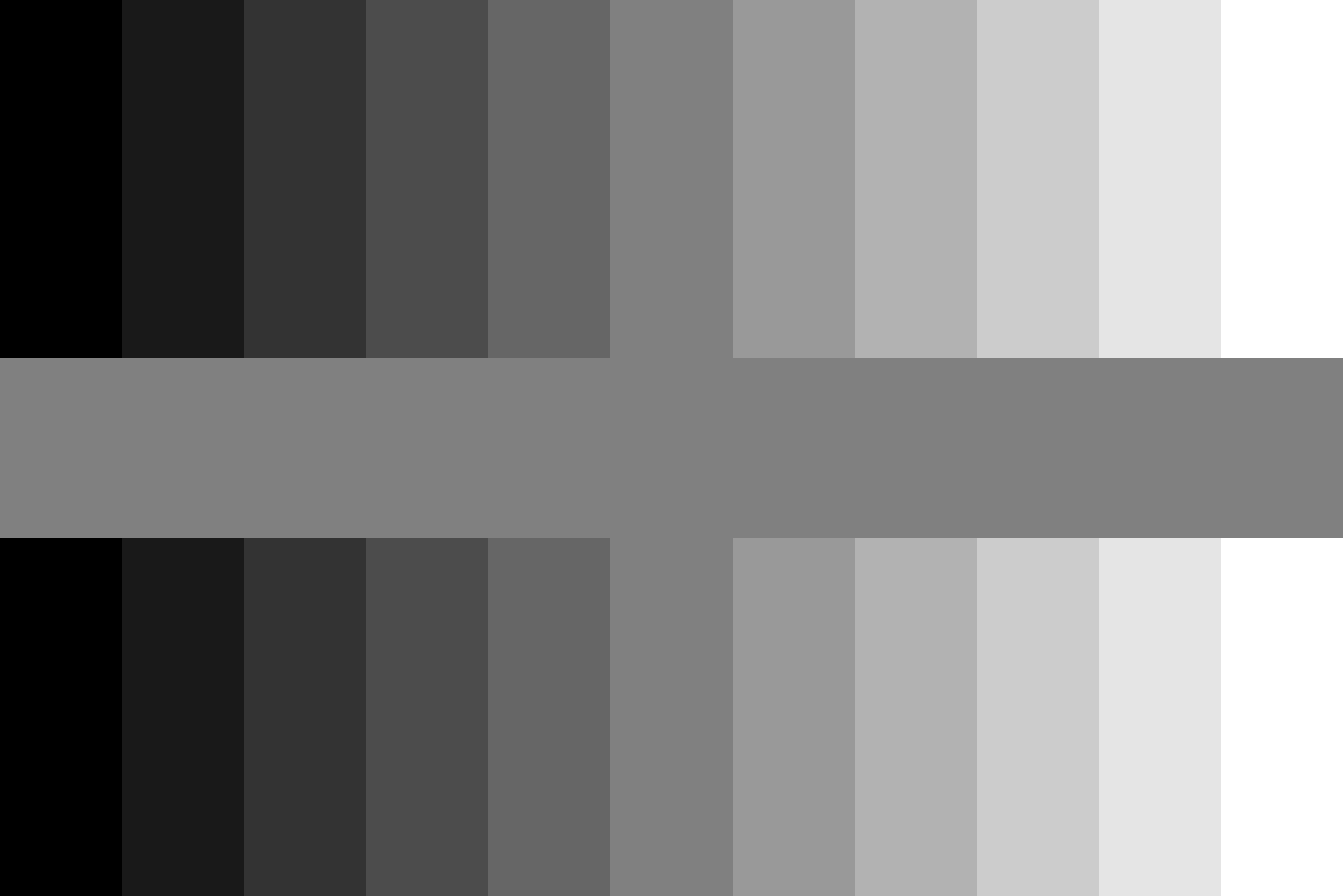}
    \caption{Constant color covering 20\% of the area in front of a graded color that consists of 20 bands. }
    \label{fig_10BandBGConstantColor/020}
 \end{subfigure}
 \hfill
      \begin{subfigure}[t]{0.32\textwidth}
\includegraphics[width=0.99\textwidth]{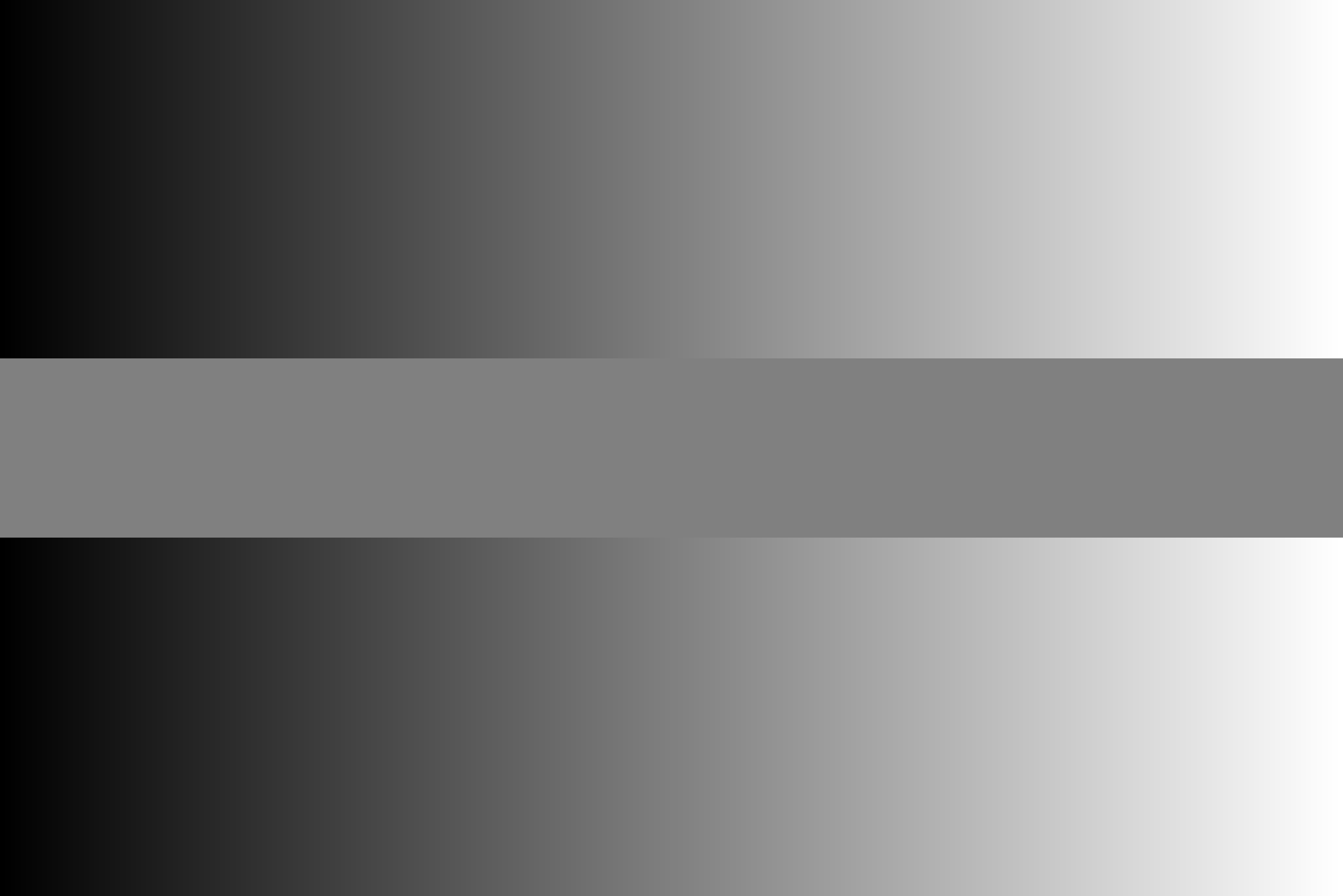}
    \caption{Constant color covering 20\% of the area in front of a continuously graded color. }
    \label{fig_ContinousBGConstantColor/020}
 \end{subfigure}
 \hfill
     \begin{subfigure}[t]{0.32\textwidth}
\includegraphics[width=0.99\textwidth]{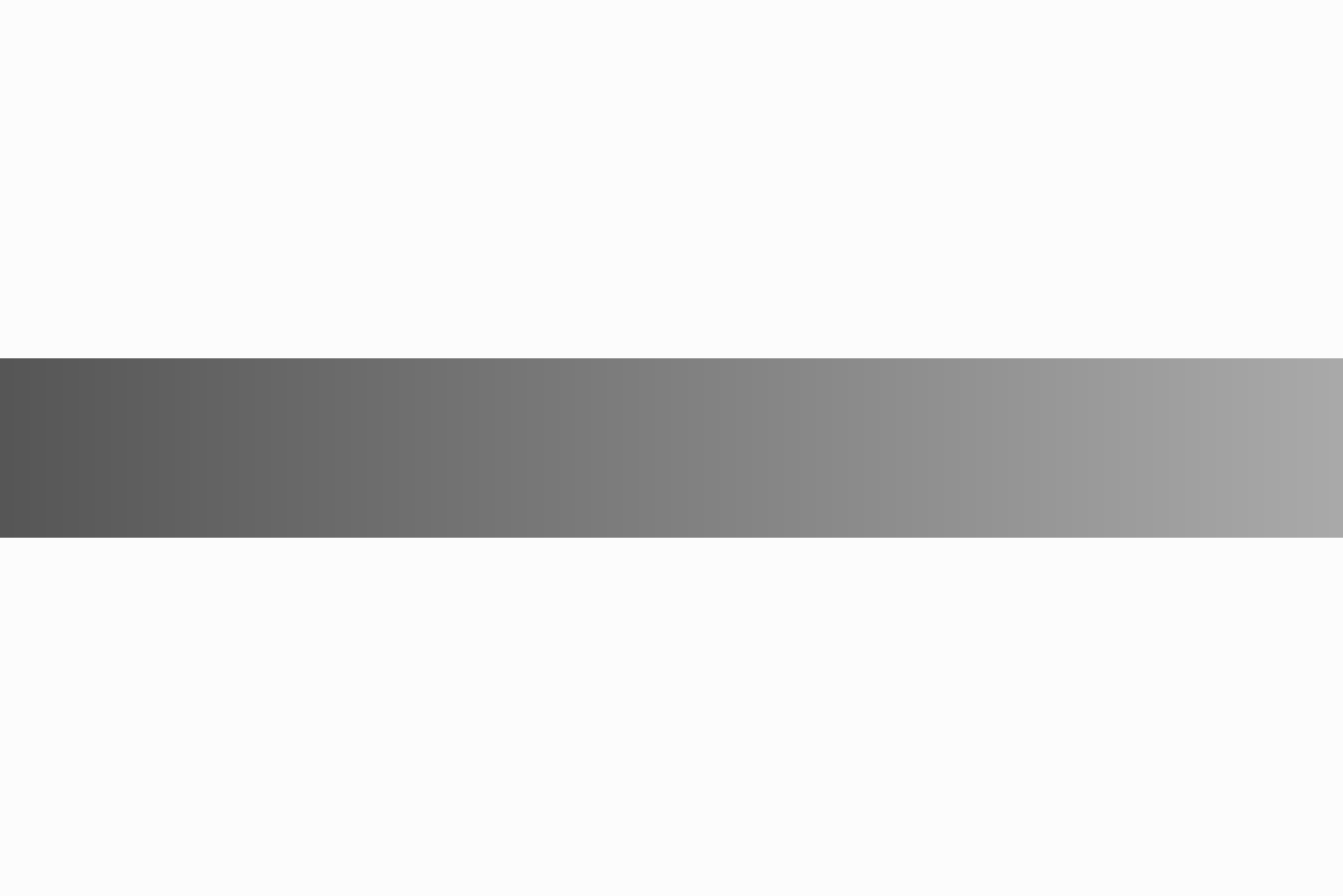}
    \caption{Our variable color band covers 20\% of the area in front of a white background. }
    \label{fig_WhiteBGConstantPerception/020}
 \end{subfigure}
 \hfill
     \begin{subfigure}[t]{0.32\textwidth}
\includegraphics[width=0.99\textwidth]{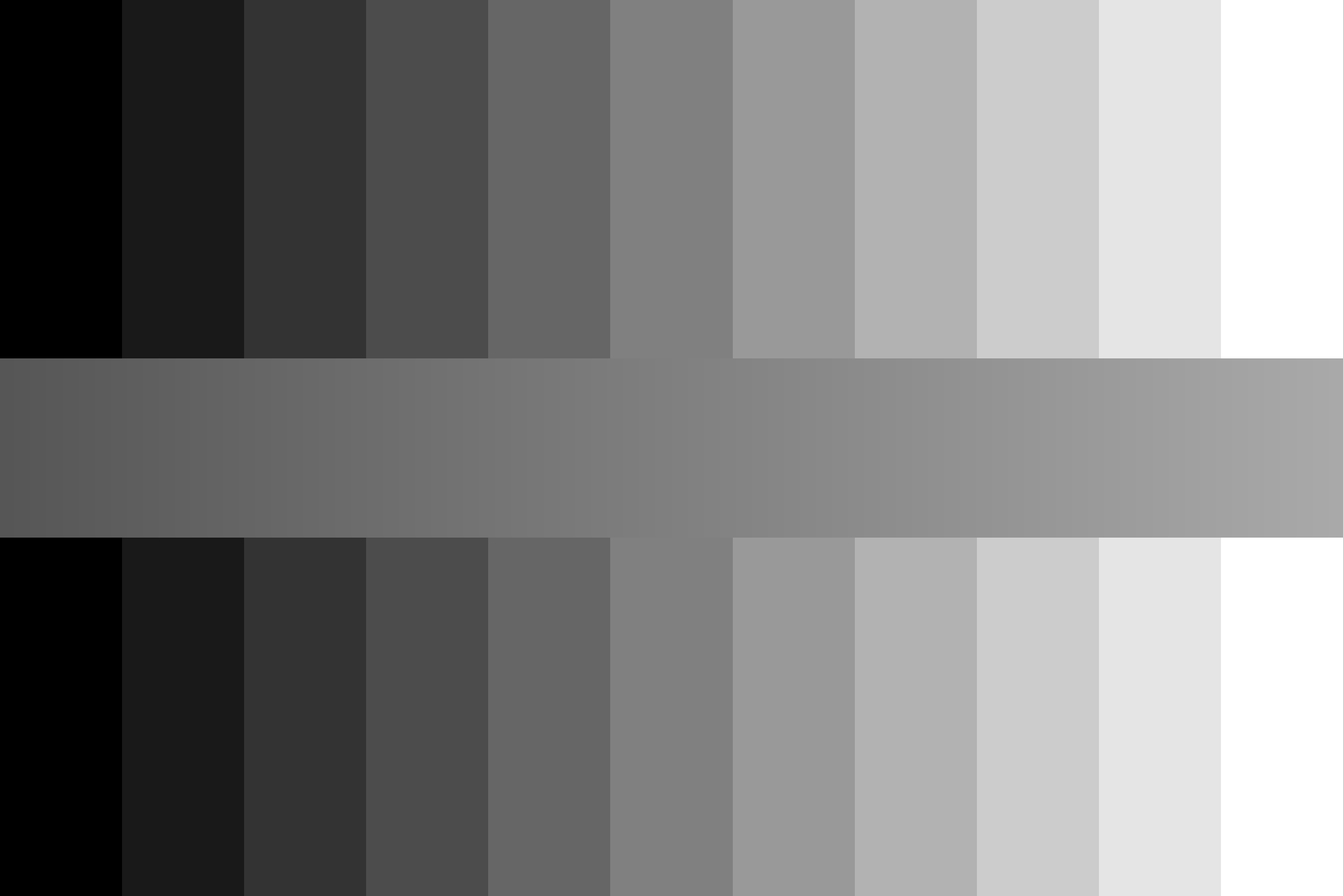}
    \caption{Our variable color band covers 20\% of the area in front of a graded color that consists of 20 bands. }
    \label{fig_10BandBGConstantPerception/020}
 \end{subfigure}
 \hfill
      \begin{subfigure}[t]{0.32\textwidth}
\includegraphics[width=0.99\textwidth]{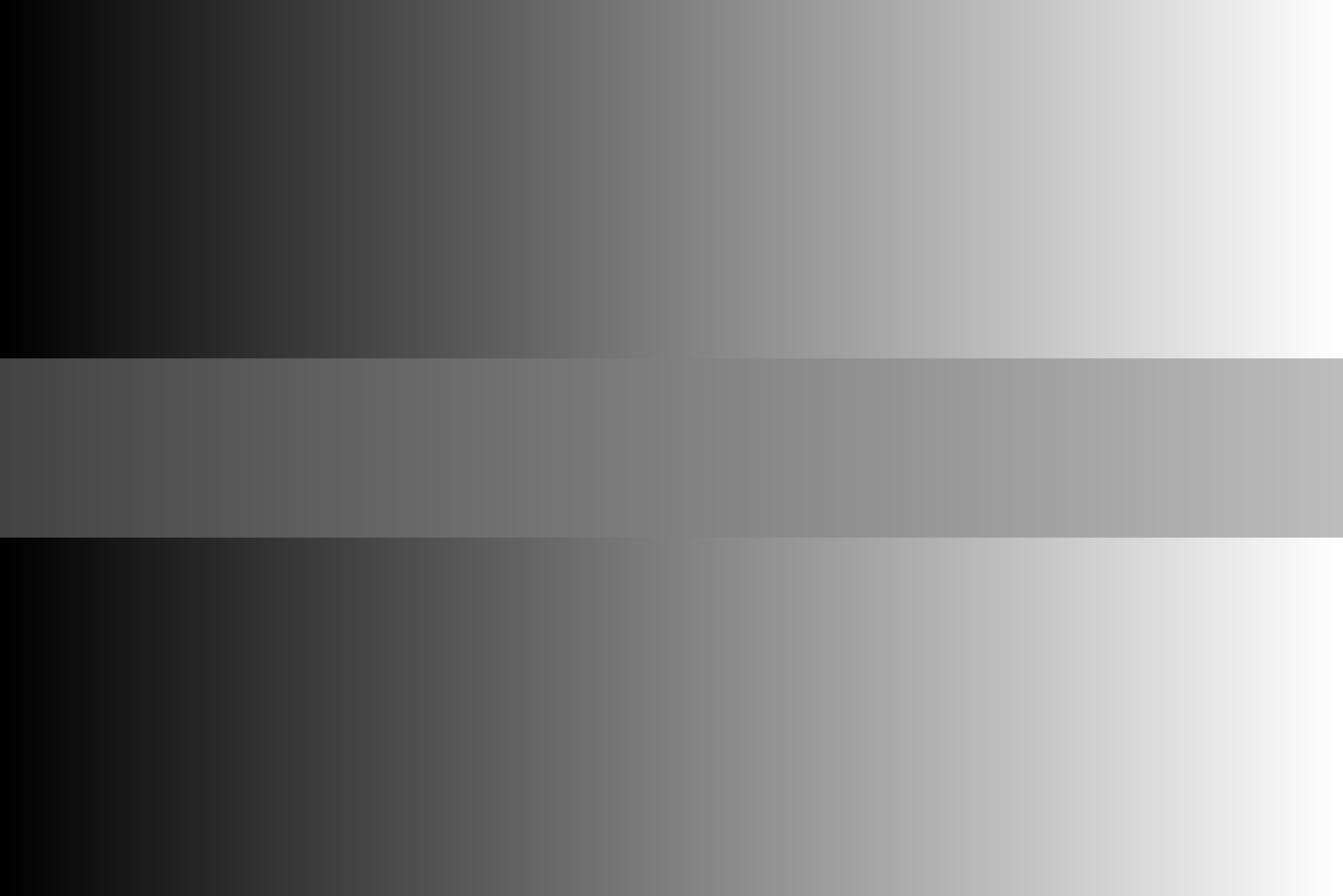}
    \caption{Our variable color band covers 20\% of the area in front of a continuously graded color. }
    \label{fig_ContinousBGBGConstantPerception/020}
 \end{subfigure}
 \hfill 
    \caption{Comparison of our method with constant color. Note that Our variable color band creates constant perception}
\label{fig_020}
\end{figure}

\begin{figure}[hbtp]
\centering
     \begin{subfigure}[t]{0.32\textwidth}
\includegraphics[width=0.99\textwidth]{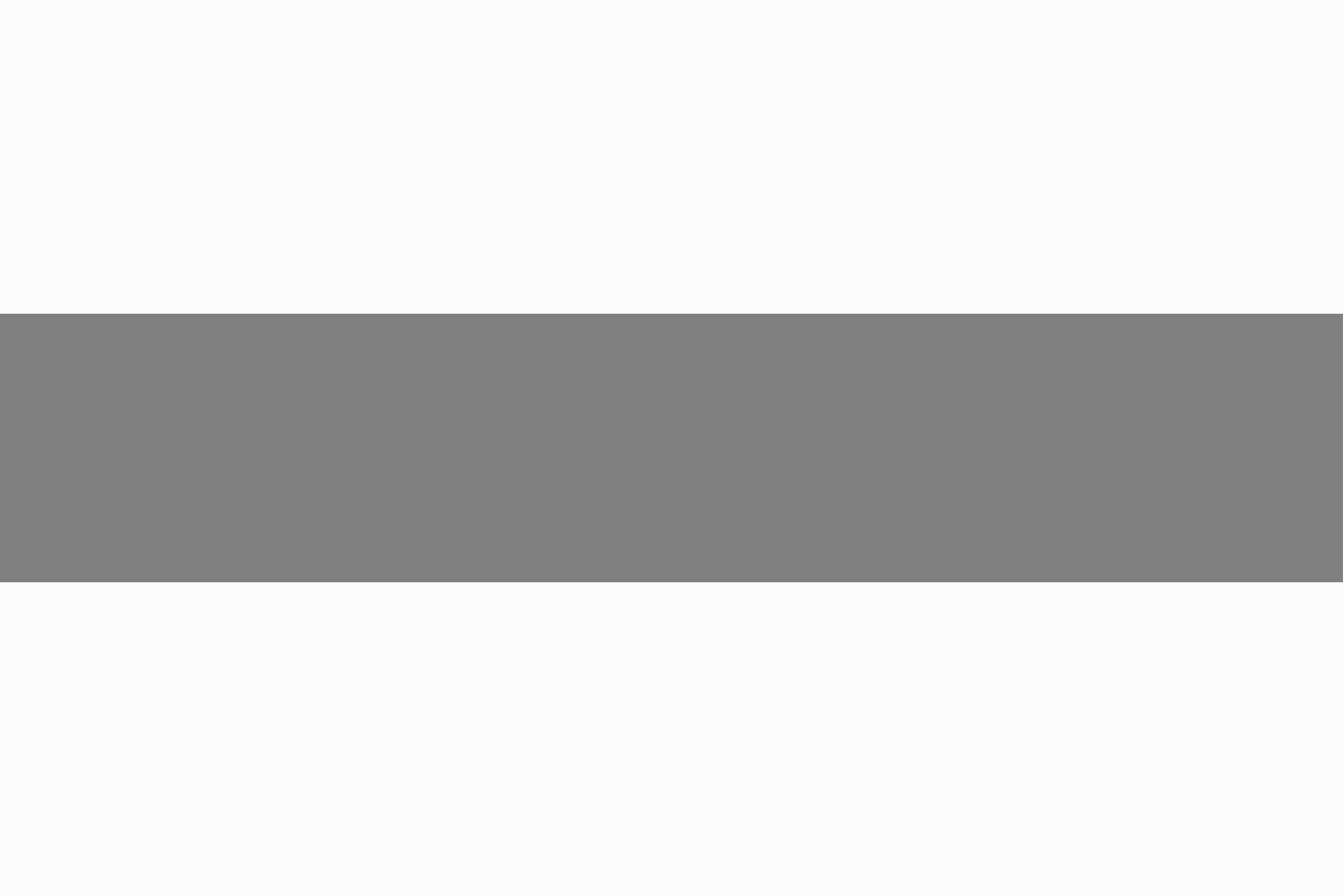}
    \caption{Constant color covering 30\% of the area in front of white background. }
    \label{fig_WhiteBGConstantColor/030}
 \end{subfigure}
 \hfill
     \begin{subfigure}[t]{0.32\textwidth}
\includegraphics[width=0.99\textwidth]{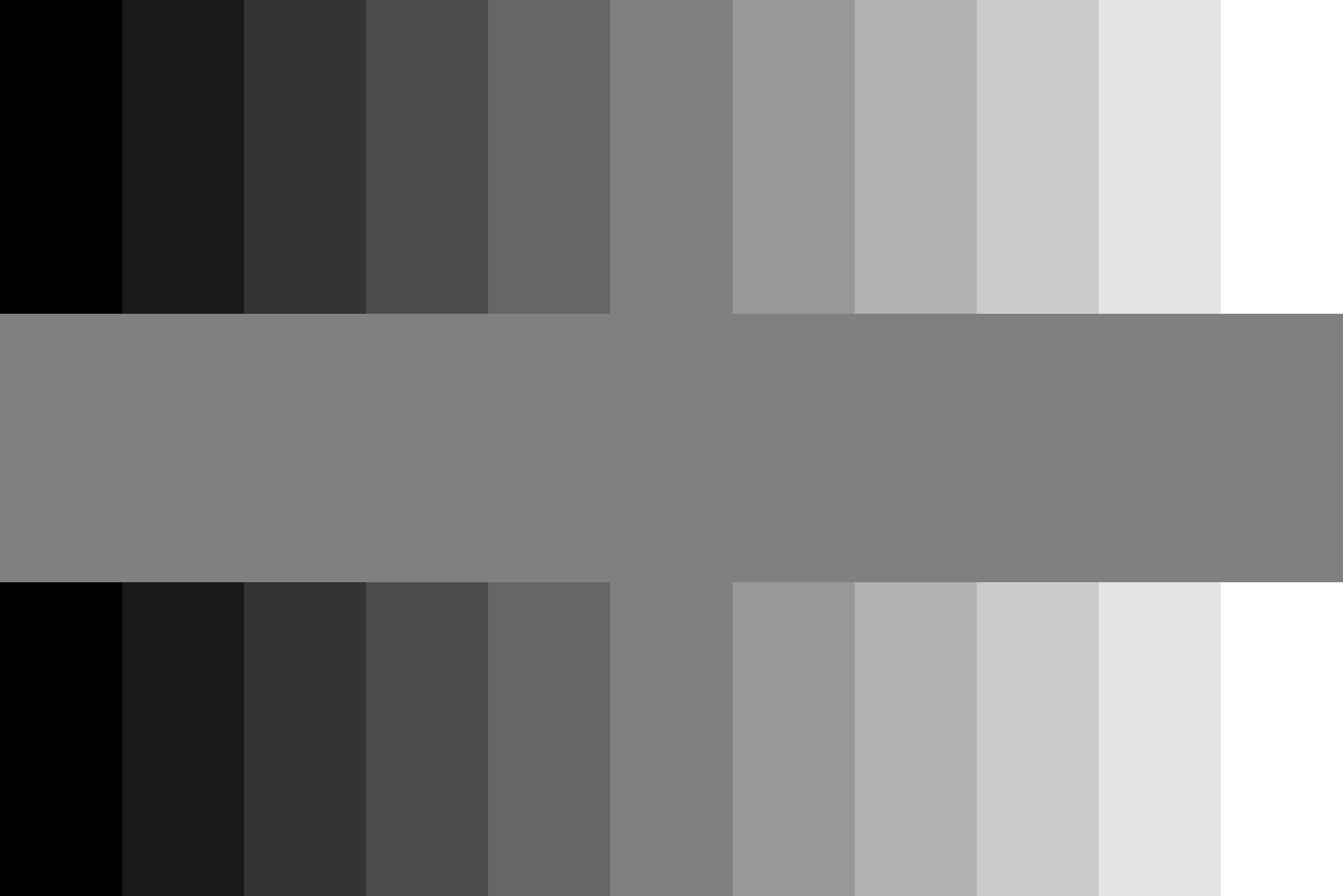}
    \caption{Constant color covering 30\% of the area in front of a graded color that consists of 30 bands. }
    \label{fig_10BandBGConstantColor/030}
 \end{subfigure}
 \hfill
      \begin{subfigure}[t]{0.32\textwidth}
\includegraphics[width=0.99\textwidth]{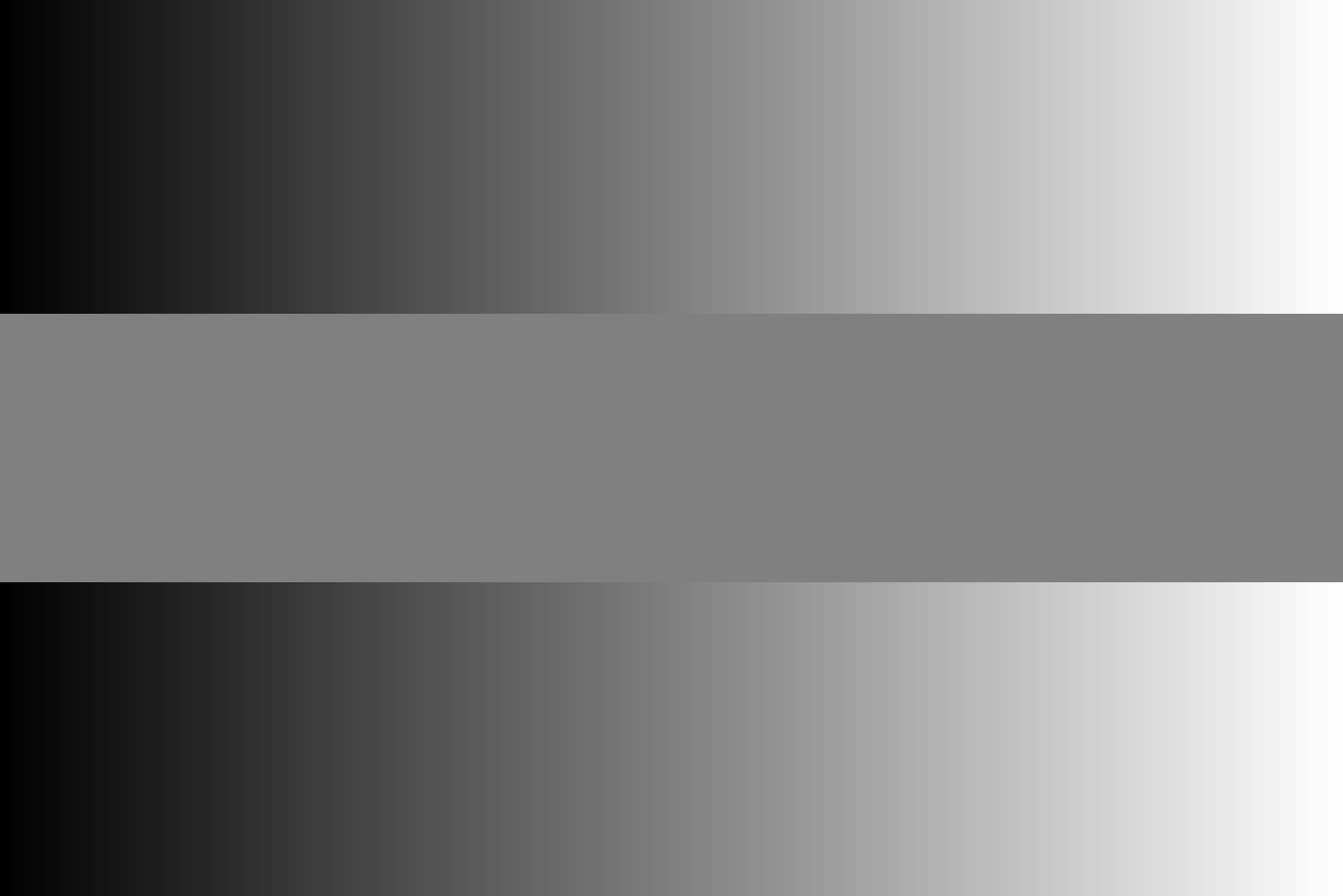}
    \caption{Constant color covering 30\% of the area in front of a continuously graded color. }
    \label{fig_ContinousBGConstantColor/030}
 \end{subfigure}
 \hfill
     \begin{subfigure}[t]{0.32\textwidth}
\includegraphics[width=0.99\textwidth]{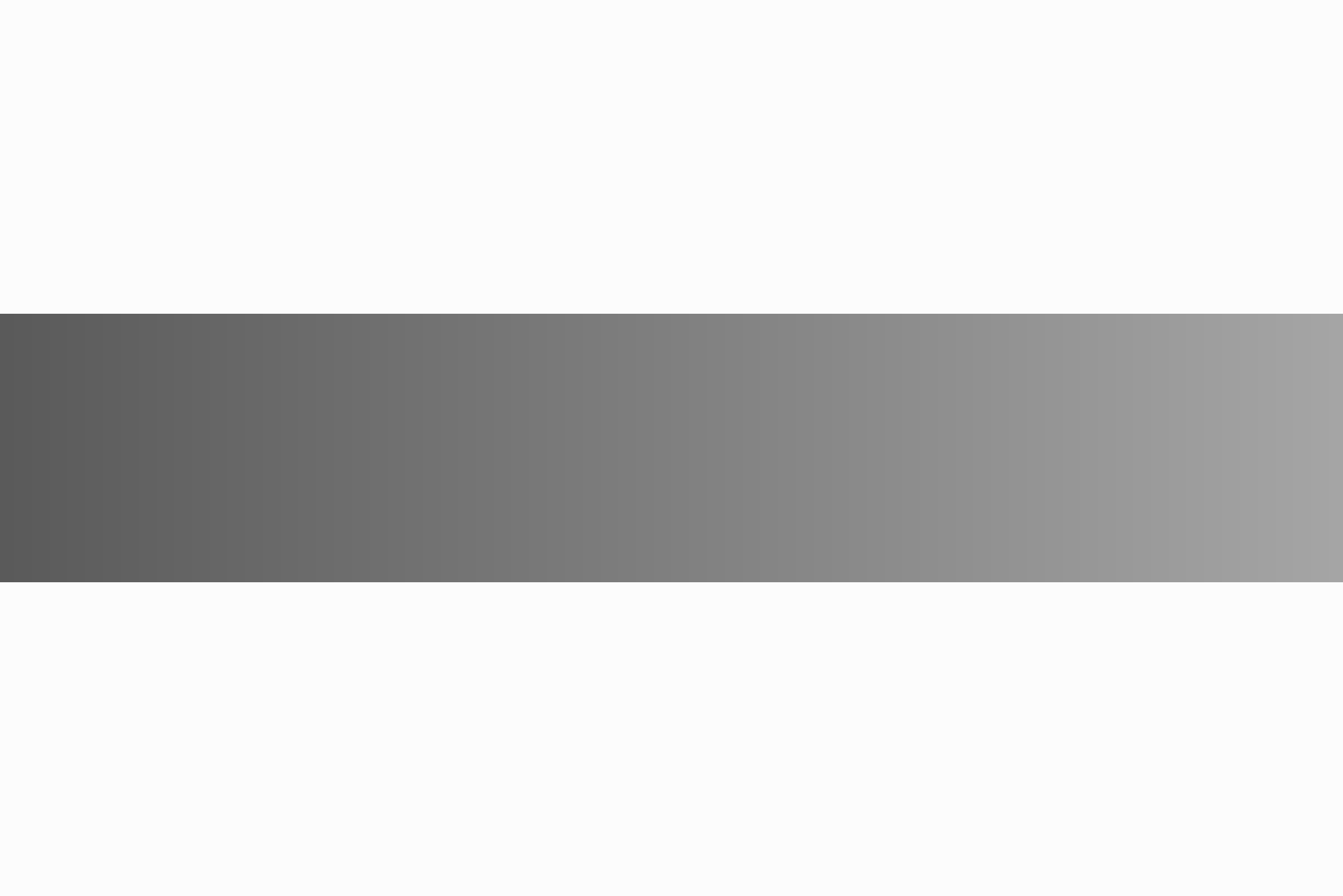}
    \caption{Our variable color band covers 30\% of the area in front of a white background. }
    \label{fig_WhiteBGConstantPerception/030}
 \end{subfigure}
 \hfill
     \begin{subfigure}[t]{0.32\textwidth}
\includegraphics[width=0.99\textwidth]{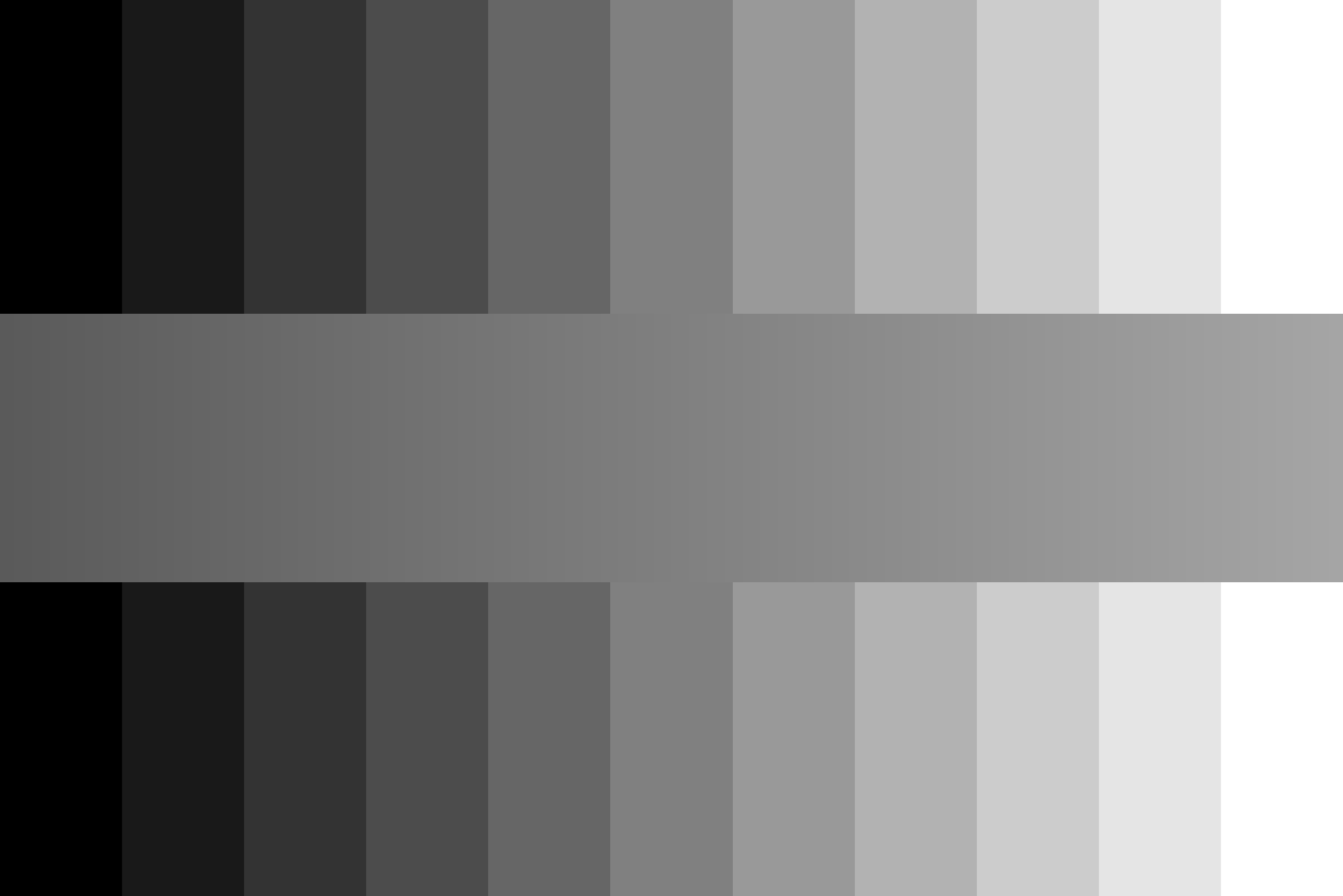}
    \caption{Our variable color band covers 30\% of the area in front of a graded color that consists of 30 bands. }
    \label{fig_10BandBGConstantPerception/030}
 \end{subfigure}
 \hfill
      \begin{subfigure}[t]{0.32\textwidth}
\includegraphics[width=0.99\textwidth]{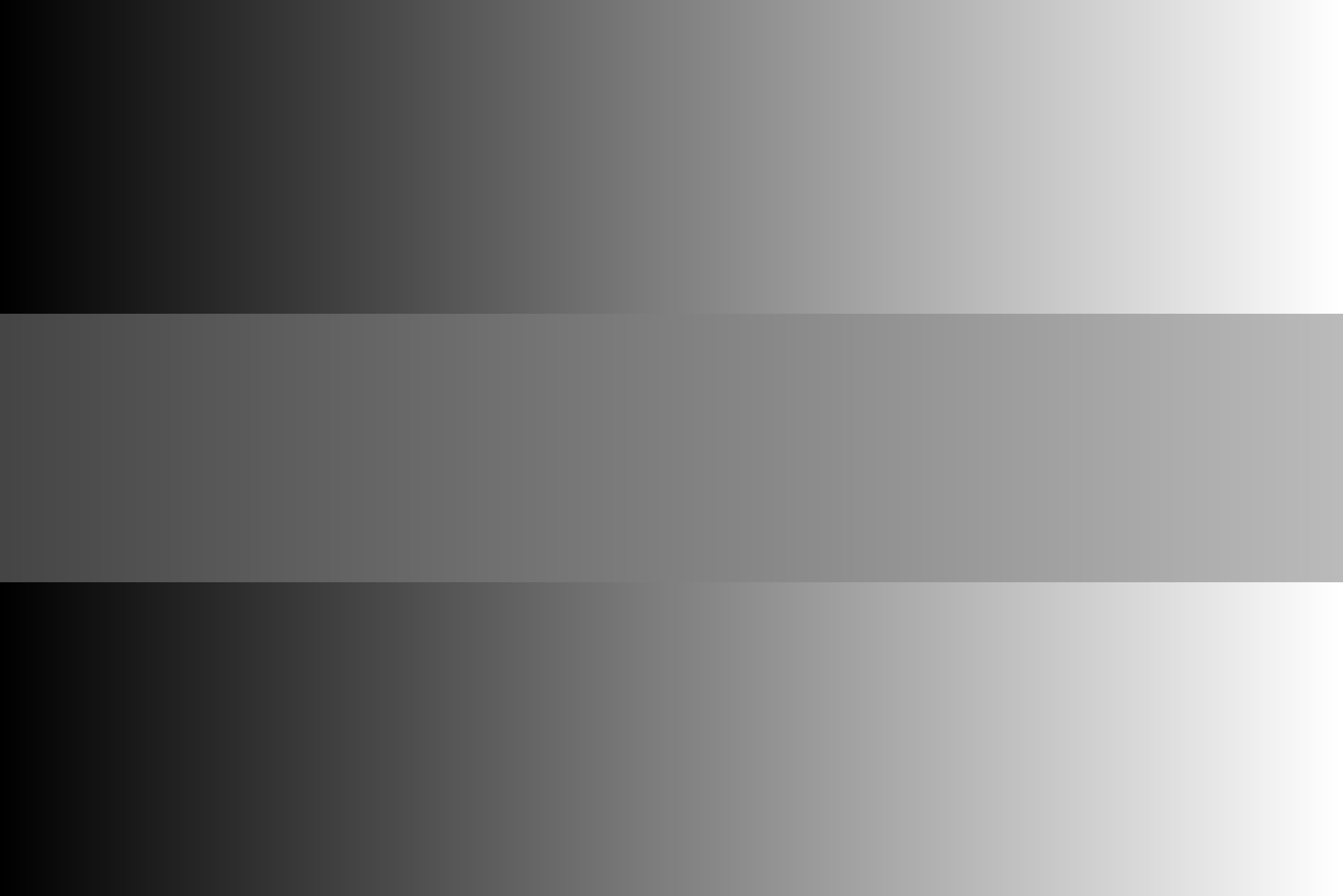}
    \caption{Our variable color band covers 30\% of the area in front of a continuously graded color. }
    \label{fig_ContinousBGBGConstantPerception/030}
 \end{subfigure}
 \hfill 
    \caption{Comparison of our method with constant color. Note that Our variable color band creates constant perception}
\label{fig_030}
\end{figure}

\begin{figure}[hbtp]
\centering
     \begin{subfigure}[t]{0.32\textwidth}
\includegraphics[width=0.99\textwidth]{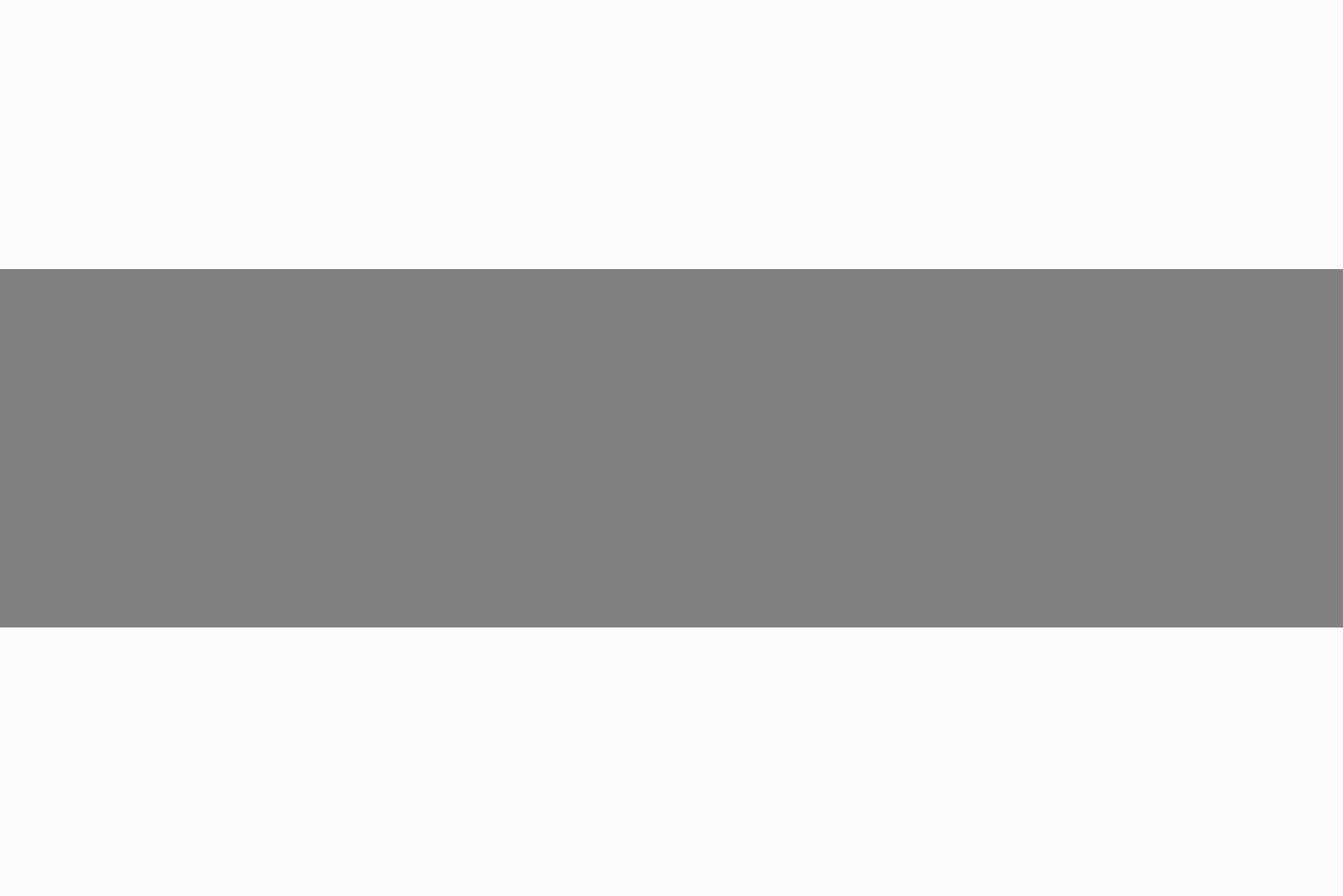}
    \caption{Constant color covering 40\% of the area in front of white background. }
    \label{fig_WhiteBGConstantColor/040}
 \end{subfigure}
 \hfill
     \begin{subfigure}[t]{0.32\textwidth}
\includegraphics[width=0.99\textwidth]{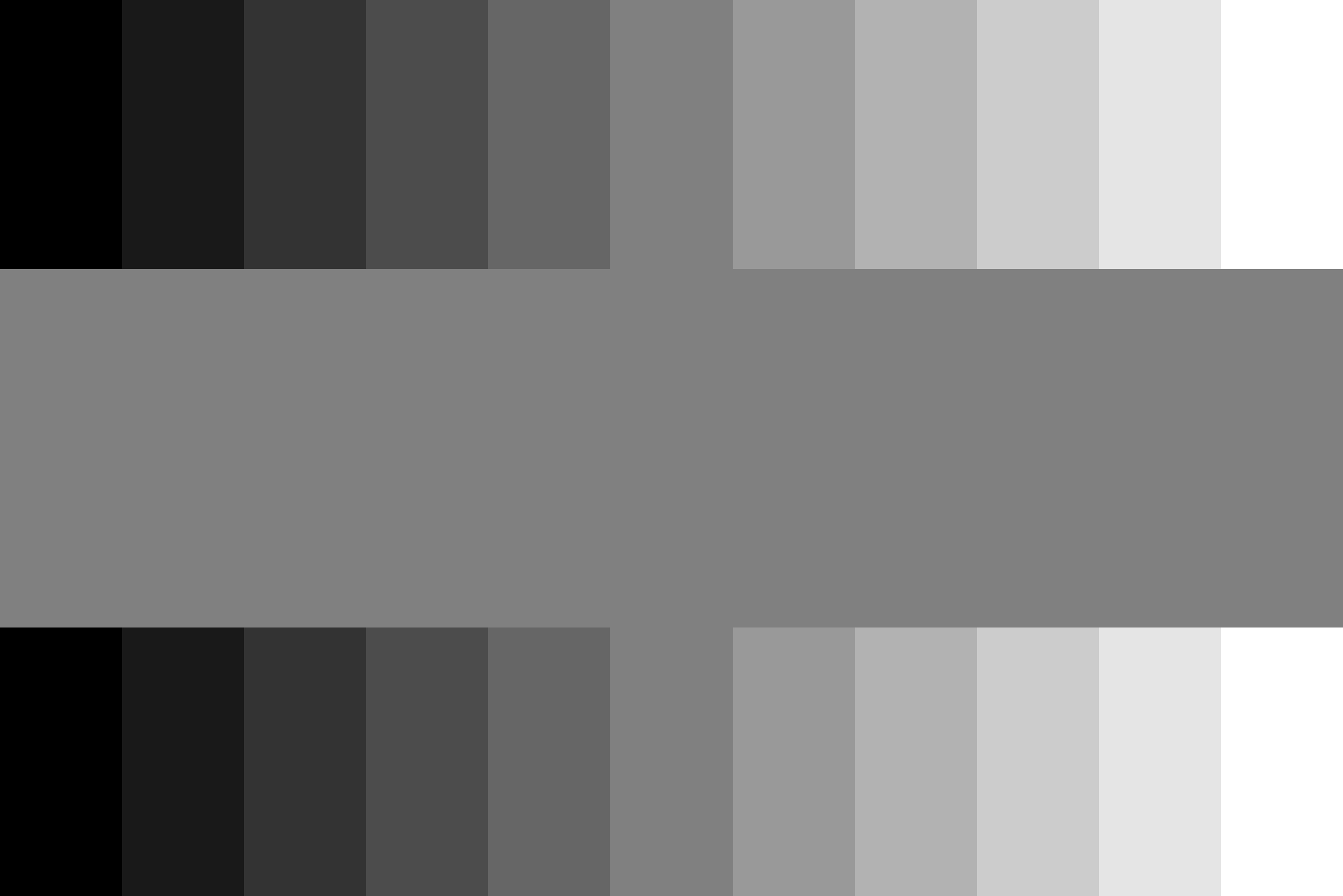}
    \caption{Constant color covering 40\% of the area in front of a graded color that consists of 40 bands. }
    \label{fig_10BandBGConstantColor/040}
 \end{subfigure}
 \hfill
      \begin{subfigure}[t]{0.32\textwidth}
\includegraphics[width=0.99\textwidth]{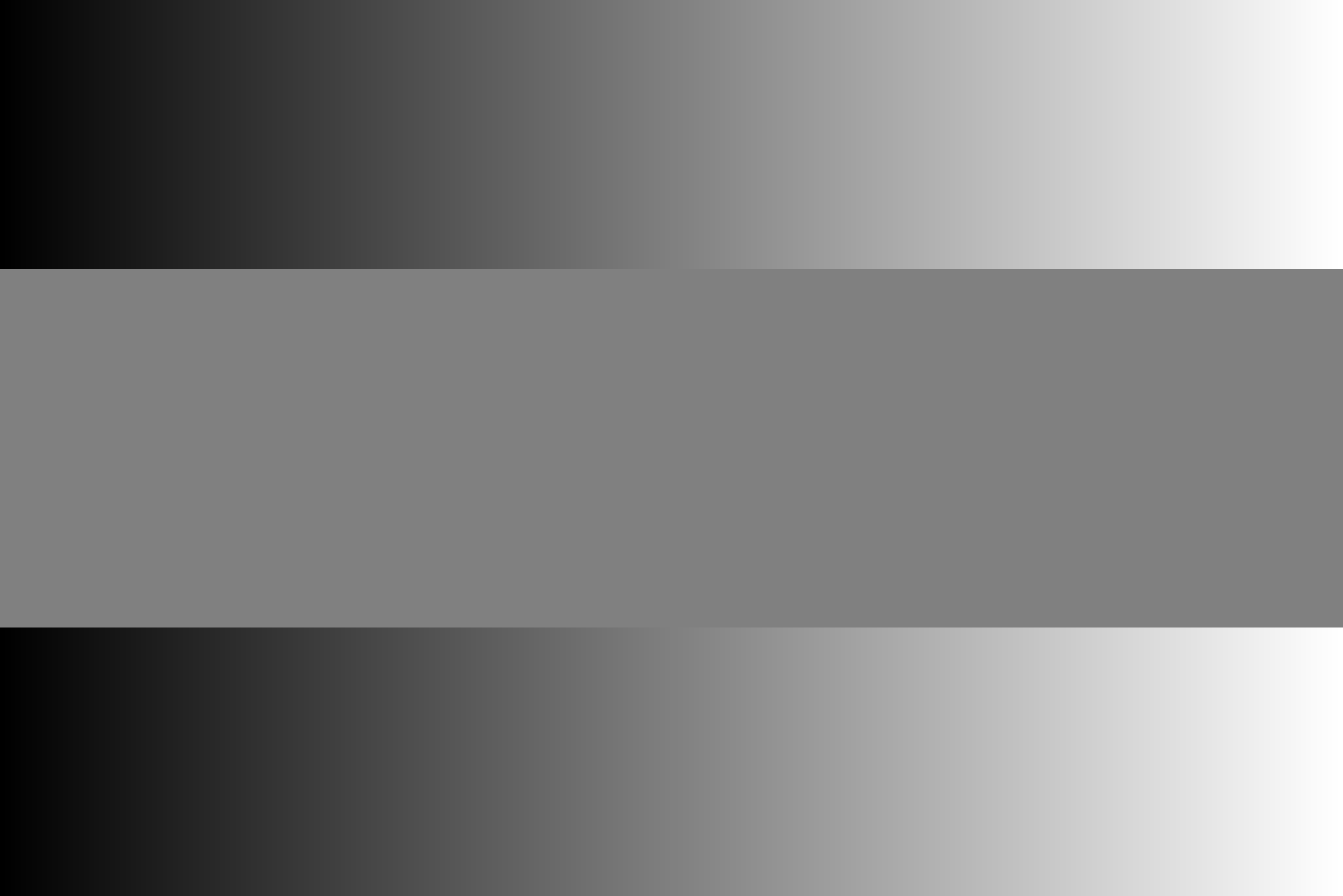}
    \caption{Constant color covering 40\% of the area in front of a continuously graded color. }
    \label{fig_ContinousBGConstantColor/040}
 \end{subfigure}
 \hfill
     \begin{subfigure}[t]{0.32\textwidth}
\includegraphics[width=0.99\textwidth]{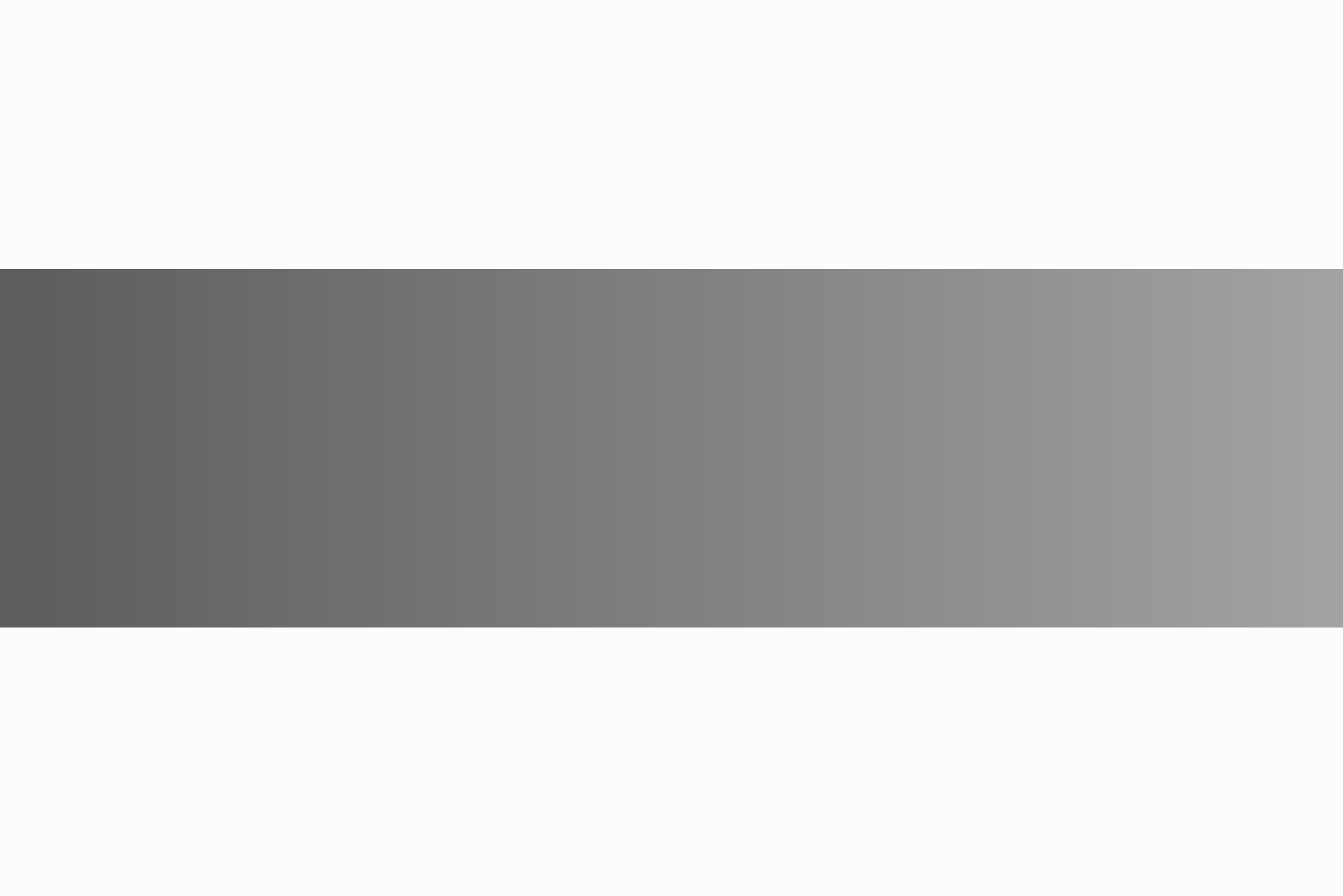}
    \caption{Our variable color band covers 40\% of the area in front of a white background. }
    \label{fig_WhiteBGConstantPerception/040}
 \end{subfigure}
 \hfill
     \begin{subfigure}[t]{0.32\textwidth}
\includegraphics[width=0.99\textwidth]{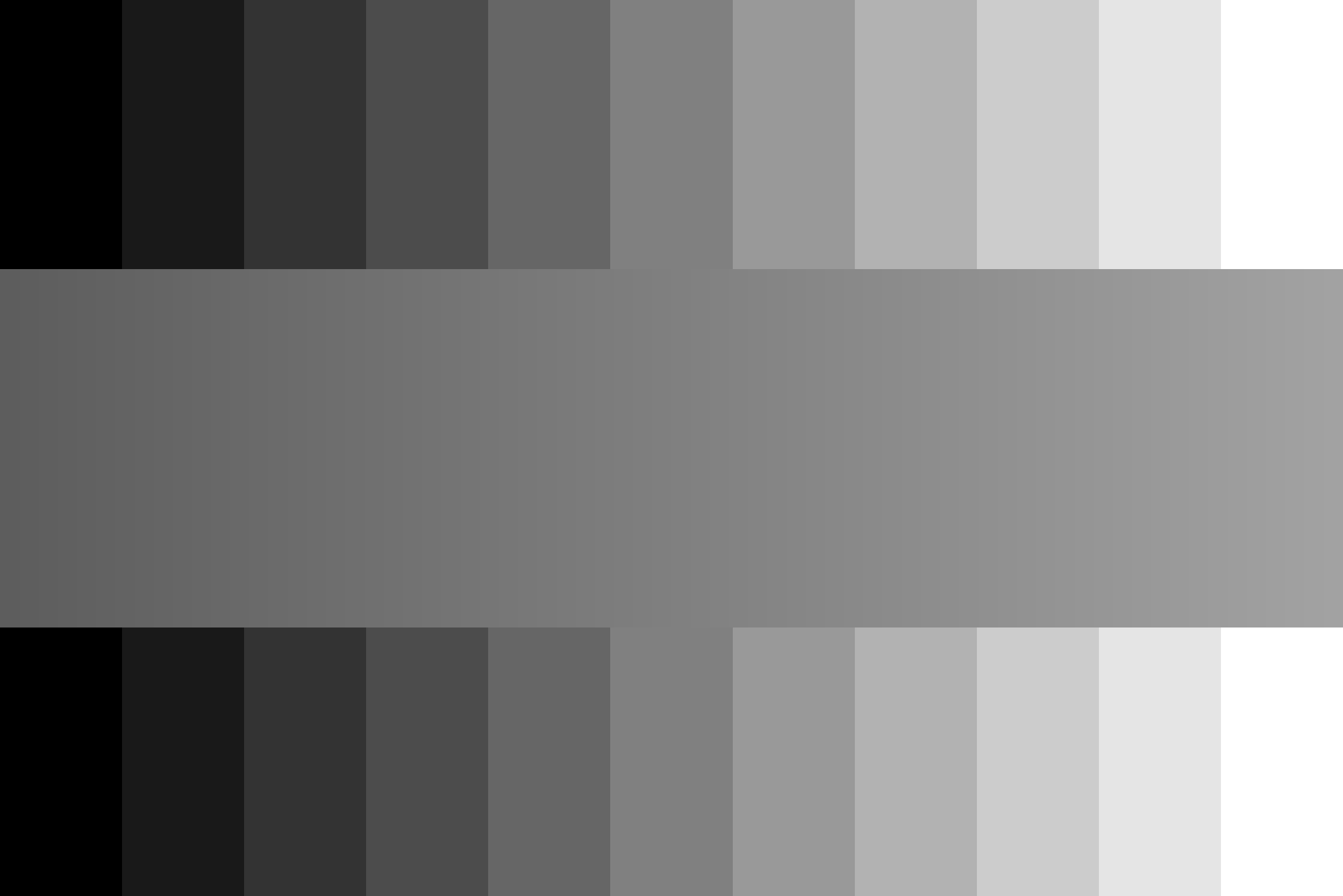}
    \caption{Our variable color band covers 40\% of the area in front of a graded color that consists of 40 bands. }
    \label{fig_10BandBGConstantPerception/040}
 \end{subfigure}
 \hfill
      \begin{subfigure}[t]{0.32\textwidth}
\includegraphics[width=0.99\textwidth]{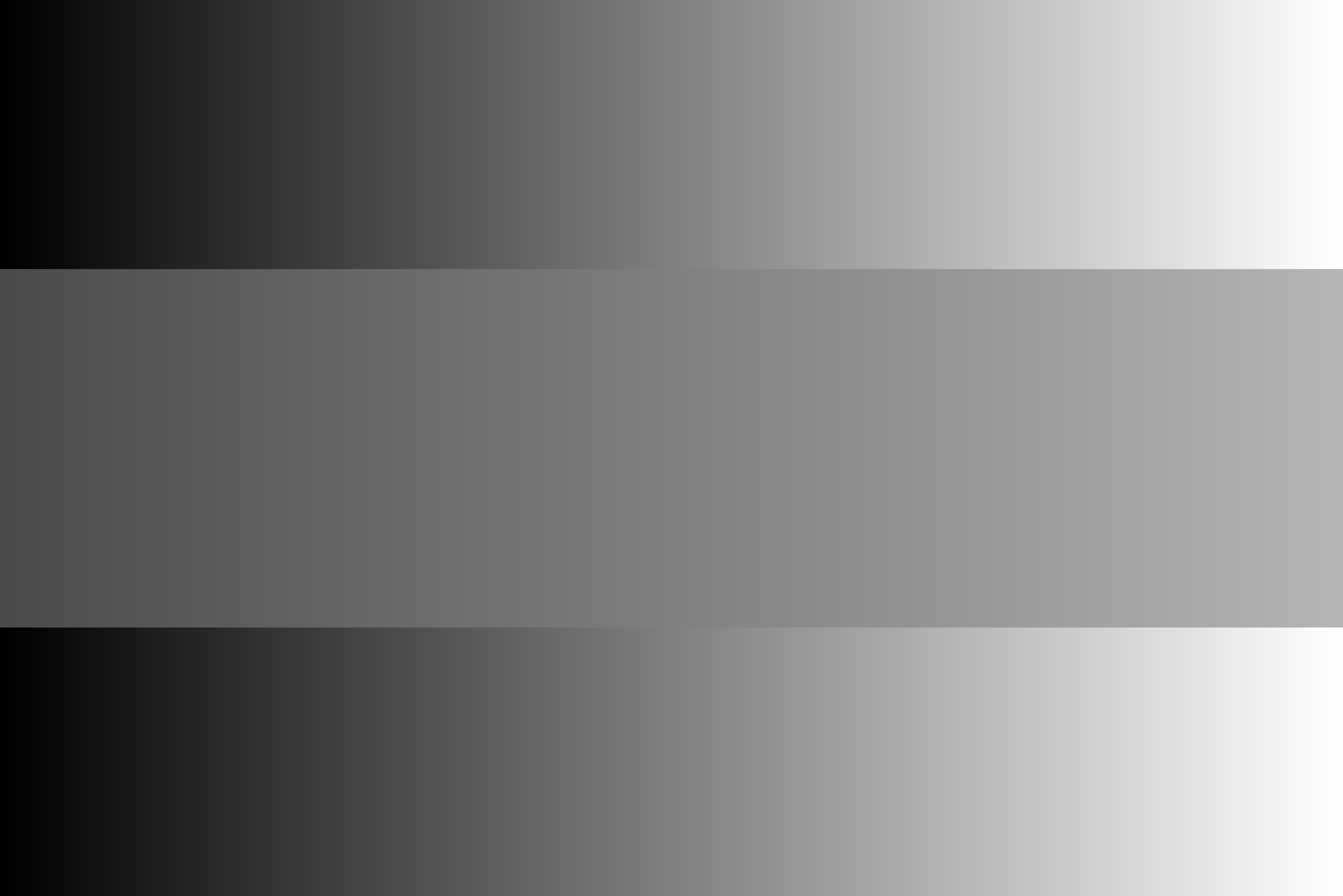}
    \caption{Our variable color band covers 40\% of the area in front of a continuously graded color. }
    \label{fig_ContinousBGBGConstantPerception/040}
 \end{subfigure}
 \hfill 
    \caption{Comparison of our method with constant color. Note that Our variable color band creates constant perception}
\label{fig_040}
\end{figure}

\begin{figure}[hbtp]
\centering
     \begin{subfigure}[t]{0.32\textwidth}
\includegraphics[width=0.99\textwidth]{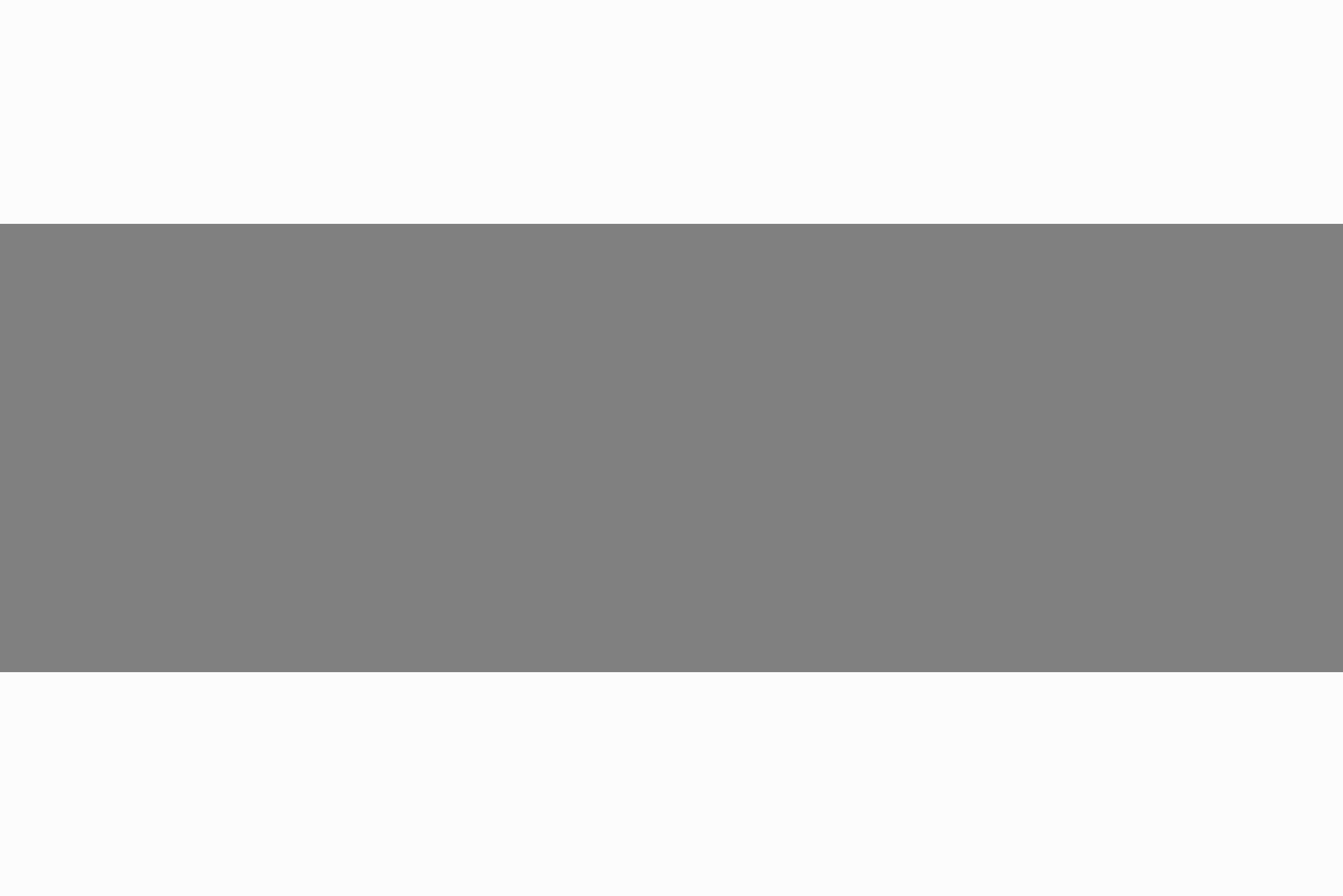}
    \caption{Constant color covering 50\% of the area in front of white background. }
    \label{fig_WhiteBGConstantColor/050}
 \end{subfigure}
 \hfill
     \begin{subfigure}[t]{0.32\textwidth}
\includegraphics[width=0.99\textwidth]{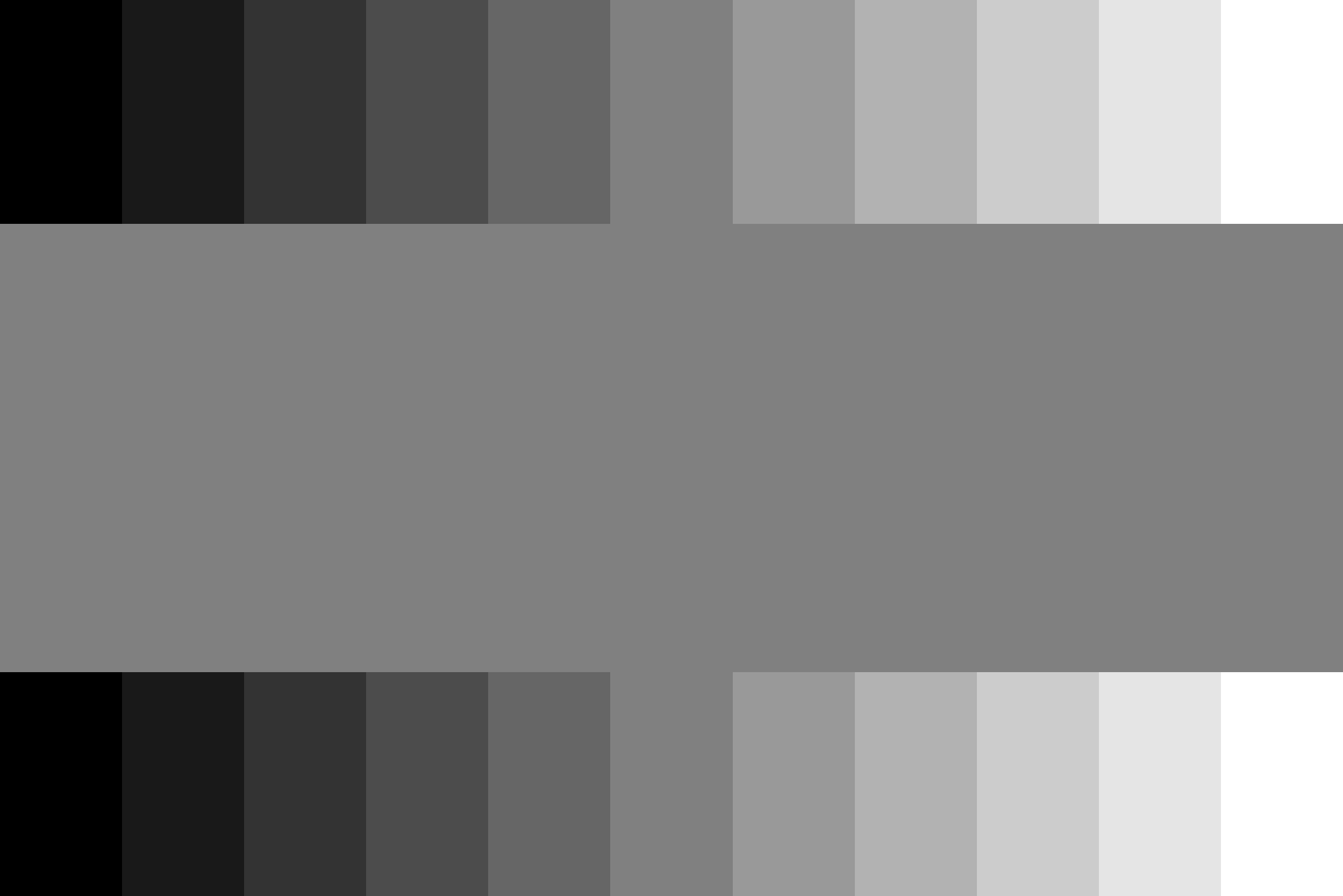}
    \caption{Constant color covering 50\% of the area in front of a graded color that consists of 10 bands. }
    \label{fig_10BandBGConstantColor/050}
 \end{subfigure}
 \hfill
      \begin{subfigure}[t]{0.32\textwidth}
\includegraphics[width=0.99\textwidth]{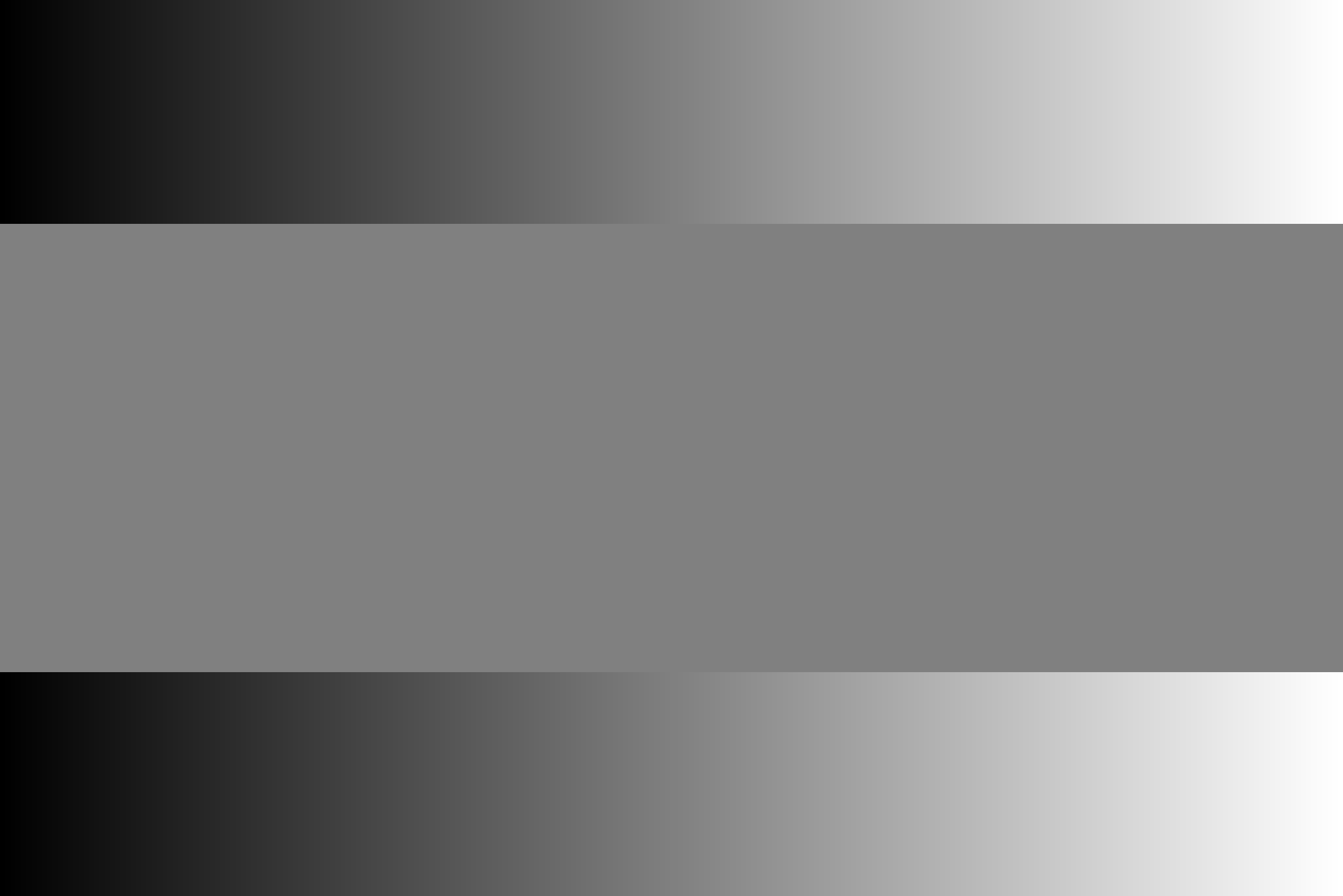}
    \caption{Constant color covering 50\% of the area in front of a continuously graded color. }
    \label{fig_ContinousBGConstantColor/050}
 \end{subfigure}
 \hfill
     \begin{subfigure}[t]{0.32\textwidth}
\includegraphics[width=0.99\textwidth]{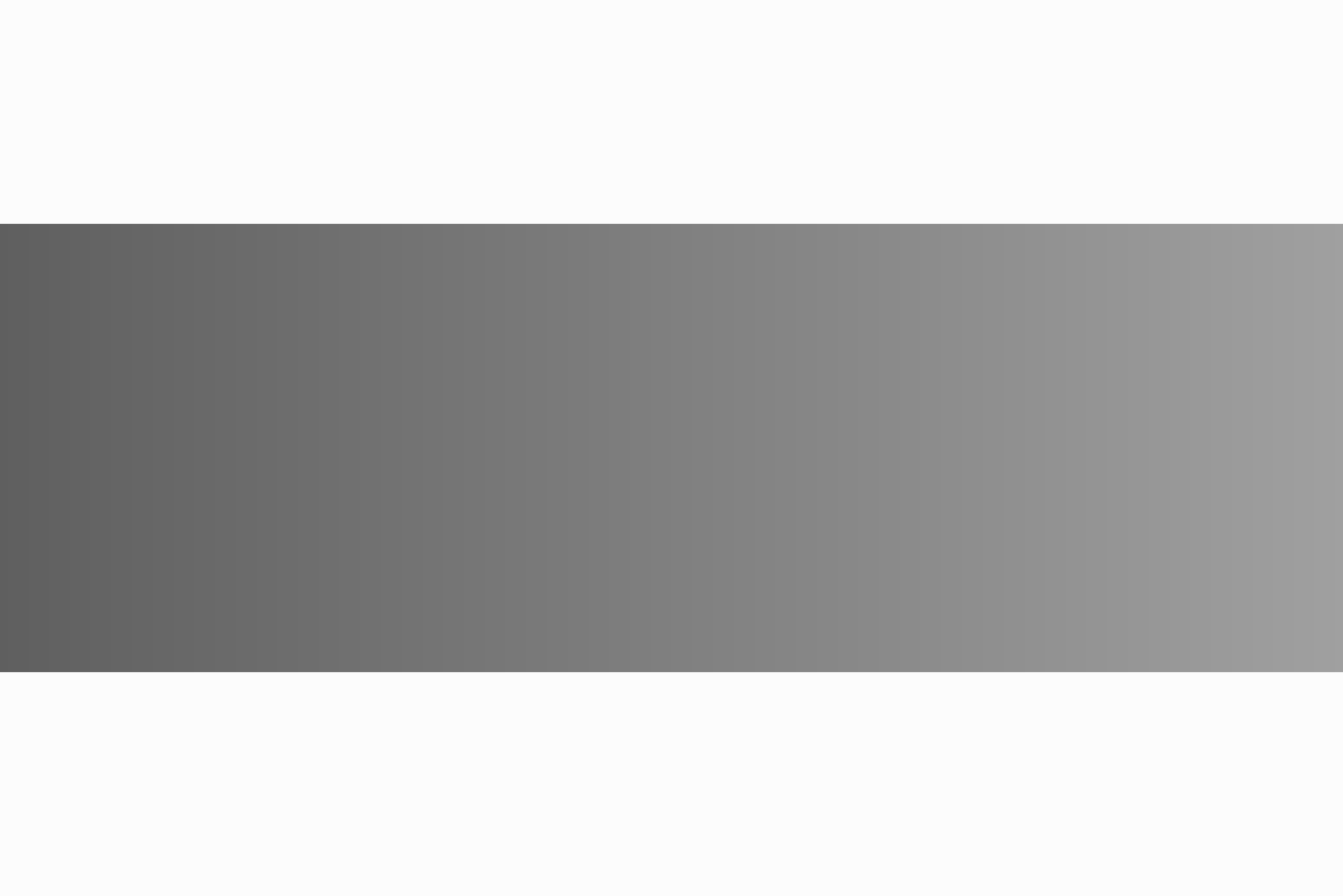}
    \caption{Our variable color band covers 50\% of the area in front of a white background. }
    \label{fig_WhiteBGConstantPerception/050}
 \end{subfigure}
 \hfill
     \begin{subfigure}[t]{0.32\textwidth}
\includegraphics[width=0.99\textwidth]{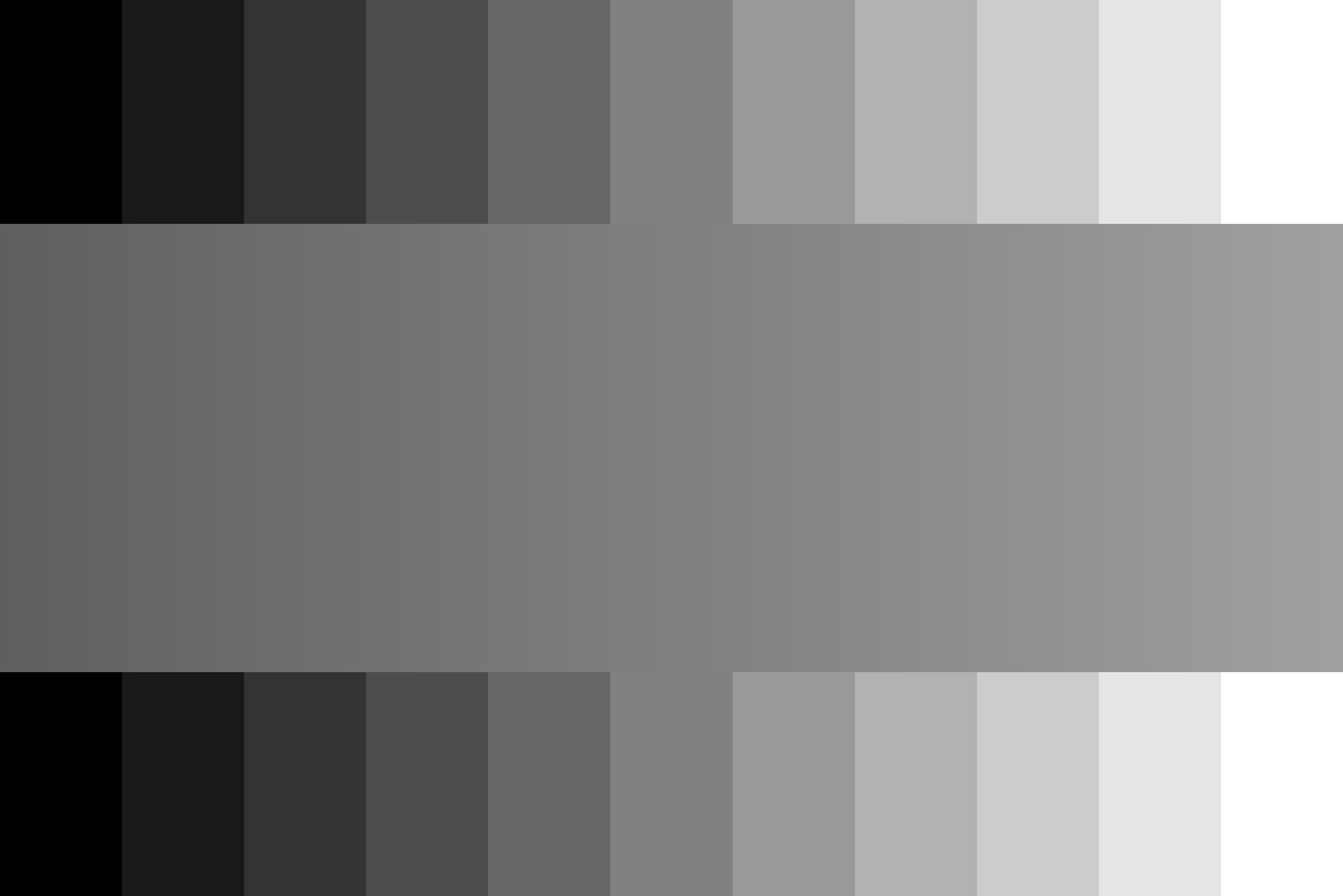}
    \caption{Our variable color band covers 50\% of the area in front of a graded color that consists of 10 bands. }
    \label{fig_10BandBGConstantPerception/050}
 \end{subfigure}
 \hfill
      \begin{subfigure}[t]{0.32\textwidth}
\includegraphics[width=0.99\textwidth]{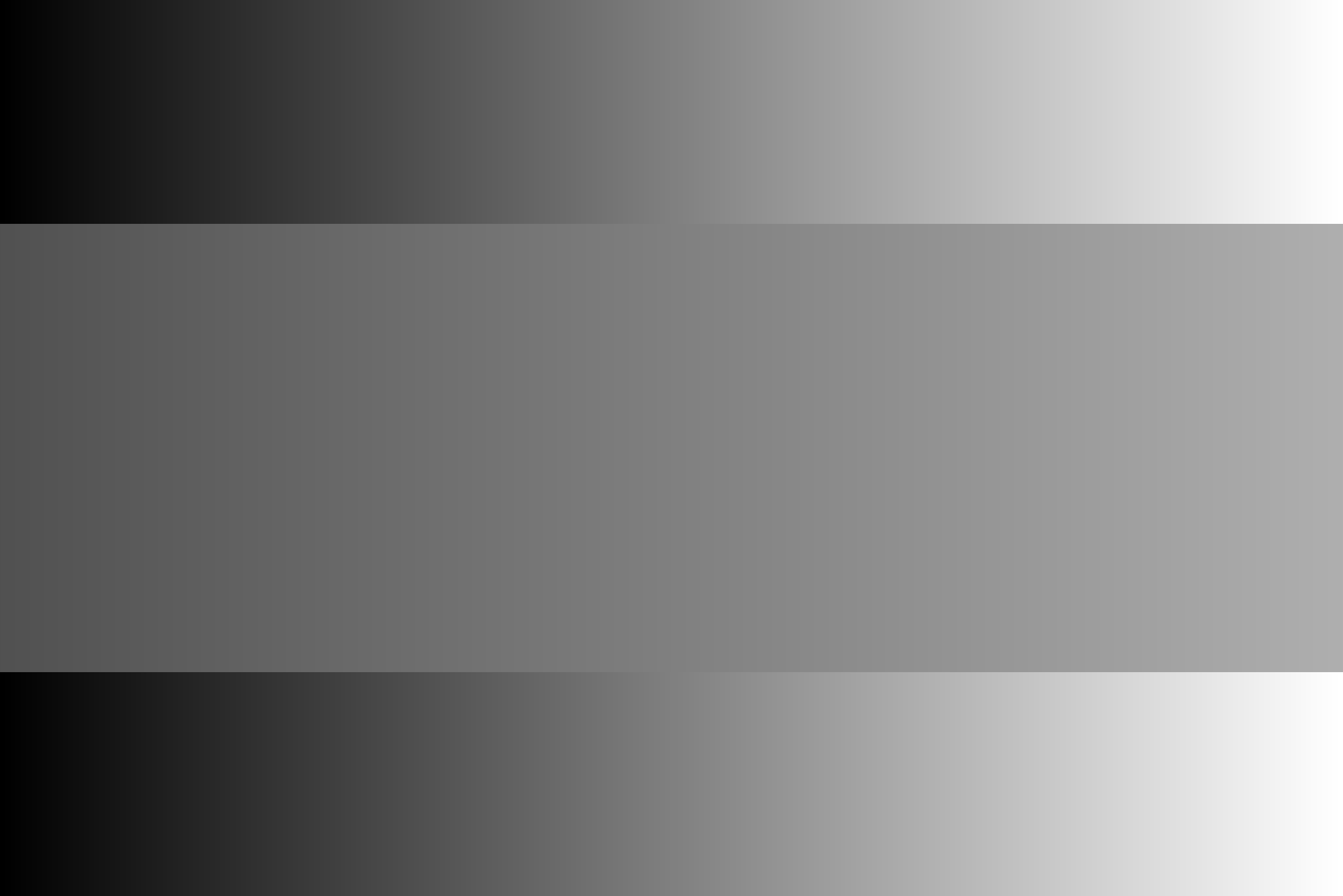}
    \caption{Our variable color band covers 50\% of the area in front of a continuously graded color. }
    \label{fig_ContinousBGBGConstantPerception/050}
 \end{subfigure}
 \hfill 
    \caption{Comparison of our method with constant color. Note that Our variable color band creates constant perception}
\label{fig_050}
\end{figure}

\begin{figure}[hbtp]
\centering
     \begin{subfigure}[t]{0.32\textwidth}
\includegraphics[width=0.99\textwidth]{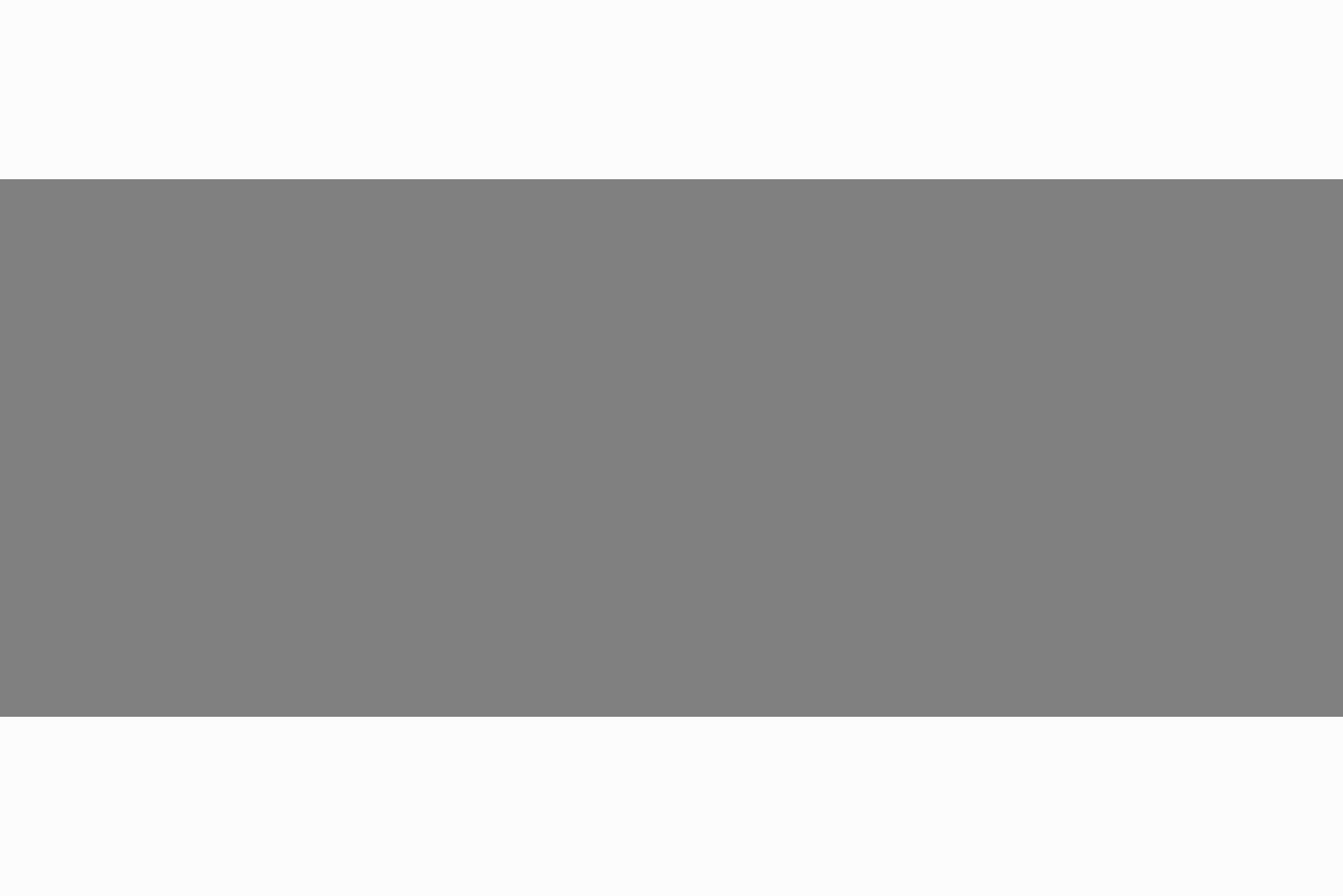}
    \caption{Constant color covering 60\% of the area in front of white background. }
    \label{fig_WhiteBGConstantColor/060}
 \end{subfigure}
 \hfill
     \begin{subfigure}[t]{0.32\textwidth}
\includegraphics[width=0.99\textwidth]{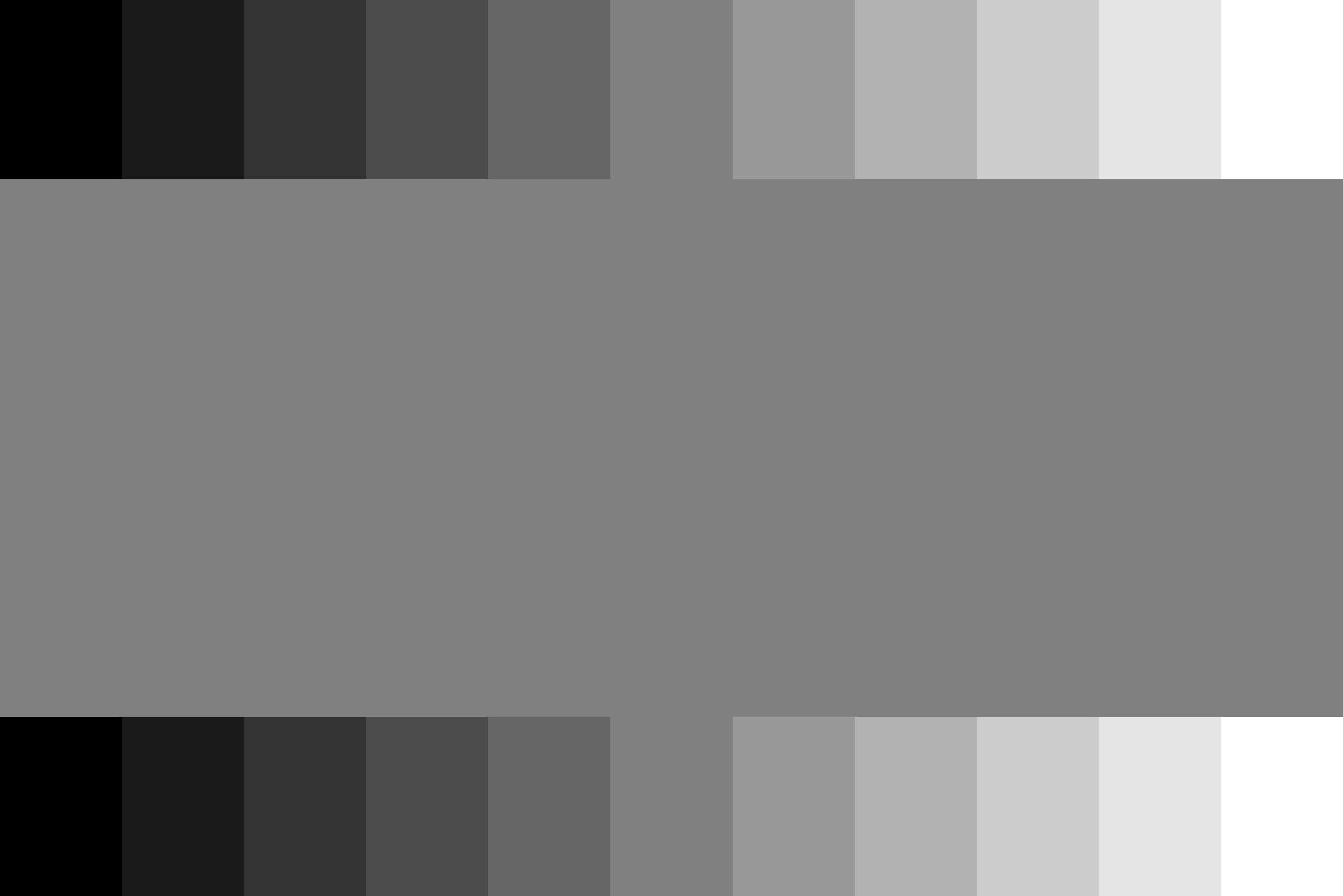}
    \caption{Constant color covering 60\% of the area in front of a graded color that consists of 60 bands. }
    \label{fig_10BandBGConstantColor/060}
 \end{subfigure}
 \hfill
      \begin{subfigure}[t]{0.32\textwidth}
\includegraphics[width=0.99\textwidth]{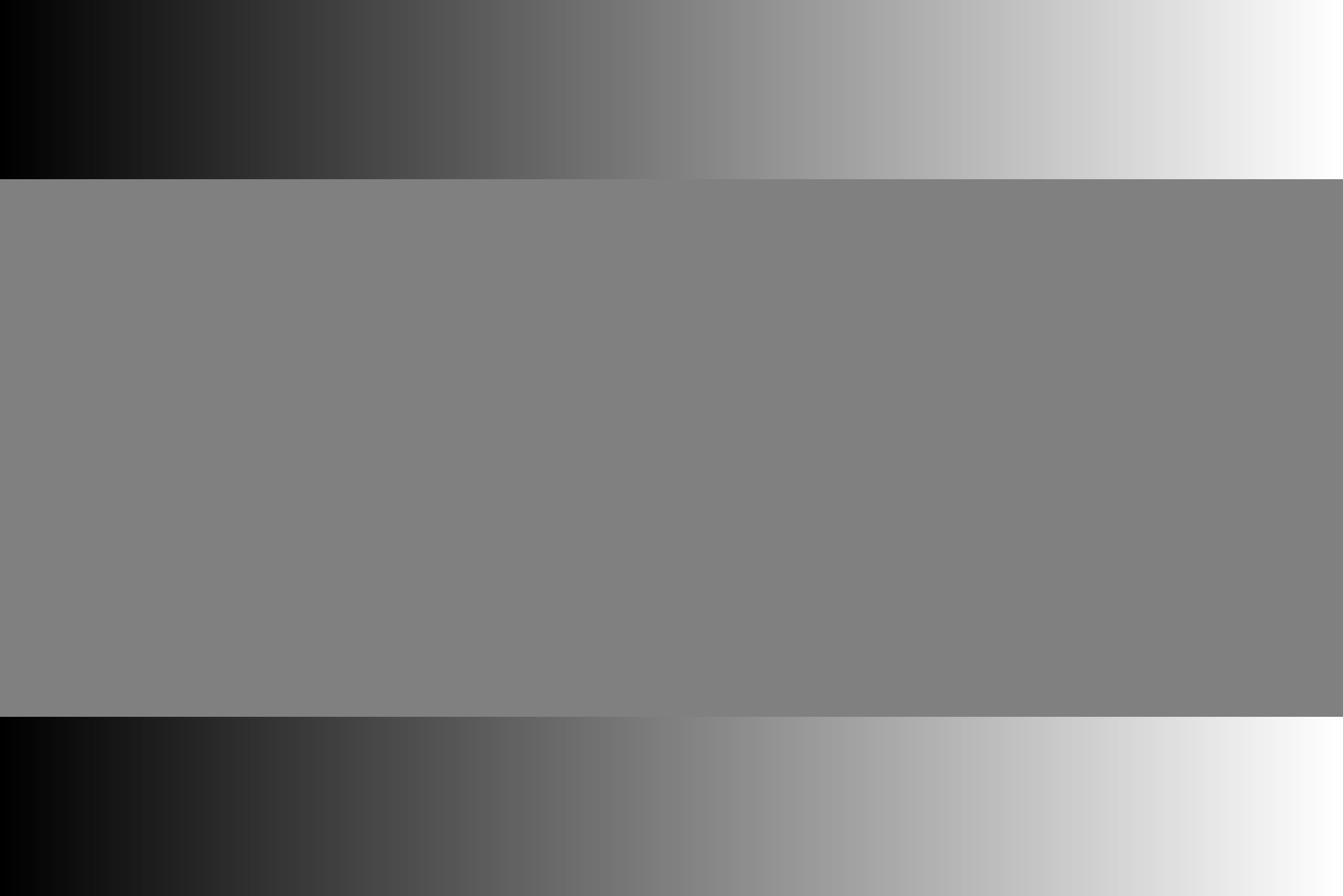}
    \caption{Constant color covering 60\% of the area in front of a continuously graded color. }
    \label{fig_ContinousBGConstantColor/060}
 \end{subfigure}
 \hfill
     \begin{subfigure}[t]{0.32\textwidth}
\includegraphics[width=0.99\textwidth]{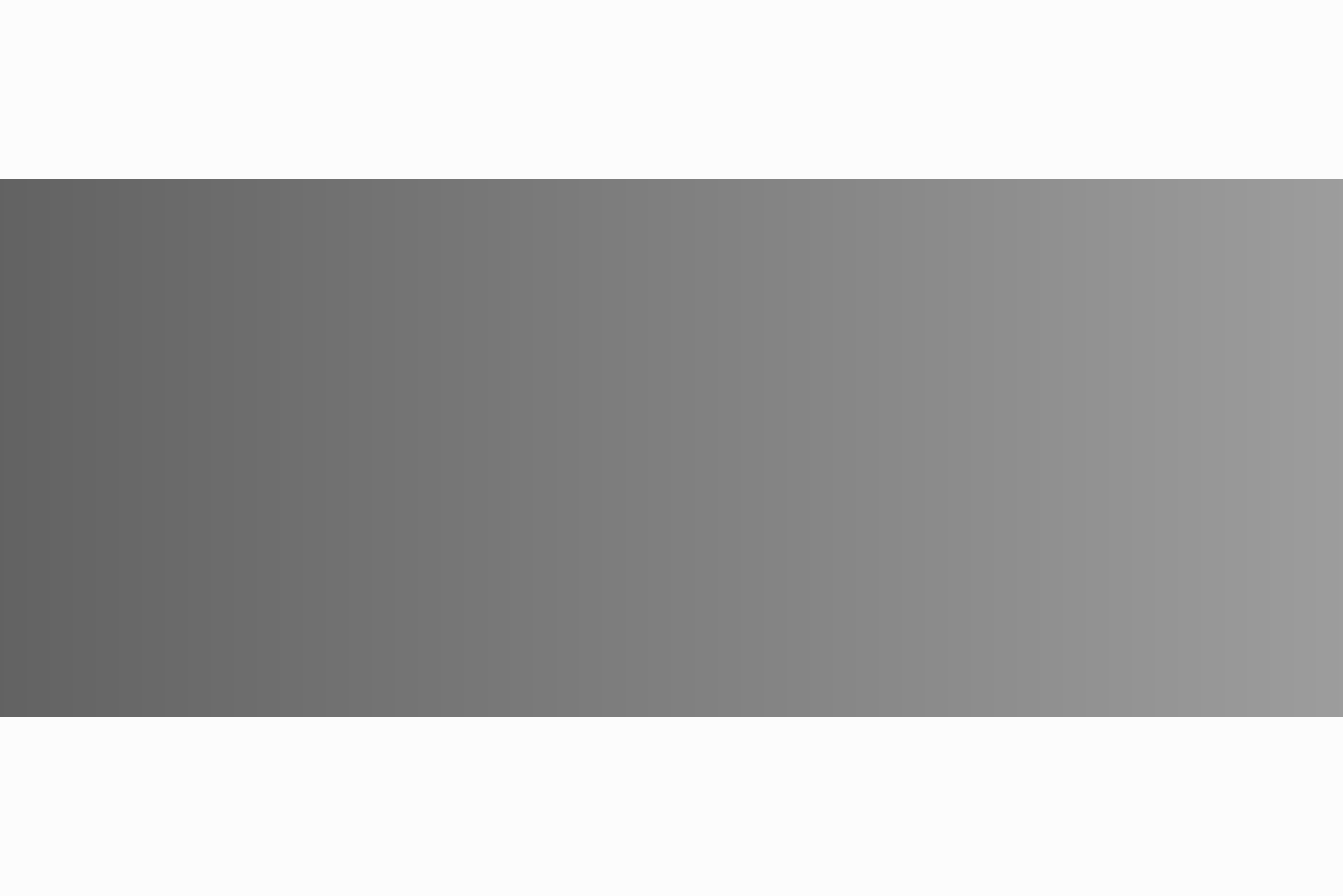}
    \caption{Our variable color band covers 60\% of the area in front of a white background. }
    \label{fig_WhiteBGConstantPerception/060}
 \end{subfigure}
 \hfill
     \begin{subfigure}[t]{0.32\textwidth}
\includegraphics[width=0.99\textwidth]{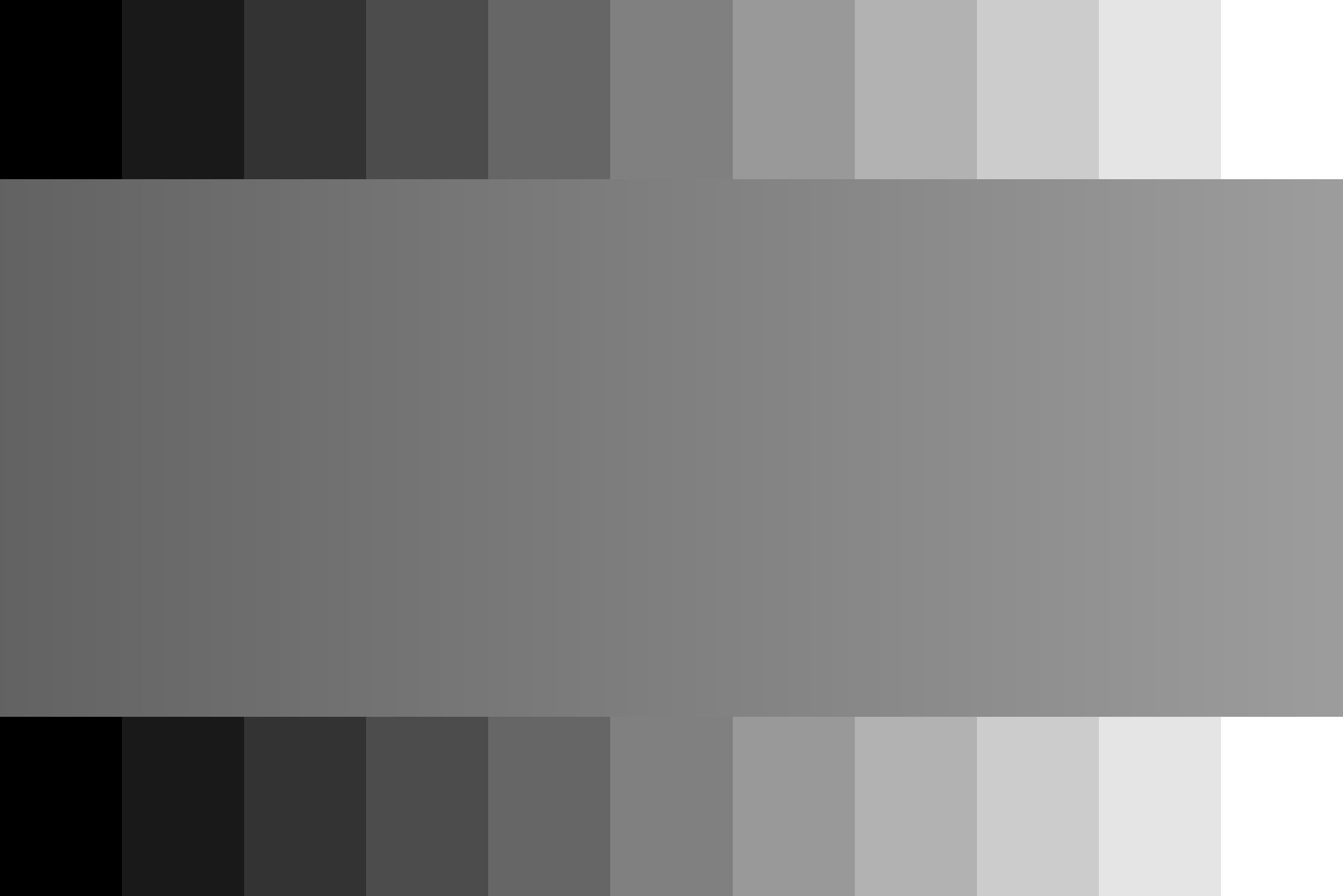}
    \caption{Our variable color band covers 60\% of the area in front of a graded color that consists of 60 bands. }
    \label{fig_10BandBGConstantPerception/060}
 \end{subfigure}
 \hfill
      \begin{subfigure}[t]{0.32\textwidth}
\includegraphics[width=0.99\textwidth]{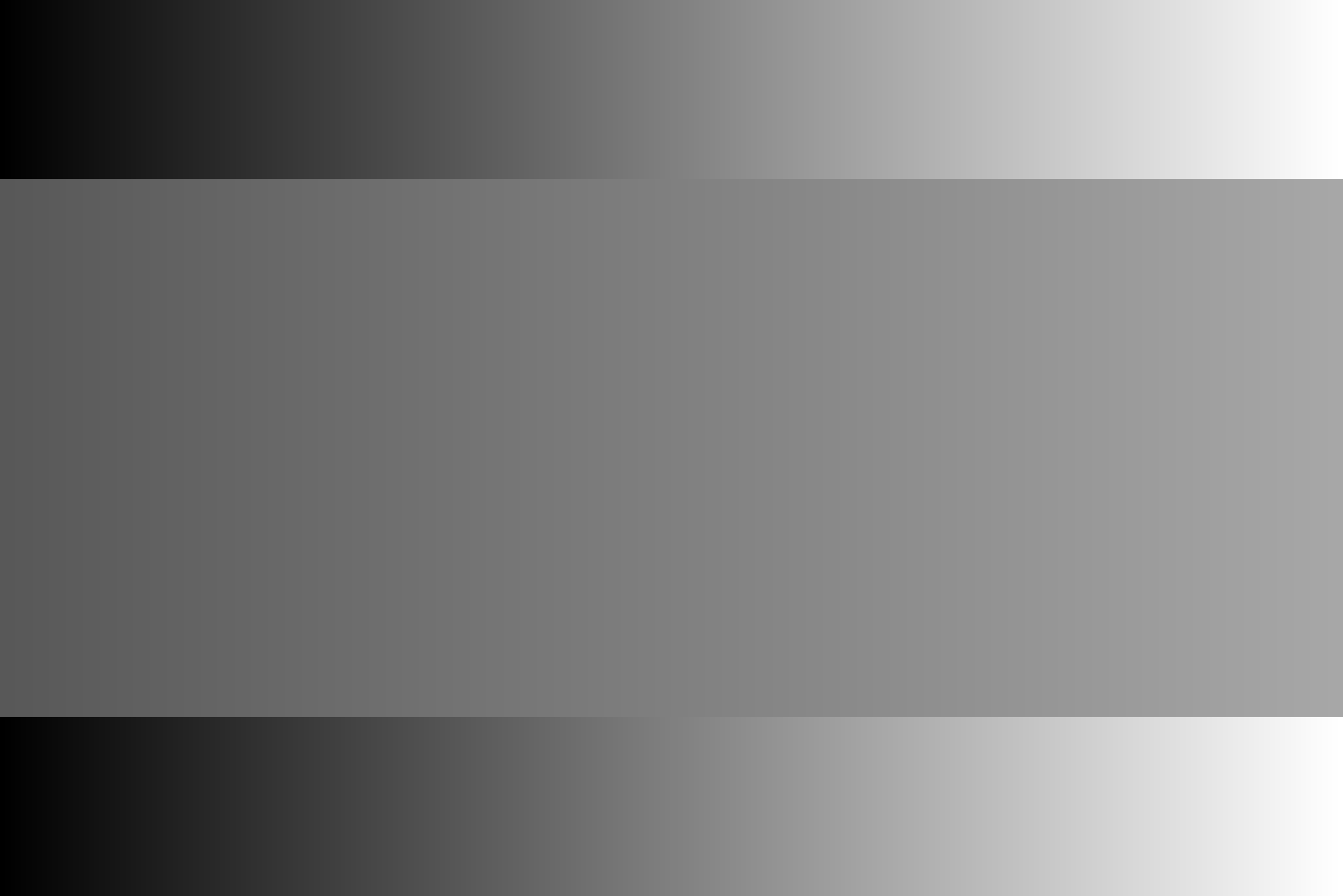}
    \caption{Our variable color band covers 60\% of the area in front of a continuously graded color. }
    \label{fig_ContinousBGBGConstantPerception/060}
 \end{subfigure}
 \hfill 
    \caption{Comparison of our method with constant color. Note that Our variable color band creates constant perception}
\label{fig_060}
\end{figure}

\begin{figure}[hbtp]
\centering
     \begin{subfigure}[t]{0.32\textwidth}
\includegraphics[width=0.99\textwidth]{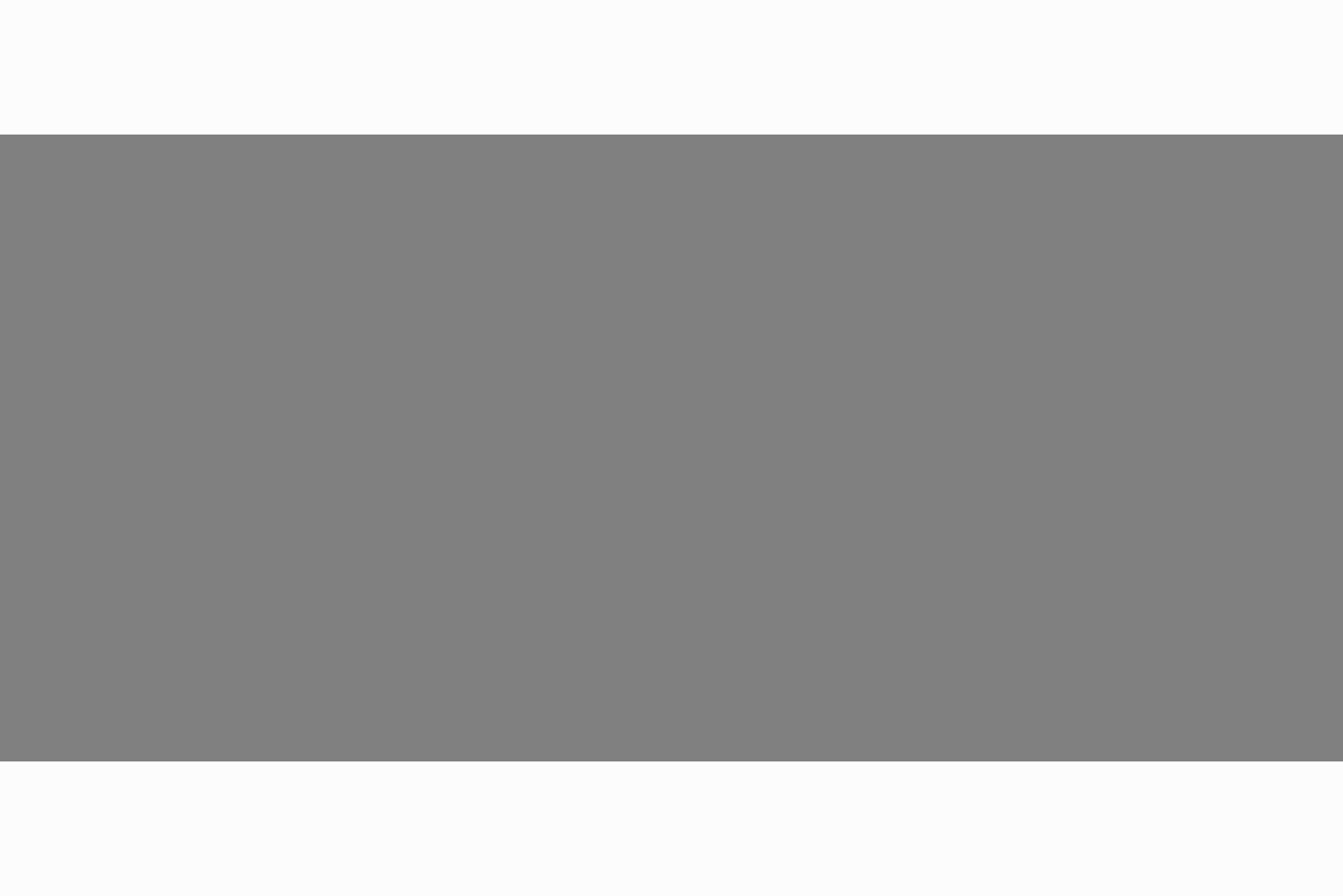}
    \caption{Constant color covering 70\% of the area in front of white background. }
    \label{fig_WhiteBGConstantColor/070}
 \end{subfigure}
 \hfill
     \begin{subfigure}[t]{0.32\textwidth}
\includegraphics[width=0.99\textwidth]{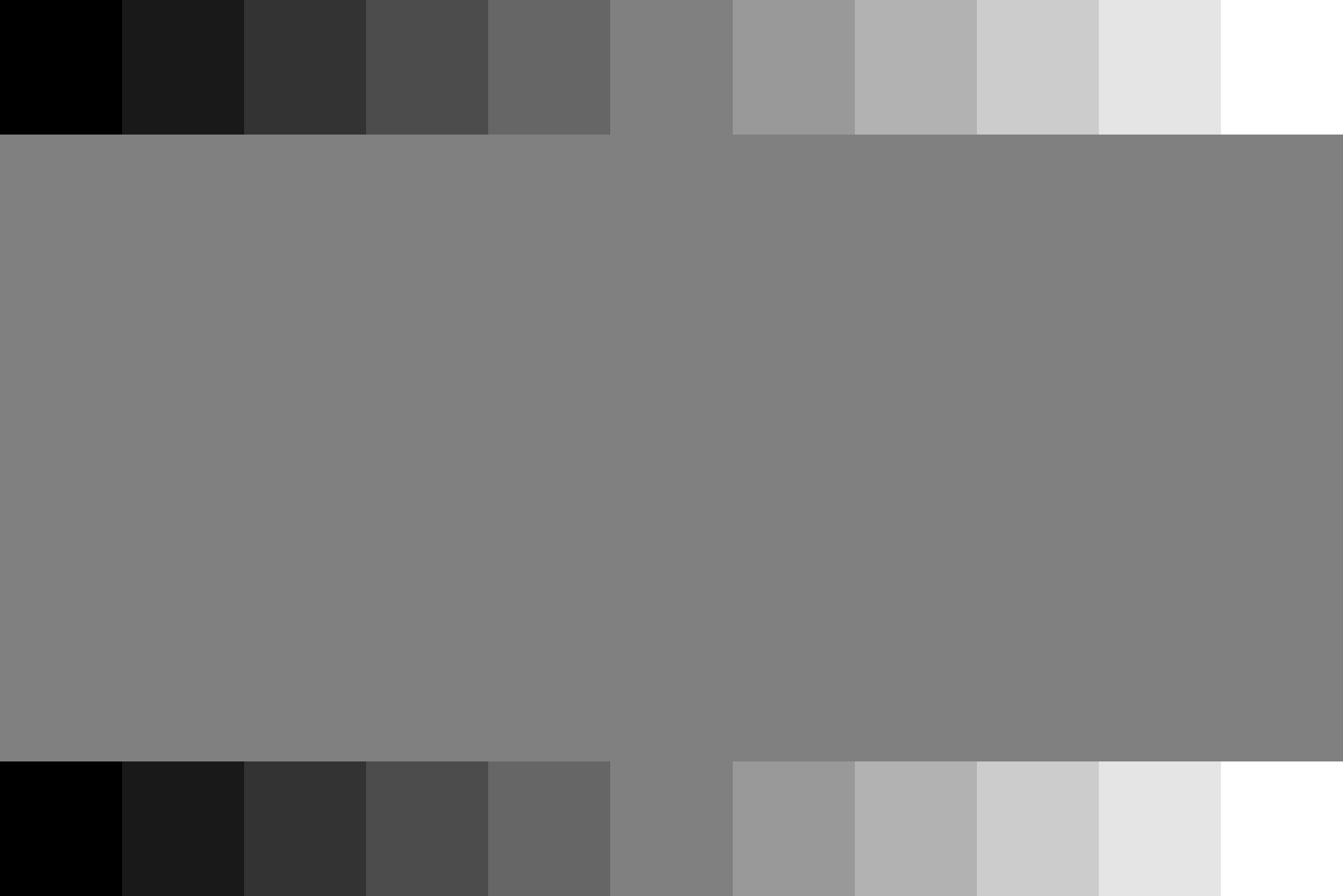}
    \caption{Constant color covering 70\% of the area in front of a graded color that consists of 70 bands. }
    \label{fig_10BandBGConstantColor/070}
 \end{subfigure}
 \hfill
      \begin{subfigure}[t]{0.32\textwidth}
\includegraphics[width=0.99\textwidth]{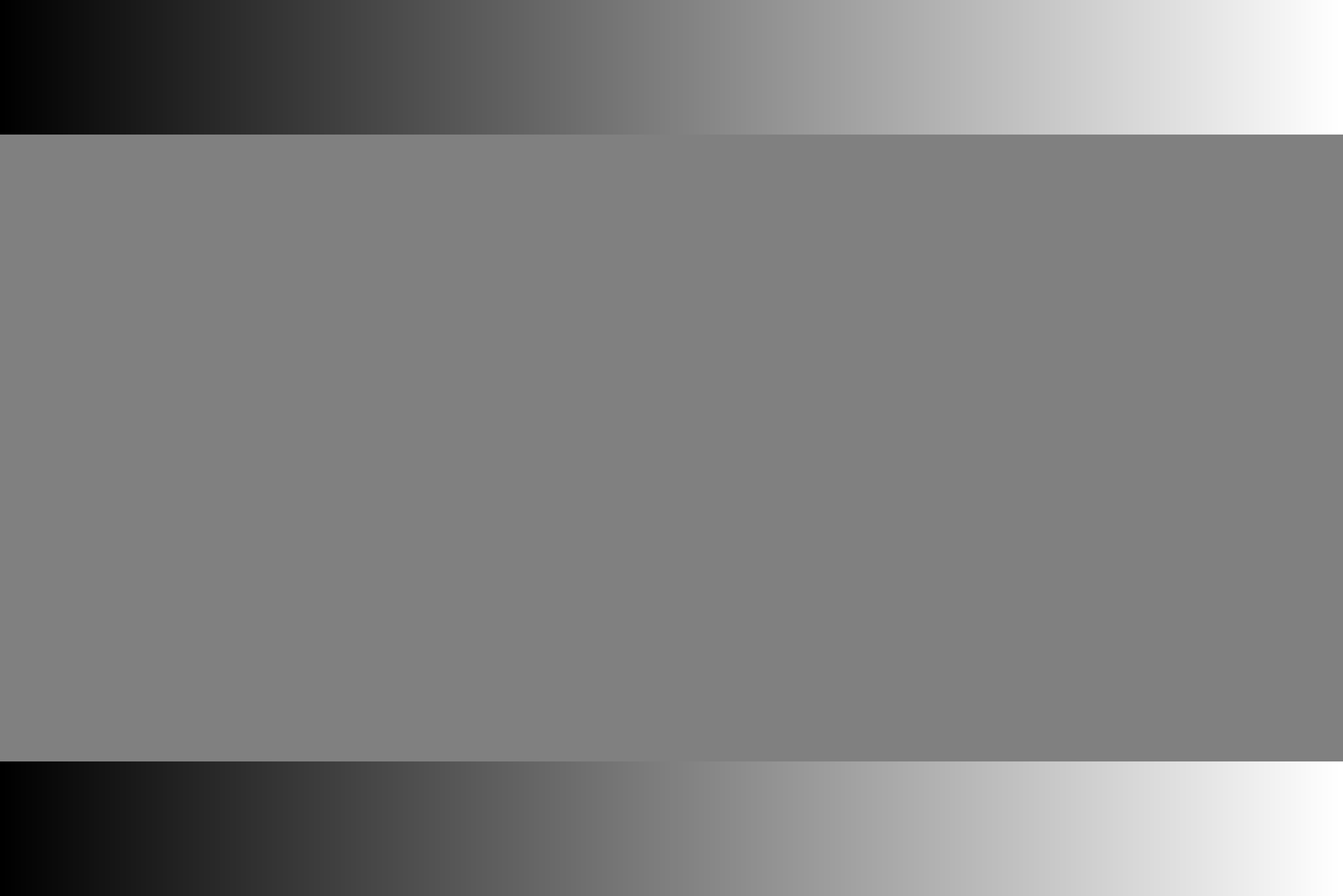}
    \caption{Constant color covering 70\% of the area in front of a continuously graded color. }
    \label{fig_ContinousBGConstantColor/070}
 \end{subfigure}
 \hfill
     \begin{subfigure}[t]{0.32\textwidth}
\includegraphics[width=0.99\textwidth]{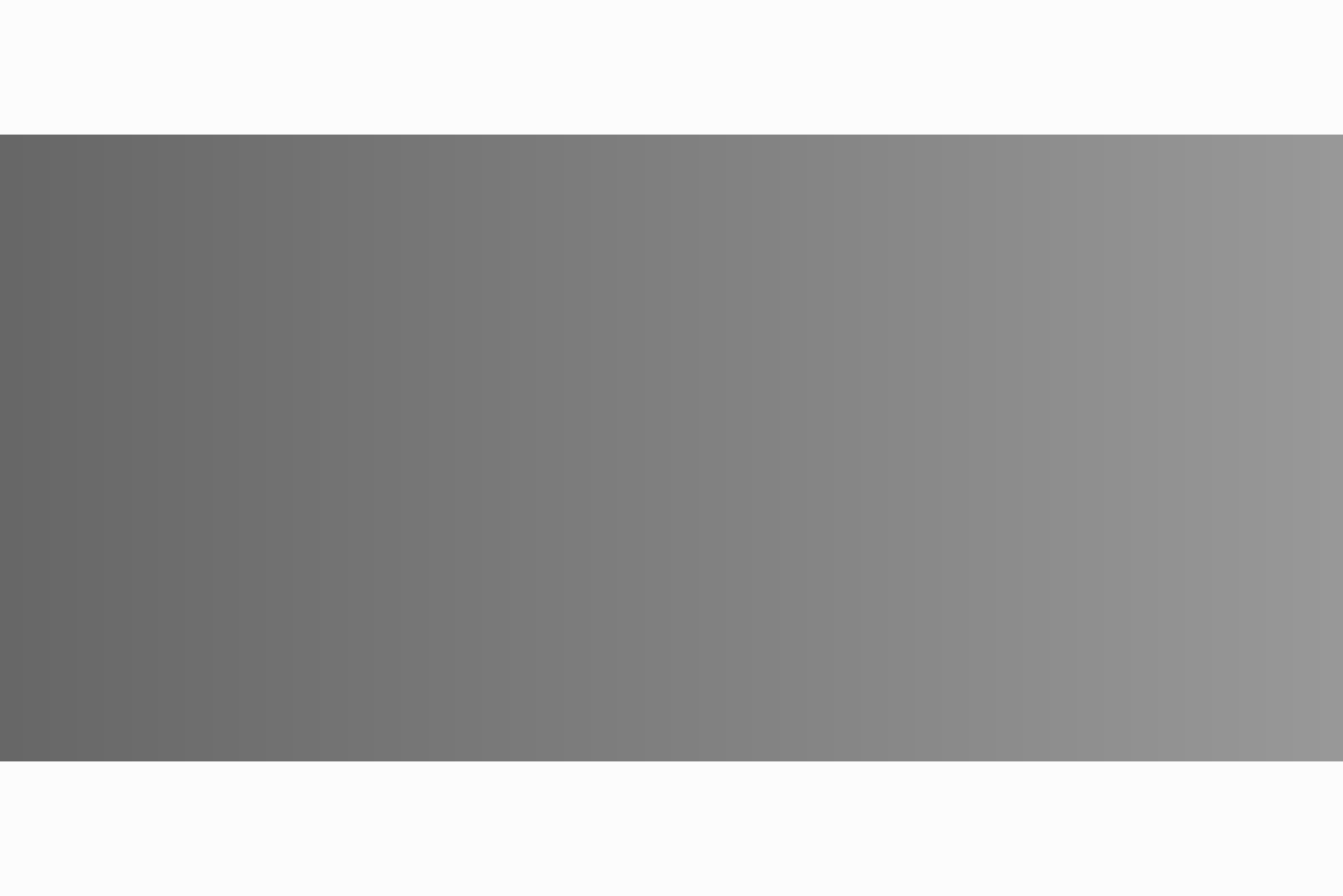}
    \caption{Our variable color band covers 70\% of the area in front of a white background. }
    \label{fig_WhiteBGConstantPerception/070}
 \end{subfigure}
 \hfill
     \begin{subfigure}[t]{0.32\textwidth}
\includegraphics[width=0.99\textwidth]{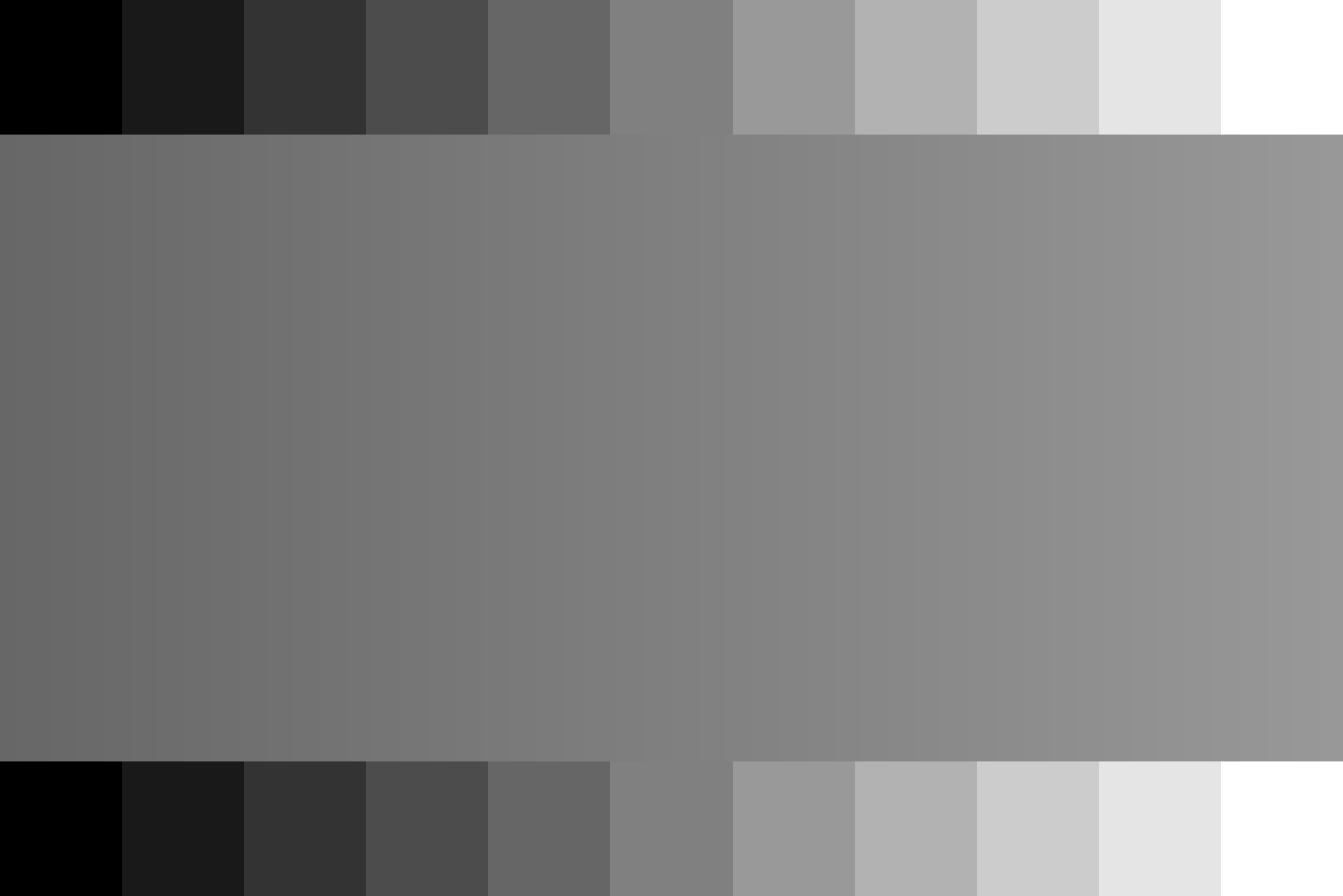}
    \caption{Our variable color band covers 70\% of the area in front of a graded color that consists of 70 bands. }
    \label{fig_10BandBGConstantPerception/070}
 \end{subfigure}
 \hfill
      \begin{subfigure}[t]{0.32\textwidth}
\includegraphics[width=0.99\textwidth]{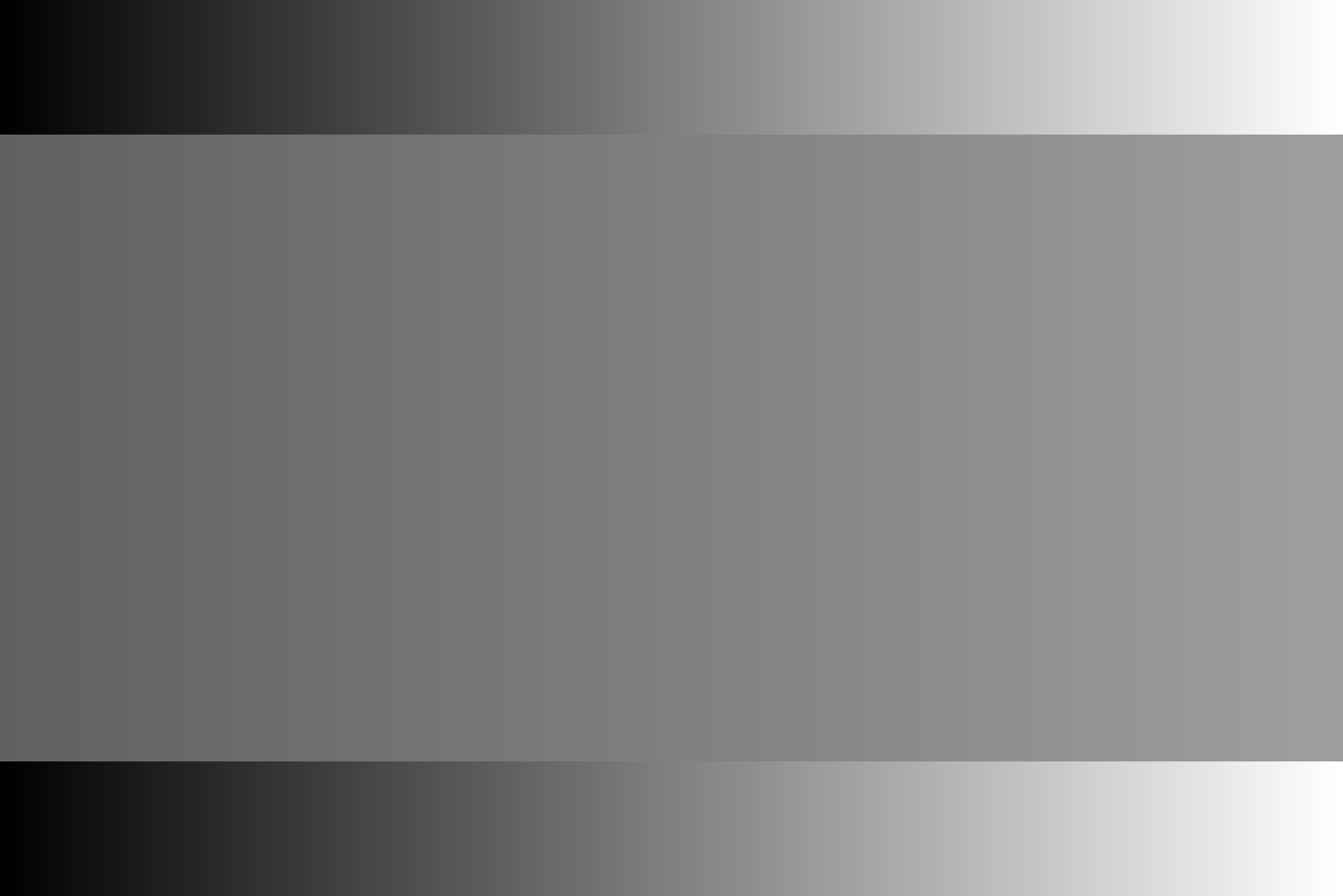}
    \caption{Our variable color band covers 70\% of the area in front of a continuously graded color. }
    \label{fig_ContinousBGBGConstantPerception/070}
 \end{subfigure}
 \hfill 
    \caption{Comparison of our method with constant color. Note that Our variable color band creates constant perception}
\label{fig_070}
\end{figure}

\begin{figure}[hbtp]
\centering
     \begin{subfigure}[t]{0.32\textwidth}
\includegraphics[width=0.99\textwidth]{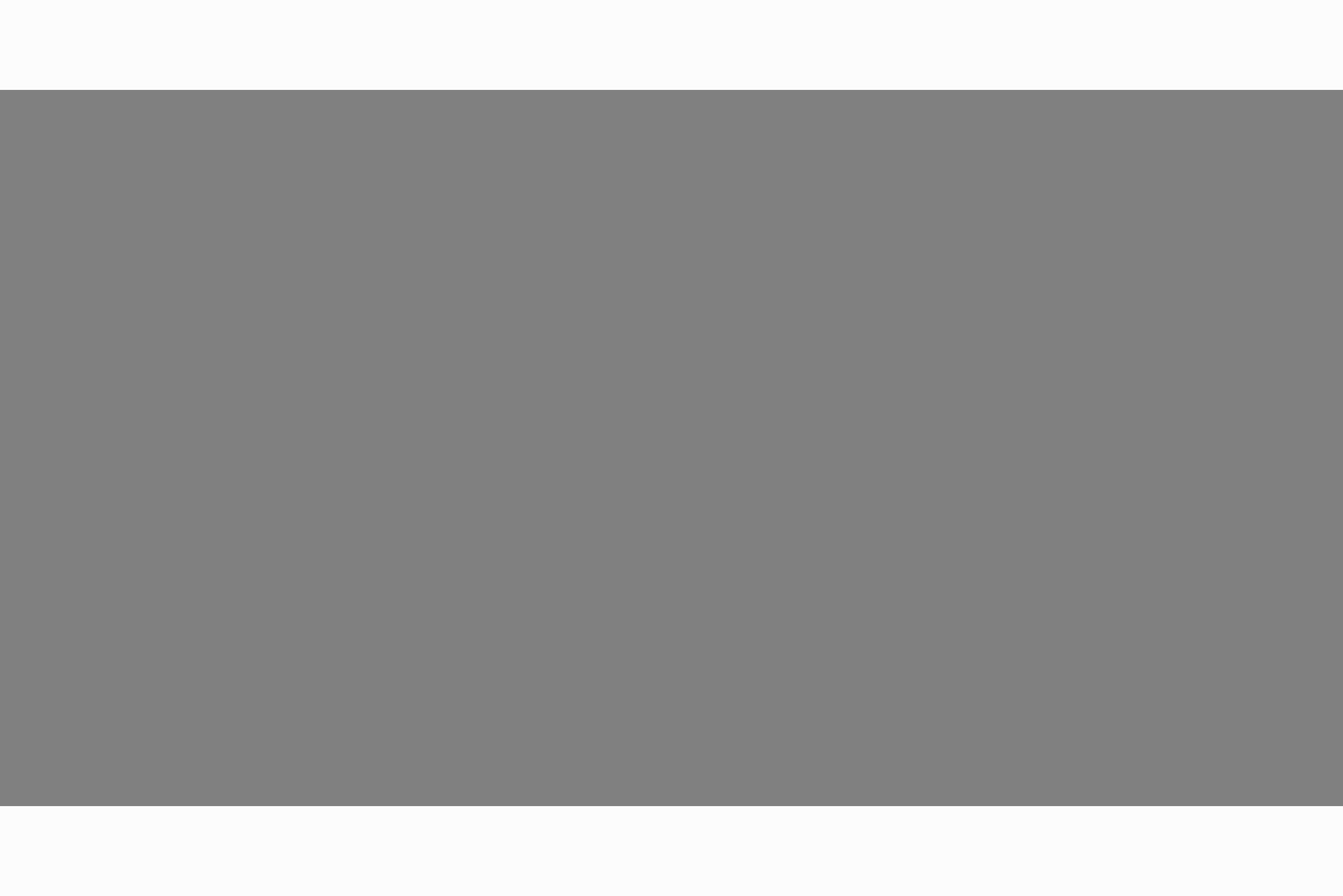}
    \caption{Constant color covering 80\% of the area in front of white background. }
    \label{fig_WhiteBGConstantColor/080}
 \end{subfigure}
 \hfill
     \begin{subfigure}[t]{0.32\textwidth}
\includegraphics[width=0.99\textwidth]{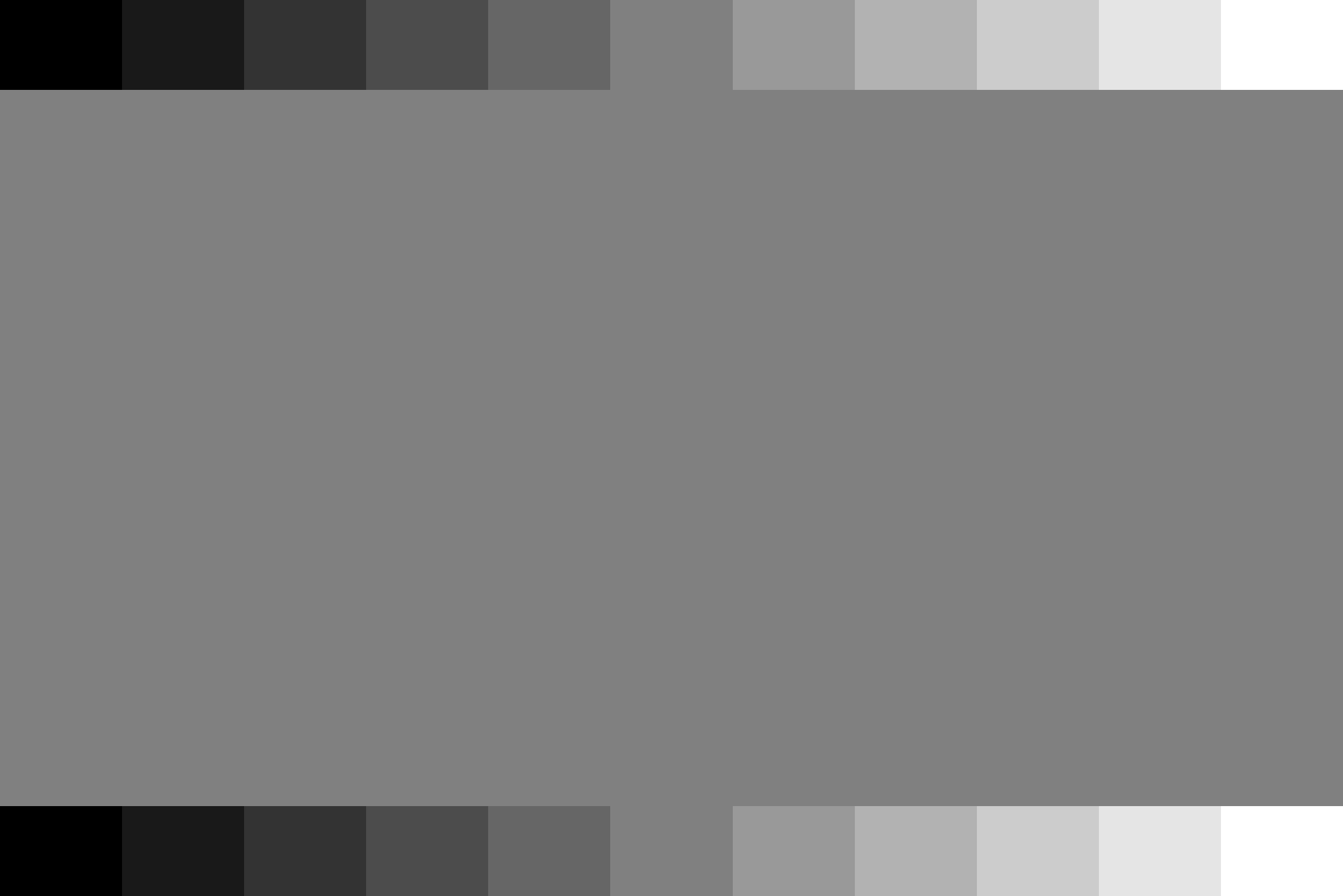}
    \caption{Constant color covering 80\% of the area in front of a graded color that consists of 80 bands. }
    \label{fig_10BandBGConstantColor/080}
 \end{subfigure}
 \hfill
      \begin{subfigure}[t]{0.32\textwidth}
\includegraphics[width=0.99\textwidth]{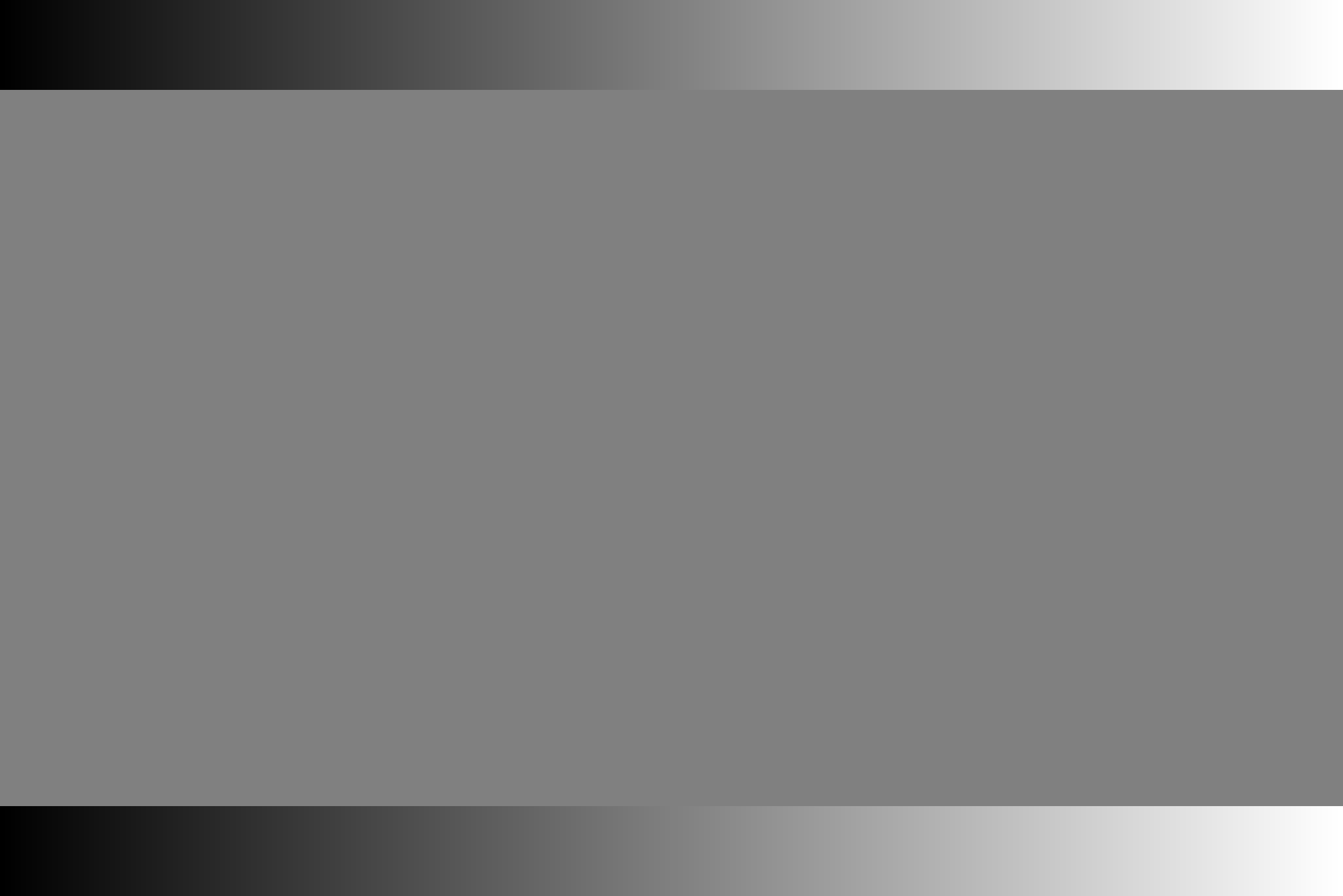}
    \caption{Constant color covering 80\% of the area in front of a continuously graded color. }
    \label{fig_ContinousBGConstantColor/080}
 \end{subfigure}
 \hfill
     \begin{subfigure}[t]{0.32\textwidth}
\includegraphics[width=0.99\textwidth]{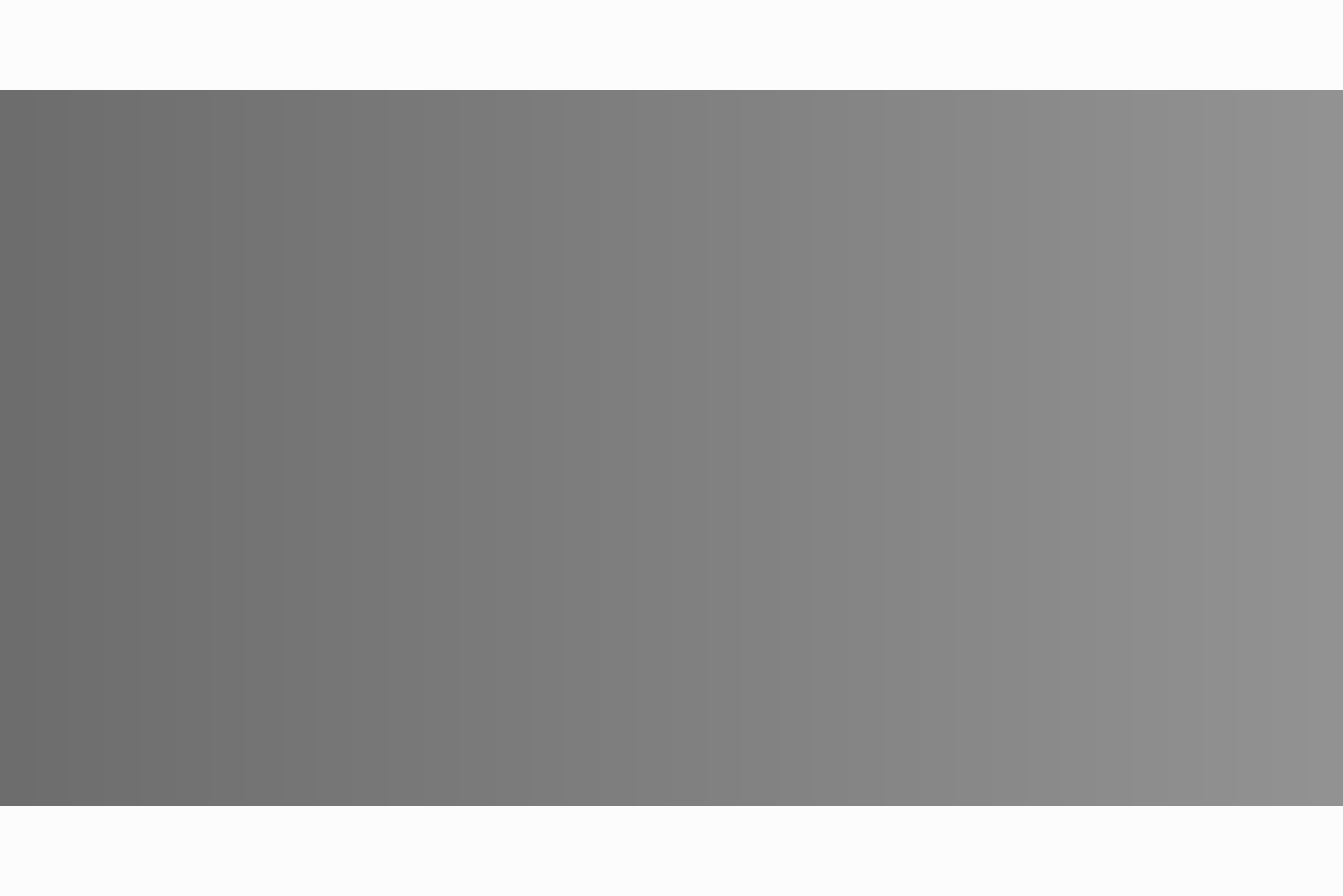}
    \caption{Our variable color band covers 80\% of the area in front of a white background. }
    \label{fig_WhiteBGConstantPerception/080}
 \end{subfigure}
 \hfill
     \begin{subfigure}[t]{0.32\textwidth}
\includegraphics[width=0.99\textwidth]{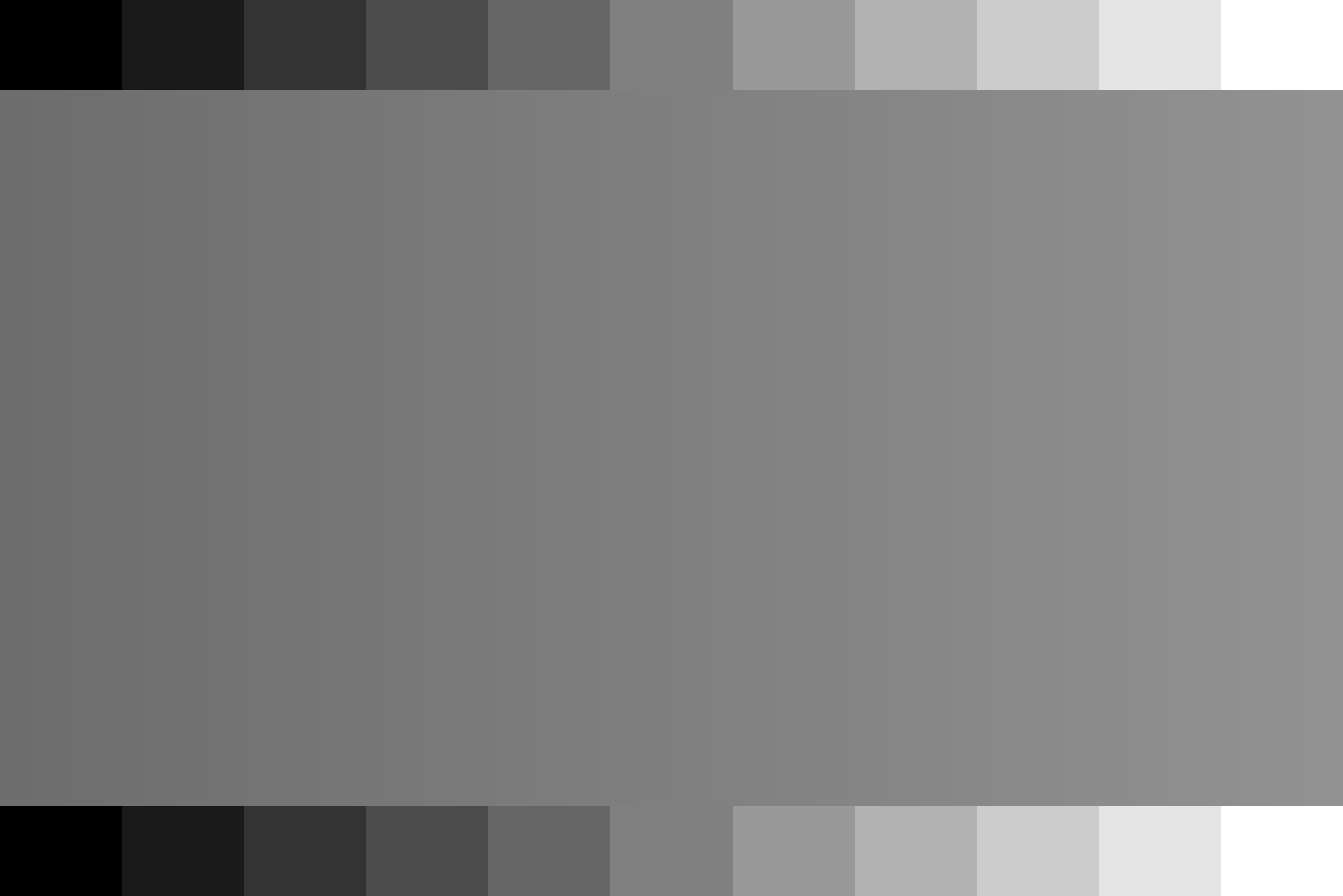}
    \caption{Our variable color band covers 80\% of the area in front of a graded color that consists of 80 bands. }
    \label{fig_10BandBGConstantPerception/080}
 \end{subfigure}
 \hfill
      \begin{subfigure}[t]{0.32\textwidth}
\includegraphics[width=0.99\textwidth]{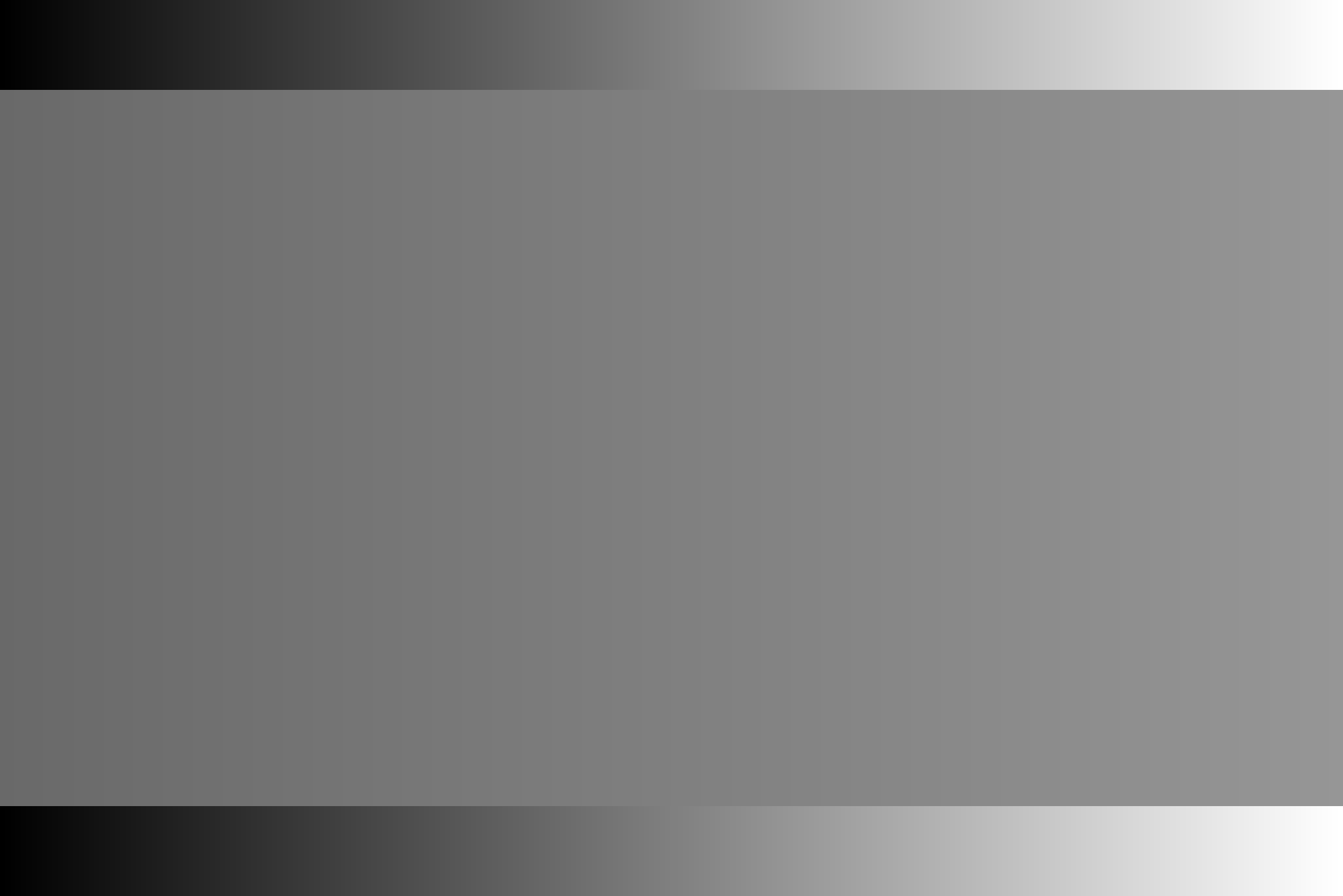}
    \caption{Our variable color band covers 80\% of the area in front of a continuously graded color. }
    \label{fig_ContinousBGBGConstantPerception/080}
 \end{subfigure}
 \hfill 
    \caption{Comparison of our method with constant color. Note that Our variable color band creates constant perception}
\label{fig_080}
\end{figure}

\begin{figure}[hbtp]
\centering
     \begin{subfigure}[t]{0.32\textwidth}
\includegraphics[width=0.99\textwidth]{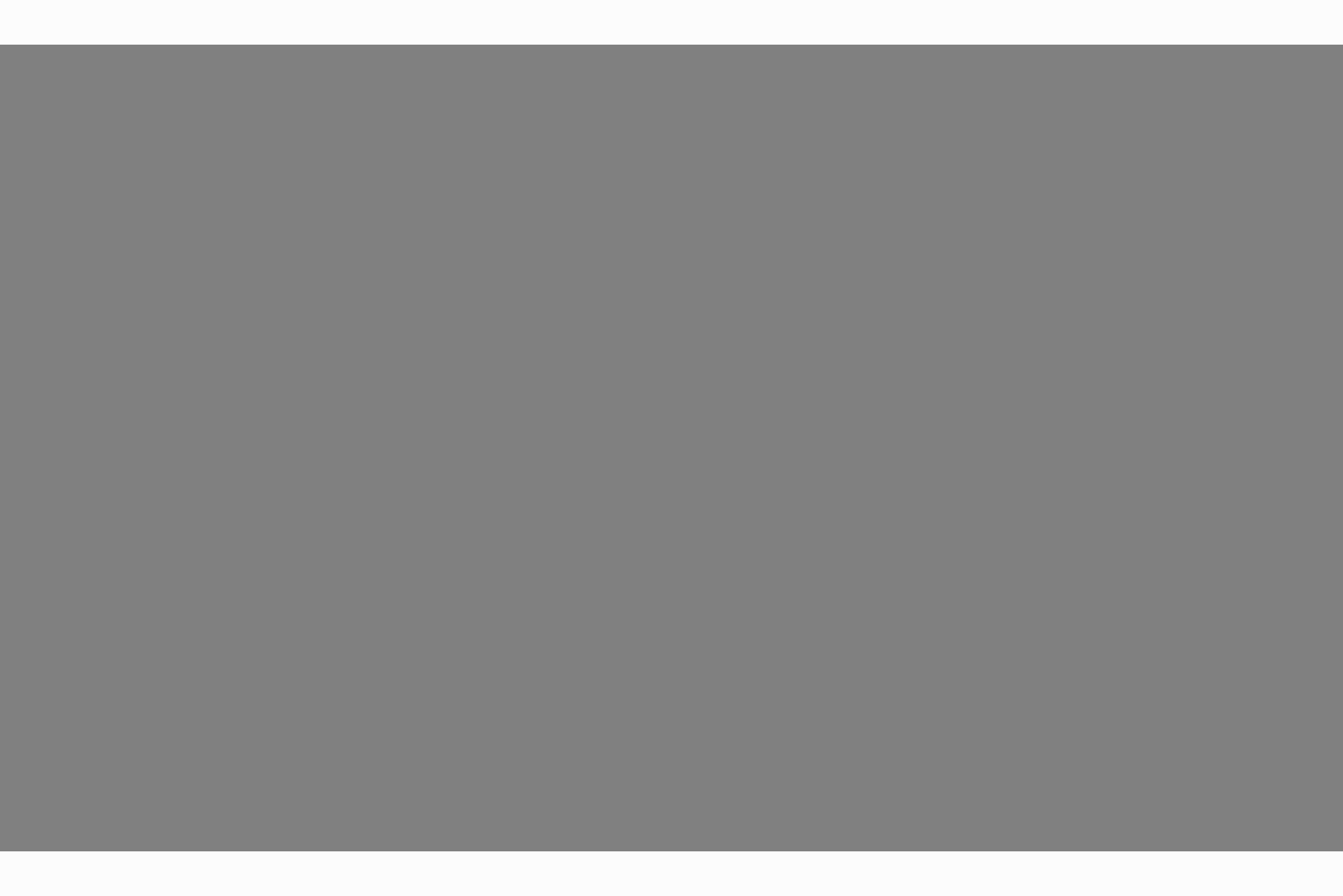}
    \caption{Constant color covering 90\% of the area in front of white background. }
    \label{fig_WhiteBGConstantColor/090}
 \end{subfigure}
 \hfill
     \begin{subfigure}[t]{0.32\textwidth}
\includegraphics[width=0.99\textwidth]{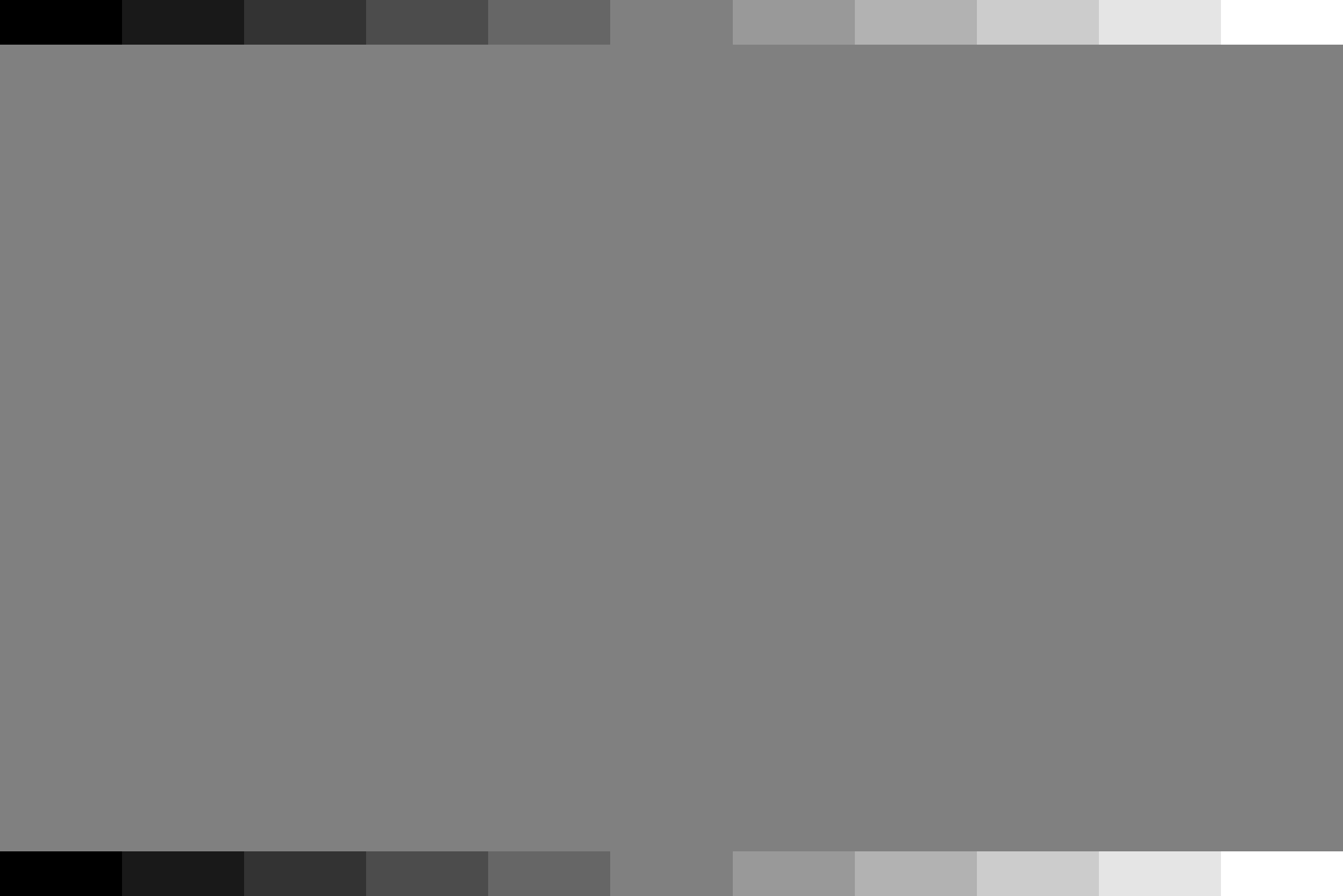}
    \caption{Constant color covering 90\% of the area in front of a graded color that consists of 90 bands. }
    \label{fig_10BandBGConstantColor/090}
 \end{subfigure}
 \hfill
      \begin{subfigure}[t]{0.32\textwidth}
\includegraphics[width=0.99\textwidth]{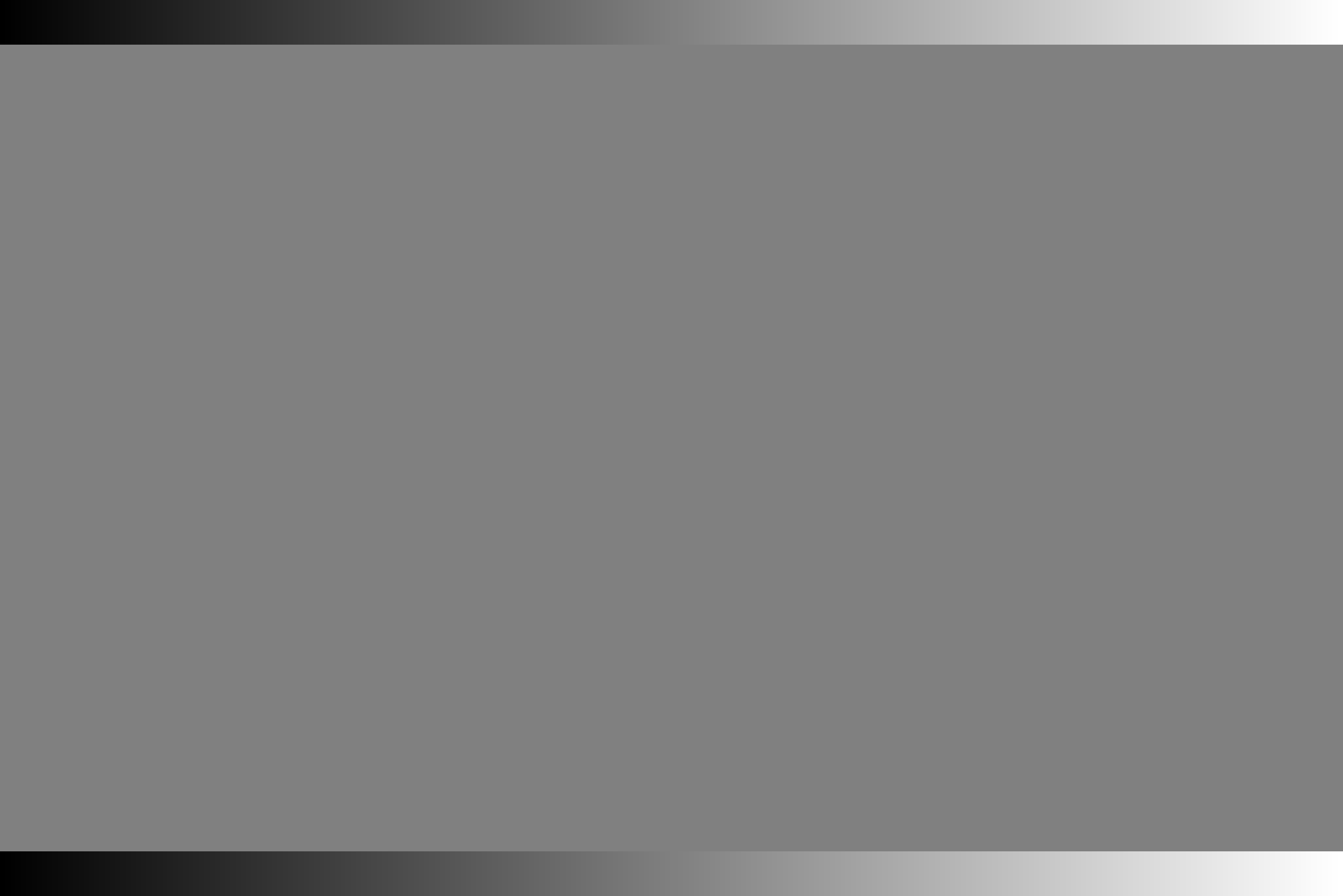}
    \caption{Constant color covering 90\% of the area in front of a continuously graded color. }
    \label{fig_ContinousBGConstantColor/090}
 \end{subfigure}
 \hfill
     \begin{subfigure}[t]{0.32\textwidth}
\includegraphics[width=0.99\textwidth]{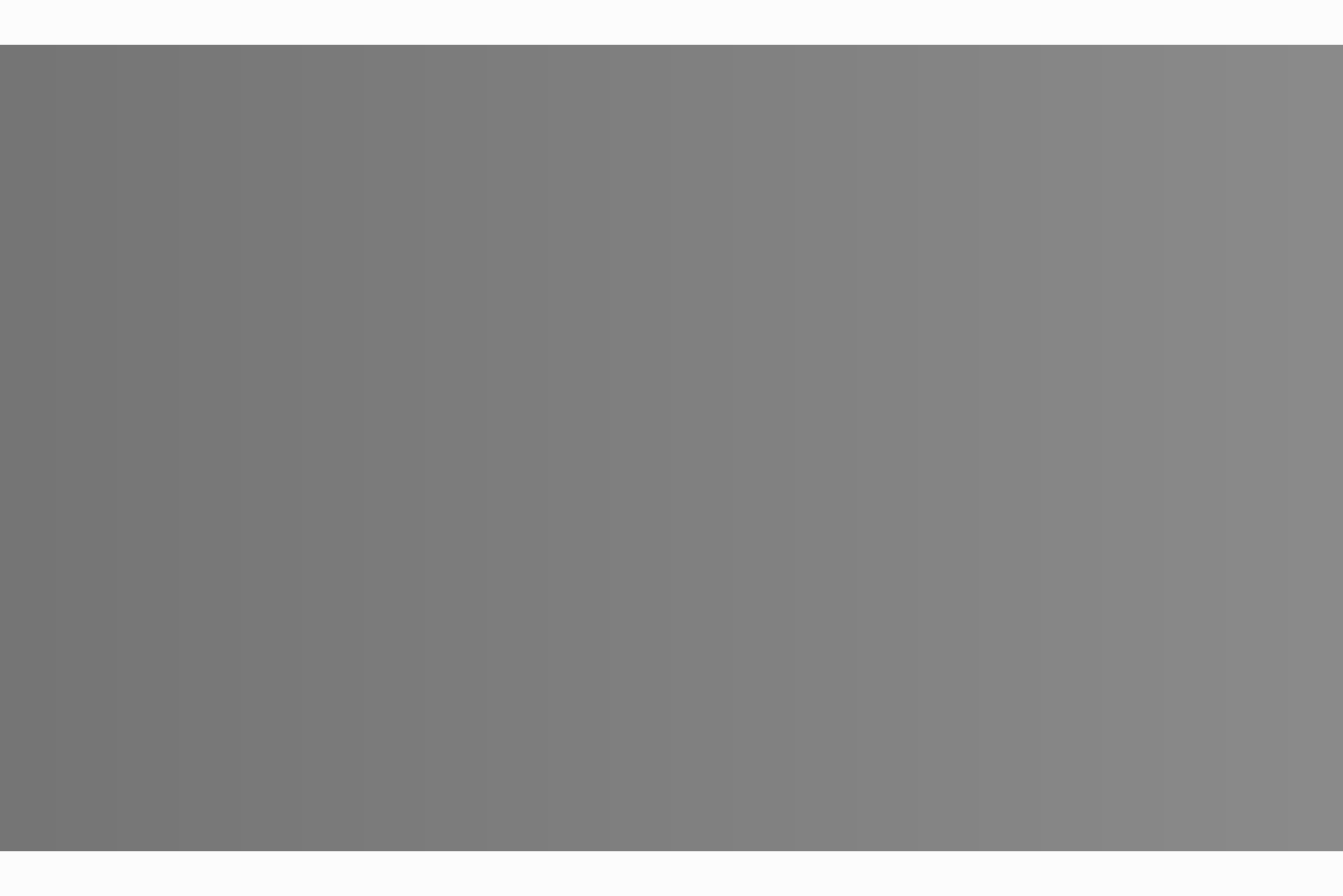}
    \caption{Our variable color band covers 90\% of the area in front of a white background. }
    \label{fig_WhiteBGConstantPerception/090}
 \end{subfigure}
 \hfill
     \begin{subfigure}[t]{0.32\textwidth}
\includegraphics[width=0.99\textwidth]{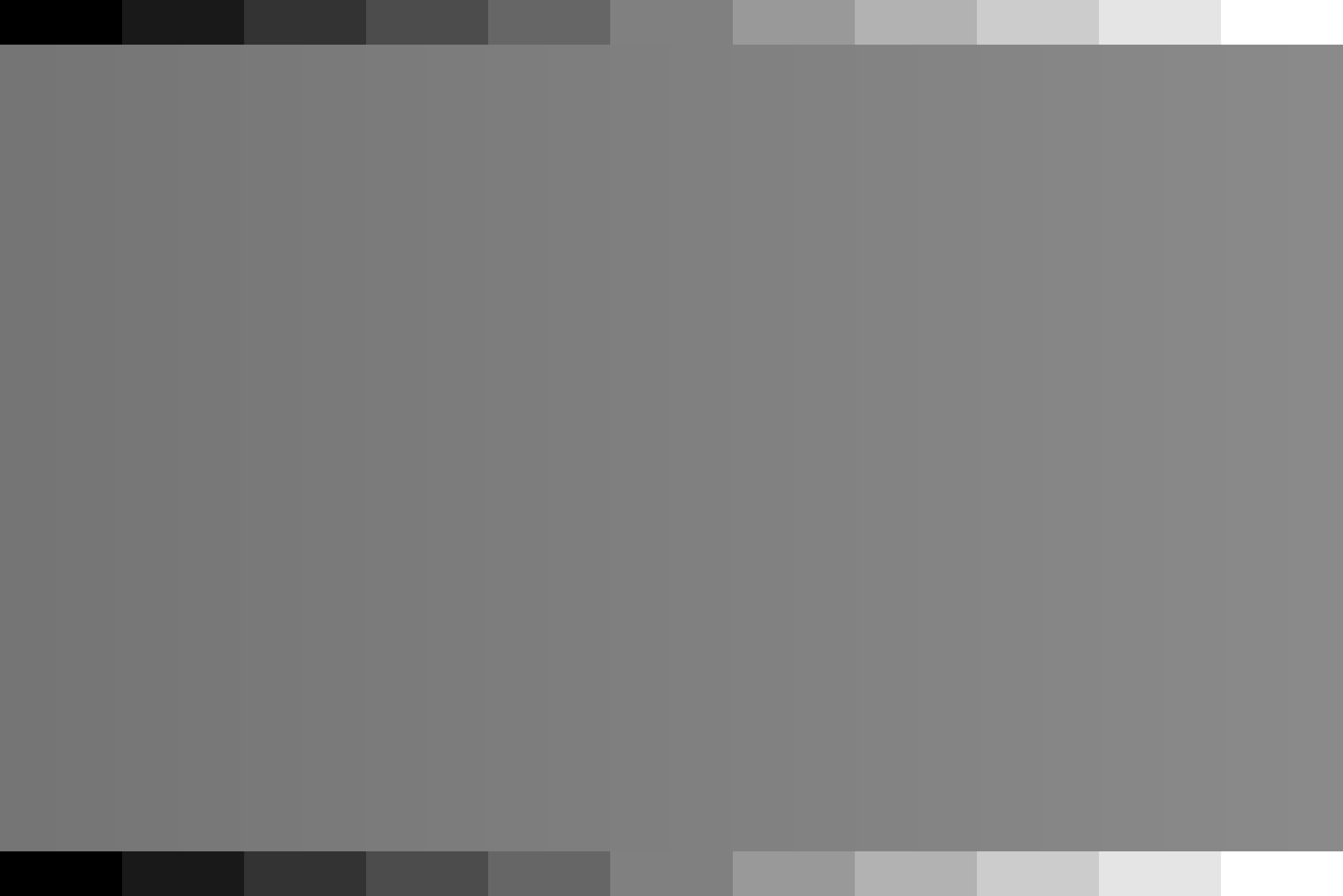}
    \caption{Our variable color band covers 90\% of the area in front of a graded color that consists of 90 bands. }
    \label{fig_10BandBGConstantPerception/090}
 \end{subfigure}
 \hfill
      \begin{subfigure}[t]{0.32\textwidth}
\includegraphics[width=0.99\textwidth]{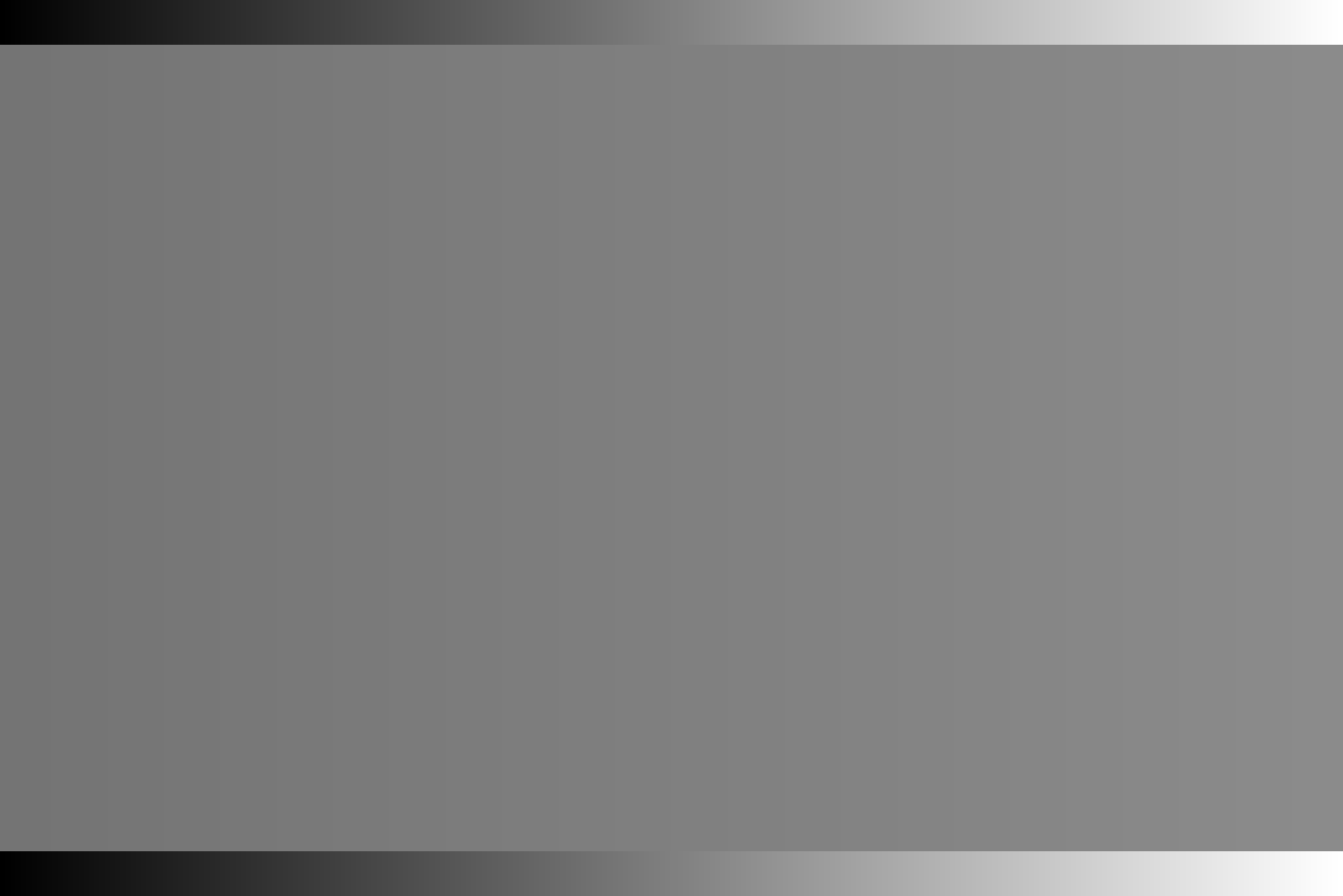}
    \caption{Our variable color band covers 90\% of the area in front of a continuously graded color. }
    \label{fig_ContinousBGBGConstantPerception/090}
 \end{subfigure}
 \hfill 
    \caption{Comparison of our method with constant color. Note that Our variable color band creates constant perception}
\label{fig_090}
\end{figure}

\begin{figure}[hbtp]
\centering
     \begin{subfigure}[t]{0.32\textwidth}
\includegraphics[width=0.99\textwidth]{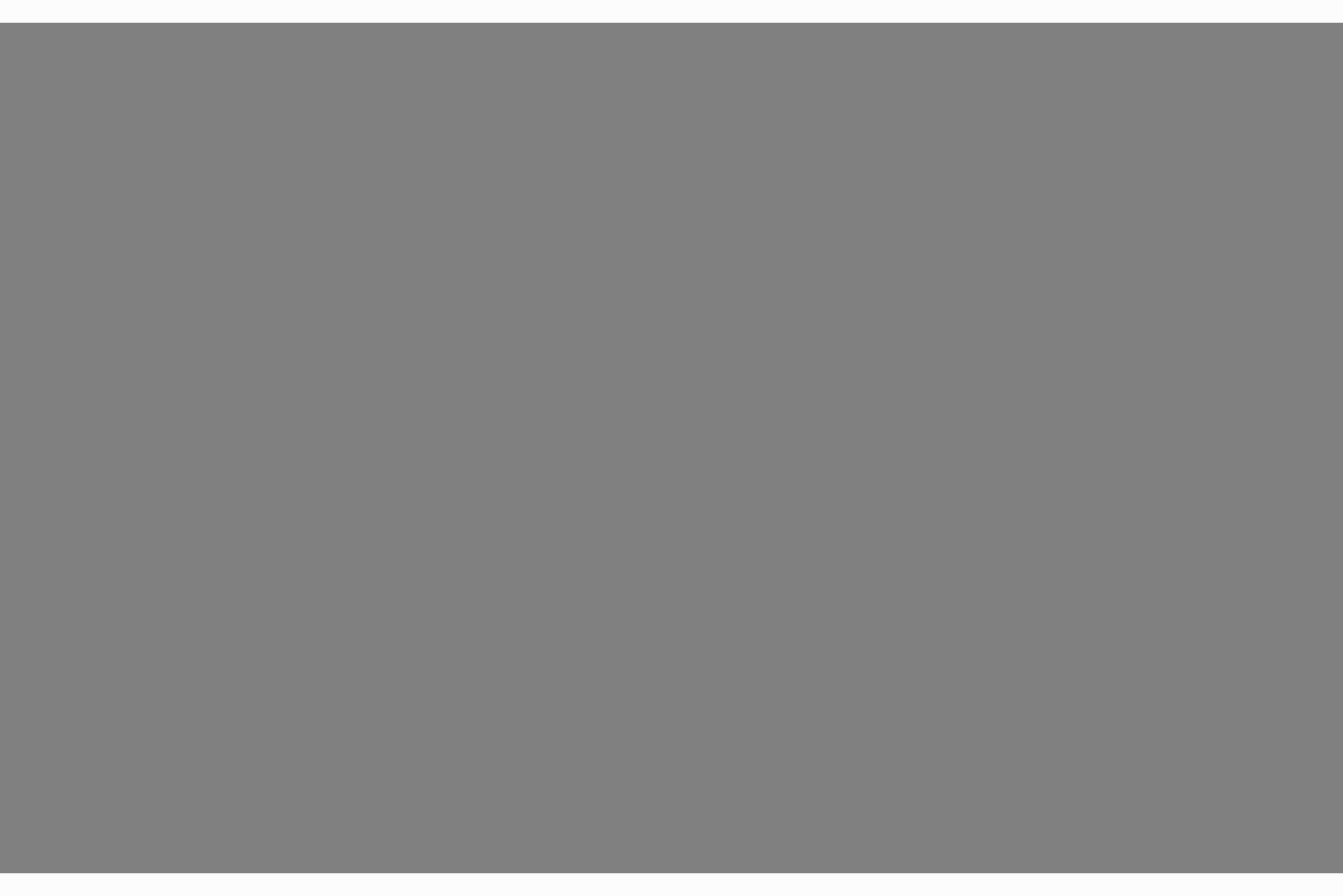}
    \caption{Constant color covering 95\% of the area in front of white background. }
    \label{fig_WhiteBGConstantColor/095}
 \end{subfigure}
 \hfill
     \begin{subfigure}[t]{0.32\textwidth}
\includegraphics[width=0.99\textwidth]{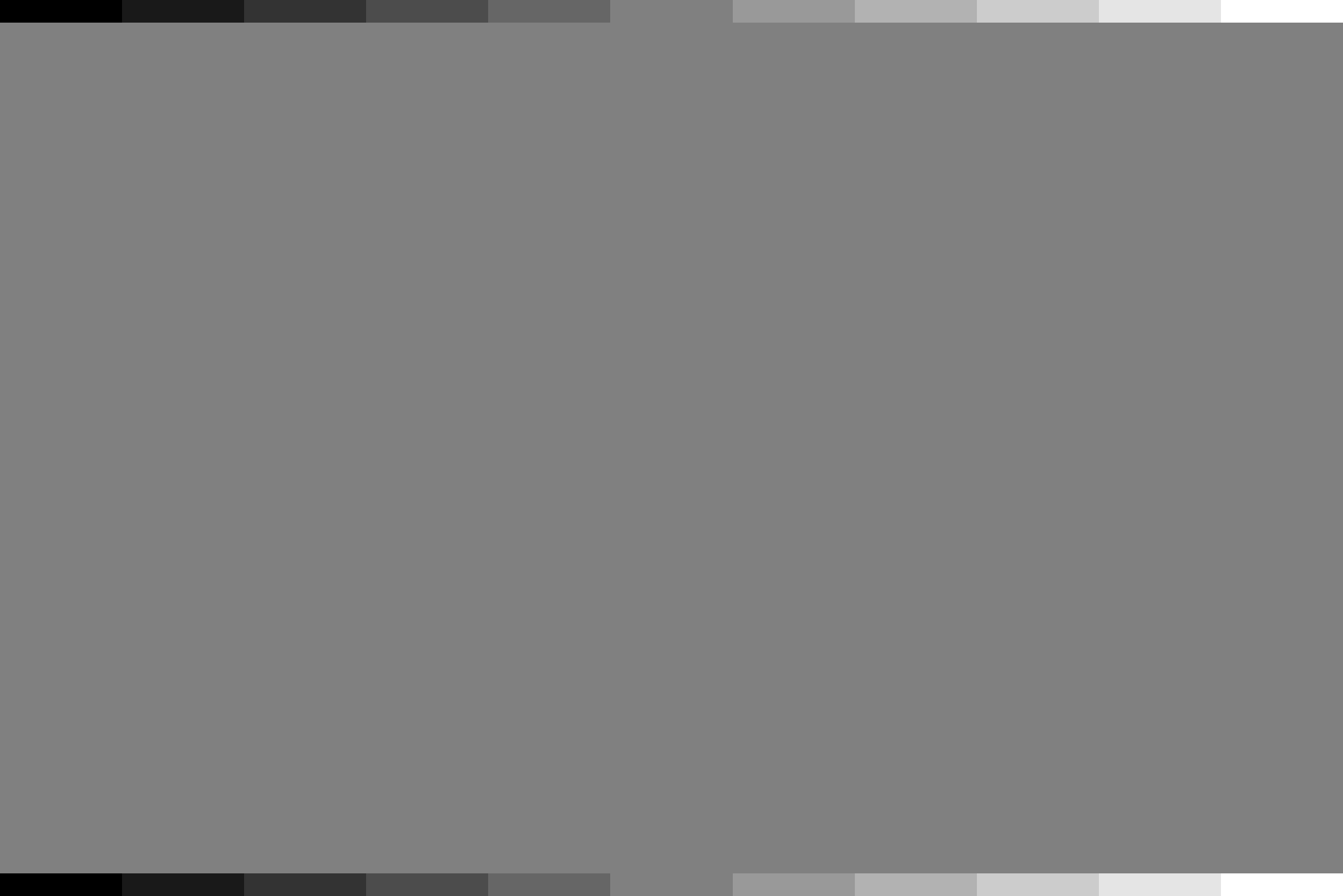}
    \caption{Constant color covering 95\% of the area in front of a graded color that consists of 95 bands. }
    \label{fig_10BandBGConstantColor/095}
 \end{subfigure}
 \hfill
      \begin{subfigure}[t]{0.32\textwidth}
\includegraphics[width=0.99\textwidth]{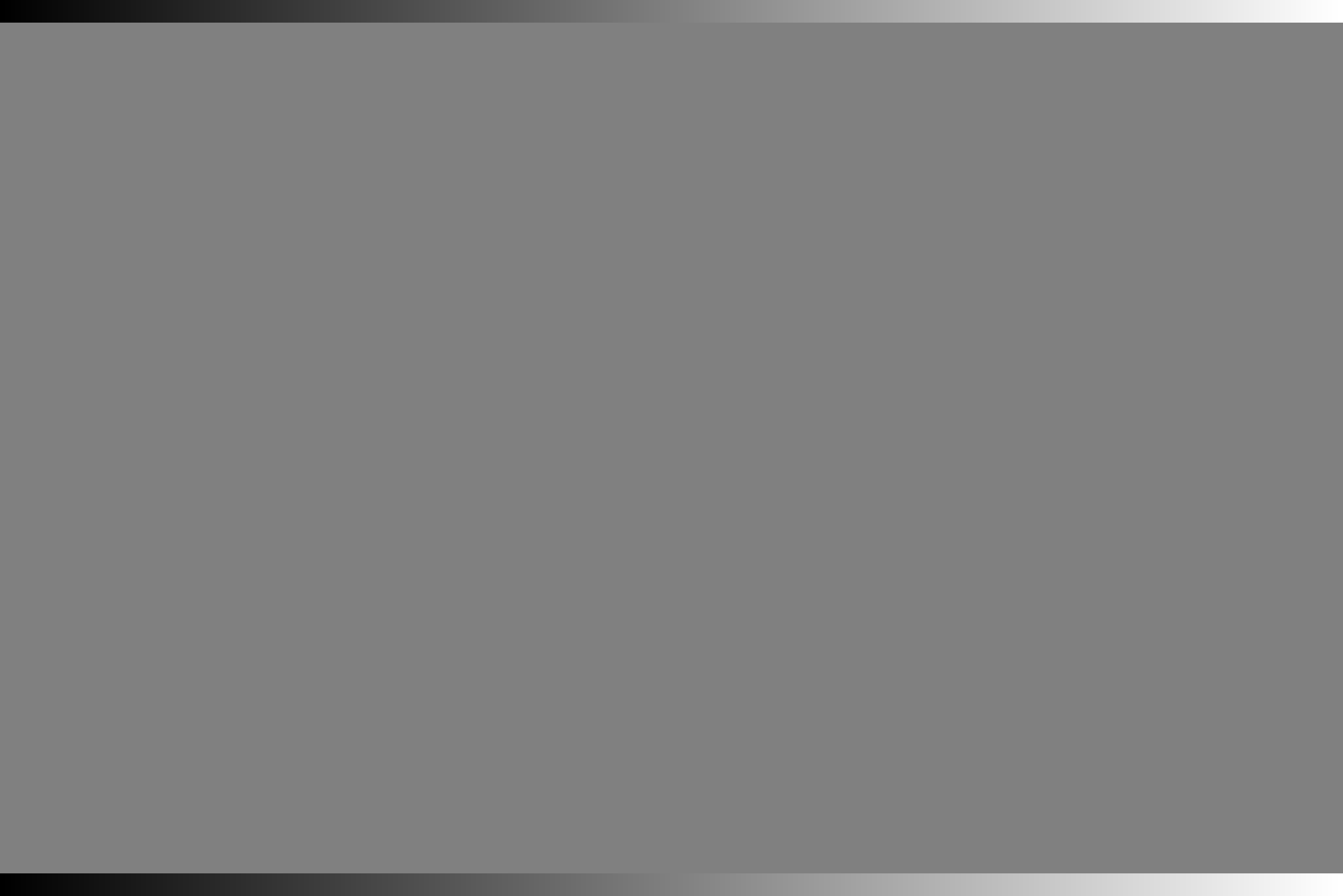}
    \caption{Constant color covering 95\% of the area in front of a continuously graded color. }
    \label{fig_ContinousBGConstantColor/095}
 \end{subfigure}
 \hfill
     \begin{subfigure}[t]{0.32\textwidth}
\includegraphics[width=0.99\textwidth]{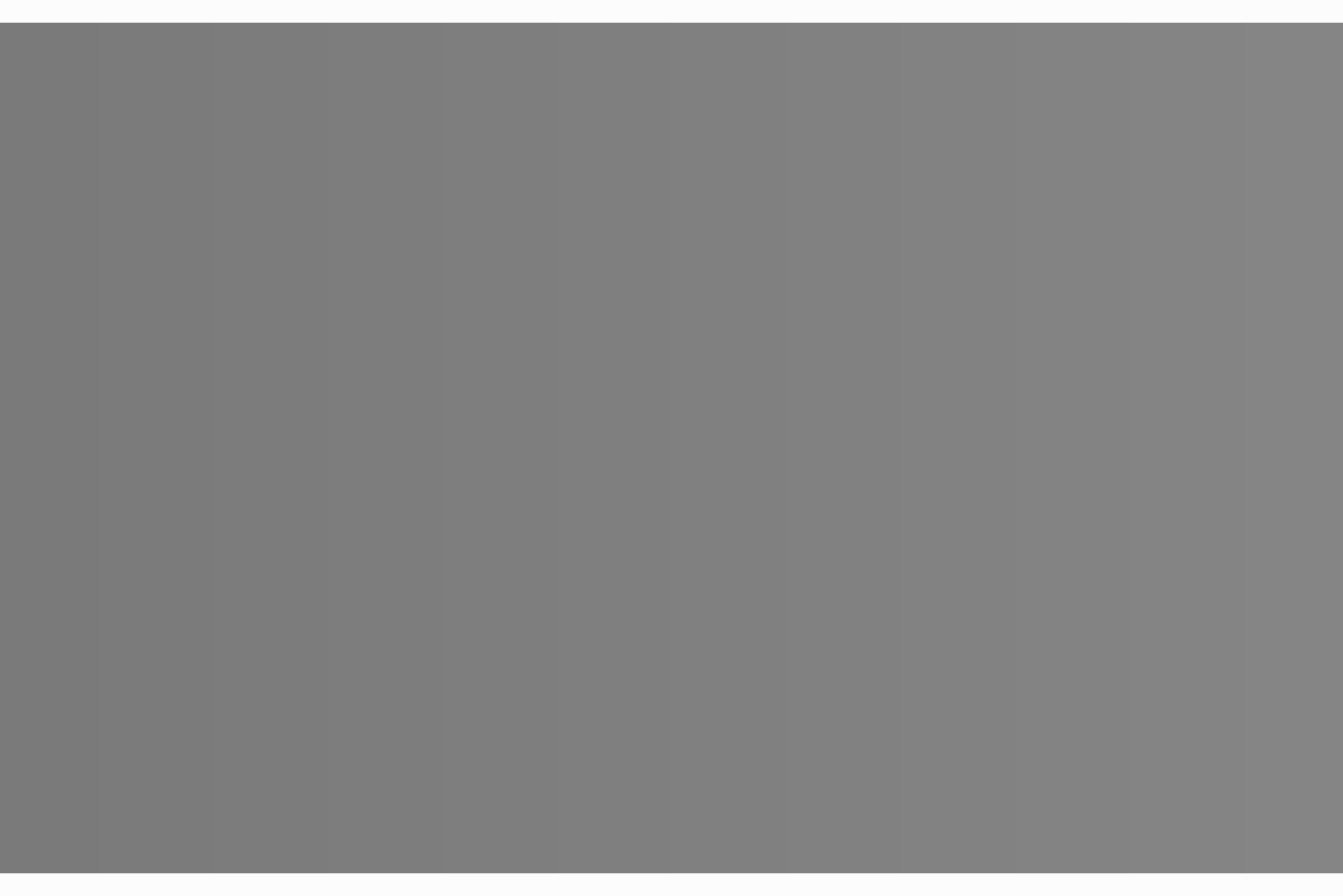}
    \caption{Our variable color band covers 95\% of the area in front of a white background. }
    \label{fig_WhiteBGConstantPerception/095}
 \end{subfigure}
 \hfill
     \begin{subfigure}[t]{0.32\textwidth}
\includegraphics[width=0.99\textwidth]{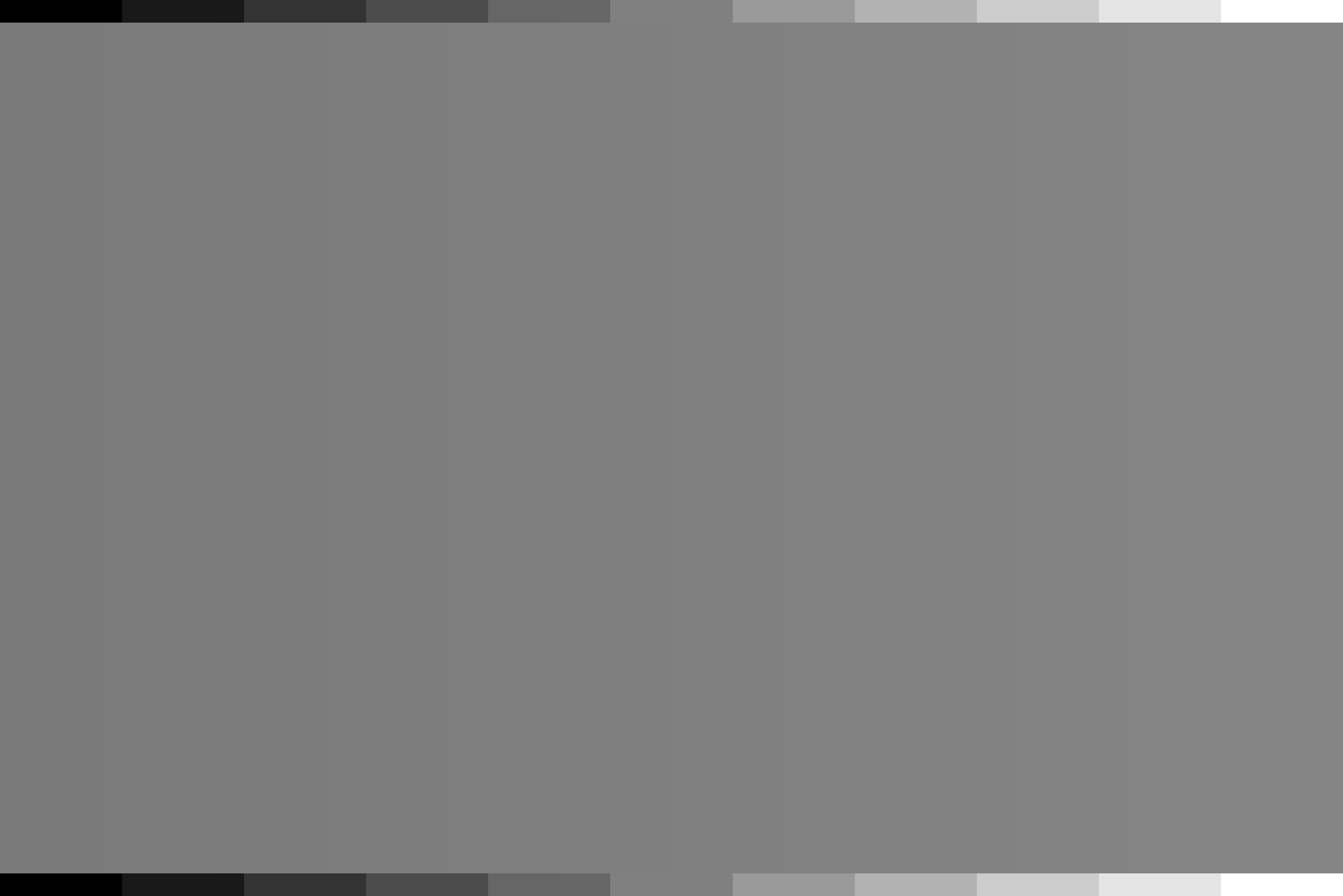}
    \caption{Our variable color band covers 95\% of the area in front of a graded color that consists of 95 bands. }
    \label{fig_10BandBGConstantPerception/095}
 \end{subfigure}
 \hfill
      \begin{subfigure}[t]{0.32\textwidth}
\includegraphics[width=0.99\textwidth]{10BandBGConstantPerception/095}
    \caption{Our variable color band covers 95\% of the area in front of a continuously graded color. }
    \label{fig_ContinousBGBGConstantPerception/095}
 \end{subfigure}
 \hfill 
    \caption{Comparison of our method with constant color. Note that Our variable color band creates constant perception}
\label{fig_095}
\end{figure}

To test our hypothesis, we have developed a web-based program in ShaderToy and made the program public: \cite{akleman2024constant}. Anyone can access the program through the following link: \href{https://www.shadertoy.com/view/XX23Dz}{\textbf{Constant Perception}} or can directly go to the next URL by typing in your browser: \url{https://www.shadertoy.com/view/XX23Dz}. The basic structure of the program is provided by Algorithm~\ref{Algol_constant_perception}. There is a need for the mouse and keyboard interaction to see all the effects. The coefficients included in Algorithm~\ref{Algol_constant_perception} are the ones we have identified. Users can change them in the ShaderToy code to see the effects. we have created images shown in Figures~\ref{fig_001},~\ref{fig_005},~\ref{fig_010},~\ref{fig_020},~\ref{fig_030},~\ref{fig_040},~\ref{fig_050},~\ref{fig_060},~\ref{fig_070},~\ref{fig_080},~\ref{fig_090},~\ref{fig_095}, and~\ref{fig_090} for values of $s$ $\{0.01, 0.05, 0.1, 0.2, 0.3, 0.4, 0.5, 0.6, 0.7,0.8, 0.9, 0.95, 0.99\}$ respectively by using non-interactive version of the same program. In the non-interactive version, we can enter precise values for relative size, which is not possible with mouse input.   

\begin{algorithm}
\caption{This is an accompanying program in ShaderToy developed specifically for this paper. It is an interactive program. Moving the mouse in the y direction changes the thickness of the rectangle, which corresponds to $x$, that is, the relative size of the foreground object. Moving the mouse in the x direction changes the perceived color. Pressing the left arrow key changes from constant perception to constant color. Pressing the right-arrow key changes the background from a lightness scale to a white image. Pressing the "Up Arrow Key" changes the background from a discrete-lightness scale to a continuous-lightness scale. Changing the Bezier coefficients changes the polynomial part of the Context-Sensitive Formula.}
\label{Algol_constant_perception}
\begin{algorithmic}
\REQUIRE $b_{2,0}=0.20, b_{2,1}=0.25, and b_{2,2}=1.00$: Coefficients of the exponent polynomial in quadratic B\'{e}zier form.
\REQUIRE $a_{0}=0.60, a_{1}=1.00$. Coefficients of the first-degree polynomial that approximate the power function.
\REQUIRE $\textbf{p}_S=(p.x,p.y) \in [0,1]^2$: the 2D position of the shading point (pixel).
\REQUIRE $L_P$: perceived luminance of the foreground. The user can change $L_P$ between $0$ and $1$ by moving the mouse in the $x$ direction within the image. 
\REQUIRE $s \in [0,1]$: Relative height of the rectangle, which corresponds to the relative size of the foreground object in the images. The user can change $s$ between $0$ and $1$ by moving the mouse in the $y$ direction within the image. Let $\cal R_F$ be the region corresponding to the foreground object. 
\REQUIRE $I_{White}$, $I_{CLS}$, $I_{DLS}$: Three background luminance images: White, continuous on the lightness scale, and discrete on the lightness scale, respectively. 
\REQUIRE $I_F(\textbf{p}_S)$ and $I_B(\textbf{p}_S)$: foreground and background images, respectively. 
\ENSURE $I_B(\textbf{p}_S)$ is selected by the user as one of the three background images by pressing the appropriate arrow keys.
\IF{ $\textbf{p}_S \in \cal R_F$ }
\IF{ y is chosen to be a power function }
\STATE  $f(s) \leftarrow b_{2,0} (1-s)^2 + b_{2,1} 2 s (1-s) + b_{2,2} s^2$ 
\STATE  $y \leftarrow s^{f(s)}$
\ENDIF
\IF{ y is chosen to be a first-degree function }
\STATE  $y \leftarrow a_{0} (1-s) + a_{1} s$ 
\ENDIF
\IF{ Constant Color }
\STATE  $I_F(\textbf{p}_S) \leftarrow (1-y) L_P + y I_{CLS}(\textbf{p}_S) $
\ENDIF
\IF{ Constant Perception }
\STATE  $I_F(\textbf{p}_S) \leftarrow L_P $
\ENDIF
\ENDIF
\end{algorithmic}
\end{algorithm}

\subsection{The Process of Obtaining the Coefficients}

The code in \ref{Algol_constant_perception} is the final form showing the working polynomials and their coefficients. To obtain these, we start with a simple structure in the form of $y=s^{b00}$. See examples in Figure~\ref{fig_functions_constant}. We quickly found that the term $b_00$ should be smaller than $b_00<1.0$. The numbers between $b_00=0.25$ (see Figure~\ref{functions/0.25}) and $b_00=0.5$ (see Figure~\ref{functions/0.5} provided good results for small values, but not for all $s$ values. We observe that for large values of $s$, it is better to use the values of $b_00 \approx 1.0$ to obtain good results. This suggested that to improve the results, one needs to use higher-degree polynomials. First, we tested first-degree polynomials. Using the intuition obtained from the constant case, we tested the range $b_10 \in [0.1, 0.5]$ and $b_11 \in [0.75, 1.5]$ (two people worked at the same time). Finally, we settled on the values $b_10 = 0.20$ and $b_11 =1.0$. However, there was still a problem with values in the middle range of $s \approx 0.5$. Here, the B\'{e}zier form becomes useful. Since it interpolated the beginning and end, we simply transferred $b_20 = 0.20$ and $b_22 =1.0$ to second-degree polynomials and only explored the values of $b_21 \in [0.2,1.0]$. We quickly satisfy the value of $b_21 \in [0.25]$ by obtaining Figure~\ref{functions/020b0025b110b2}. 

We initially did not suspect that the shape of the curve provided any additional hint. However, out of curiosity, we draw the final curve shown in Figure~\ref{functions/020b0025b110b2} and realize that going to a quadratic polynomial and adding the last term $b_21 \in [0.25]$ made the curve mostly straight. We have checked the equation of the straight line that approximates the shape shown in Figure~\ref{functions/020b0025b110b2}. We approximated it with another first-degree polynomial in  B\'{e}zier form as $y=a_0 (1-s) + a_1s$ where $a_0=0.6$ and $a_0=1.0$. We also added the new function as an option. It works almost the same as our original curve. For small values of $s$, it even appears slightly better.

This particular set of functions works well for the three of us, the authors. We showed it to a few other people, and the functions seem to work also for them. On the other hand, we need to caution that the functions may not be working for others. For them, it will be useful to have the program available. Using B\'{e}zier form they can also identify functions. They can first simply change the coefficients to obtain better constant perceptions for themselves.  If they feel that additional control is needed, they can simply increase the degree of the polynomials and change the coefficients where it is needed using the properties of the B\'{e}zier form. 

\section{Discussion}
\label{Sec_Discussion}

The basic premise of our approach is that it is essentially based on a forward system approach (computer graphics). Vision systems, on the other hand, are inverse systems. Note that it is not possible to start with the inverse approach and identify coefficients with user studies. The only caveat of our approach is that our equations need to be inverted to obtain inverse formulas as follows:
\begin{equation}L_P = \frac{ L_O + (1-y) I_a}{y} \label{Eq_barycentric_inverse}\end{equation}
Here, the good news is that the inverses are guaranteed to be meaningful as long as $y$ is not zero. Note that $y$ can go to zero in the power formula and as a result the inverse can go to $\infty$ as shown below,
\begin{equation} lim_{s \rightarrow 0} L_P = lim_{s \rightarrow 0} \frac{ L_O + (1-s^{f(s)}) I_a}{s^{f(s)}} \rightarrow \infty \label{Eq_power_inverse}\end{equation}
This suggests that the power formula is not practical. In other words, visual systems may use rational affine polynomials since the inverse formula is shown below. 
\begin{equation} L_P = \frac{ L_O + (a_0(1-s)+a_1s) I_a} {a_0(1-s)+a_1s}\label{Eq_affine_inverse}\end{equation}
will always be bounded, since the range of this family of functions is $\{0, \frac{1}{a_1} \}$. Since $a_1$ is expected to be $1$, the range is most likely to be $\{0, 1 \}$. Bounding values in this range can be useful for internal computation in the brain since the level of signals will always be bounded if the input is bounded. In conclusion, even if the coefficients could be significantly different for different cultures or groups, the inverses of the resulting formulas are expected to be rational polynomials. This is theoretically useful since it suggests that there is a corresponding visual system that can provide the observed effect, and this inverse function can be computed. 

The forward computer graphics approach also simplifies certain problems. For instance, a hidden parameter in our formula is the operation to obtain a blurred version of the background. Our formula comes from a very simple blur operation; since the background can be either a discrete or continuous lightness scale, the blurred version is simply a continuous lightness scale. For a general background, it is necessary to investigate the effects of different types of blur operations. Blur operations do not have to be low-pass filters and the foreground object does not have to have the same translucency everywhere. Such a flexible approach can allow us to employ cloning methods to match the boundaries \cite{perez2003poisson,agarwala2007efficient,farbman2009coordinates}.

We also want to point out that for our approach to working with photographs, we need an effect that is visually similar to seamless cloning to avoid a visual effect of the shower door. The examples in Figure~\ref{fig_Vienna} demonstrate that our method has conceptual similarities to cloning methods, since the foreground images change color to fit the background image \cite{farbman2009coordinates}. We think that combining with photographs requires further research by employing seamless cloning methods. However, simpler opacity changes from the center to the boundaries shown in the examples in Figure~\ref{fig_Vienna}  can work for most photographic backgrounds. 
 
Another simplification of the analysis comes from the fact that we precisely choose the value of the relative size of $s$. In real vision systems, these must also be estimated.  Most likely, these estimations are not very accurate. We do think this is a significant problem since the resulting functions do not change rapidly. If the inaccuracies are not significantly large, we expect that the coefficients of the functions will be similar. It is also possible to test this. We can simply add a random error to the values of $s$ to simulate inaccuracies in the estimates.

\subsection{Conclusion, Limitations and Future Work}
\label{Sec_CLFW}

In this article, we have developed a method to identify context-sensitive luminance correction formulas to obtain constant luminance perception for a foreground object. We have implemented our method in Shadertoy as a web-based interactive program. The program is available in public code in Shadertoy. Anyone can access the program through the following link: \href{https://www.shadertoy.com/view/XX23Dz}{\textbf{Constant Perception}} or go directly to the next URL by typing in your browser: \url{https://www.shadertoy.com/view/XX23Dz} \cite{akleman2024constant}. Researchers can easily improve the code to test their hypotheses.  Using the explanation of how to change the polynomial part of the formula intuitively, researchers can provide better interfaces for large-scale user studies. Individual users can change the polynomial part of the formula directly by changing coefficients to obtain their own perceptively constant luminance. Their comments can be useful and can provide a crowd-sourcing experiment for further improvement of the formula. For future work, we want to apply a similar approach employing art and computer graphics expert knowledge to explain other luminance-related perception problems. We also expect that the methods can be extended to handle colors. However, this can require one to use unusual color models that we have been developing \cite{akleman2023circular,akleman2024projective}. 

\begin{figure}[hbtp]
\centering
     \begin{subfigure}[t]{0.32\textwidth}
\includegraphics[width=0.99\textwidth]{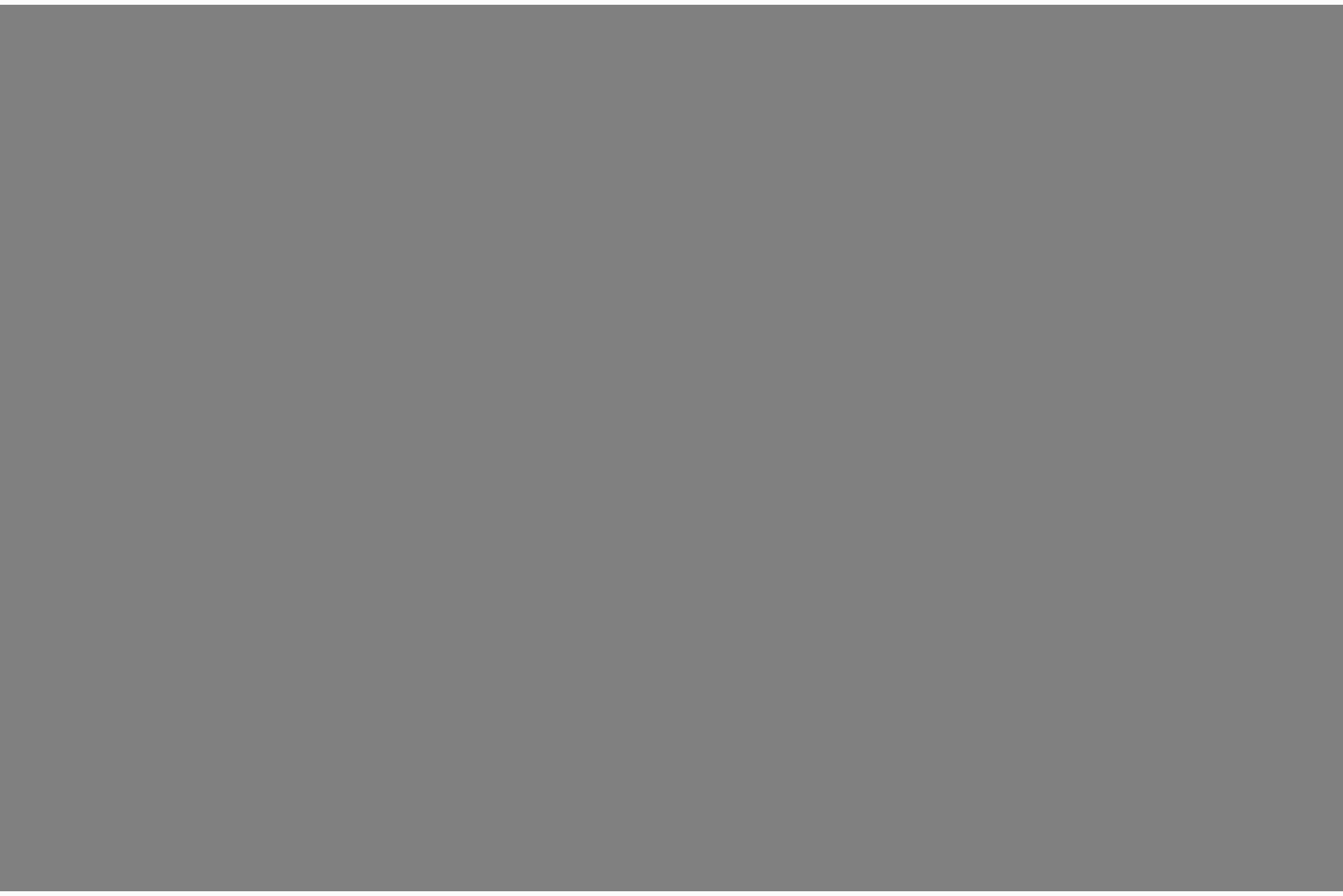}
    \caption{Constant color covering 99\% of the area in front of white background. }
    \label{fig_WhiteBGConstantColor/099}
 \end{subfigure}
 \hfill
     \begin{subfigure}[t]{0.32\textwidth}
\includegraphics[width=0.99\textwidth]{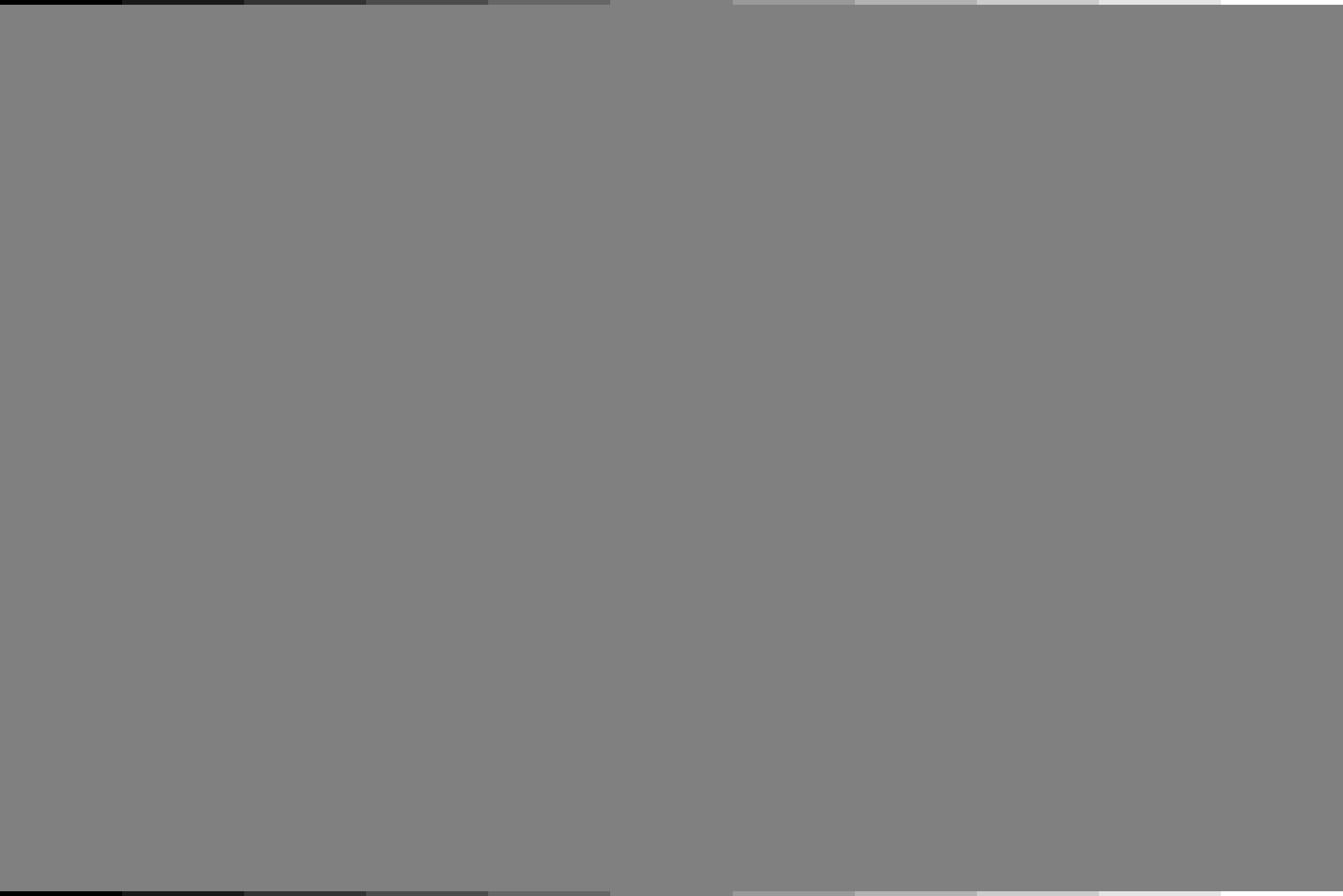}
    \caption{Constant color covering 99\% of the area in front of a graded color that consists of 99 bands. }
    \label{fig_10BandBGConstantColor/099}
 \end{subfigure}
 \hfill
      \begin{subfigure}[t]{0.32\textwidth}
\includegraphics[width=0.99\textwidth]{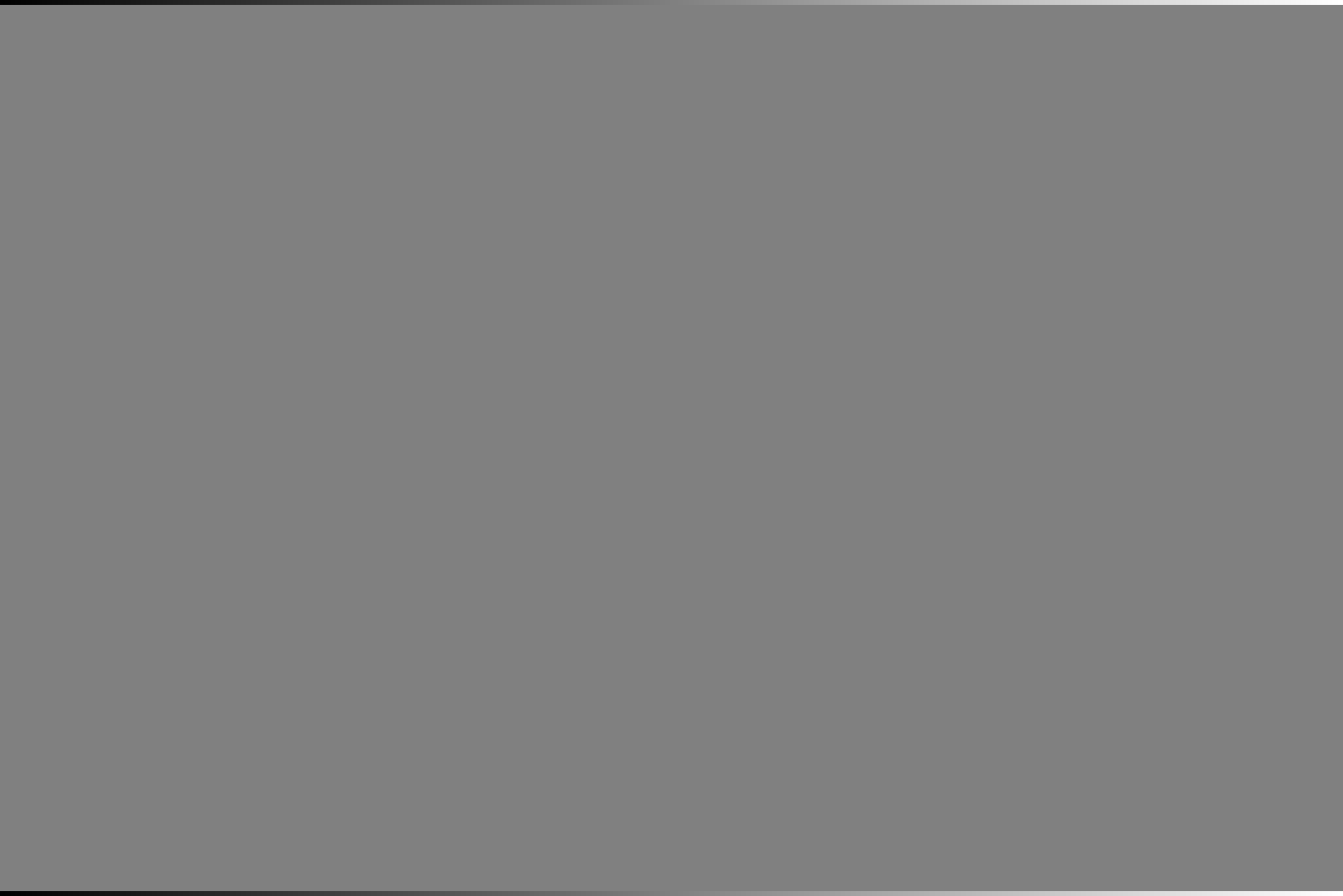}
    \caption{Constant color covering 99\% of the area in front of a continuously graded color. }
    \label{fig_ContinousBGConstantColor/099}
 \end{subfigure}
 \hfill
     \begin{subfigure}[t]{0.32\textwidth}
\includegraphics[width=0.99\textwidth]{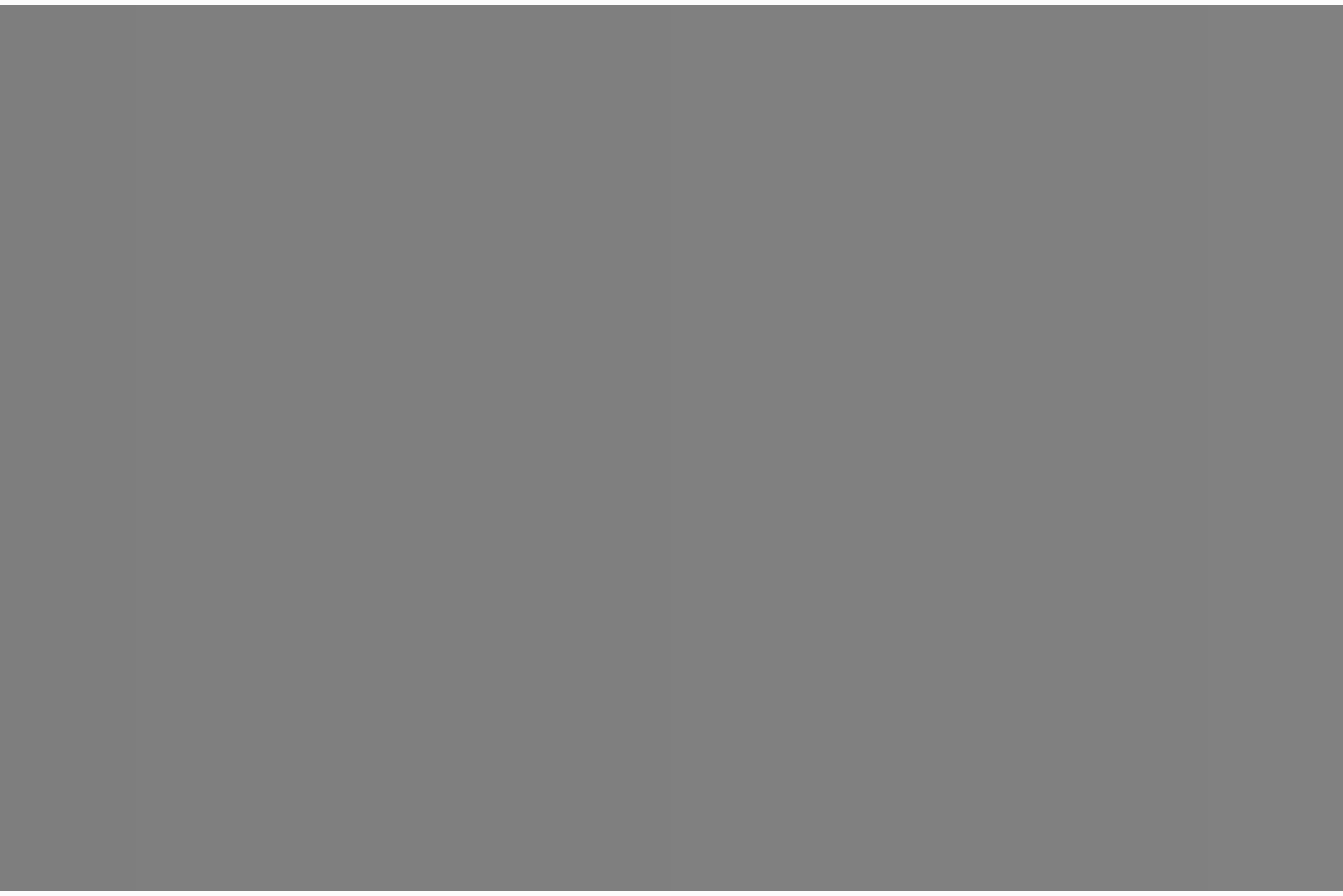}
    \caption{Our variable color band covers 99\% of the area in front of a white background. }
    \label{fig_WhiteBGConstantPerception/099}
 \end{subfigure}
 \hfill
     \begin{subfigure}[t]{0.32\textwidth}
\includegraphics[width=0.99\textwidth]{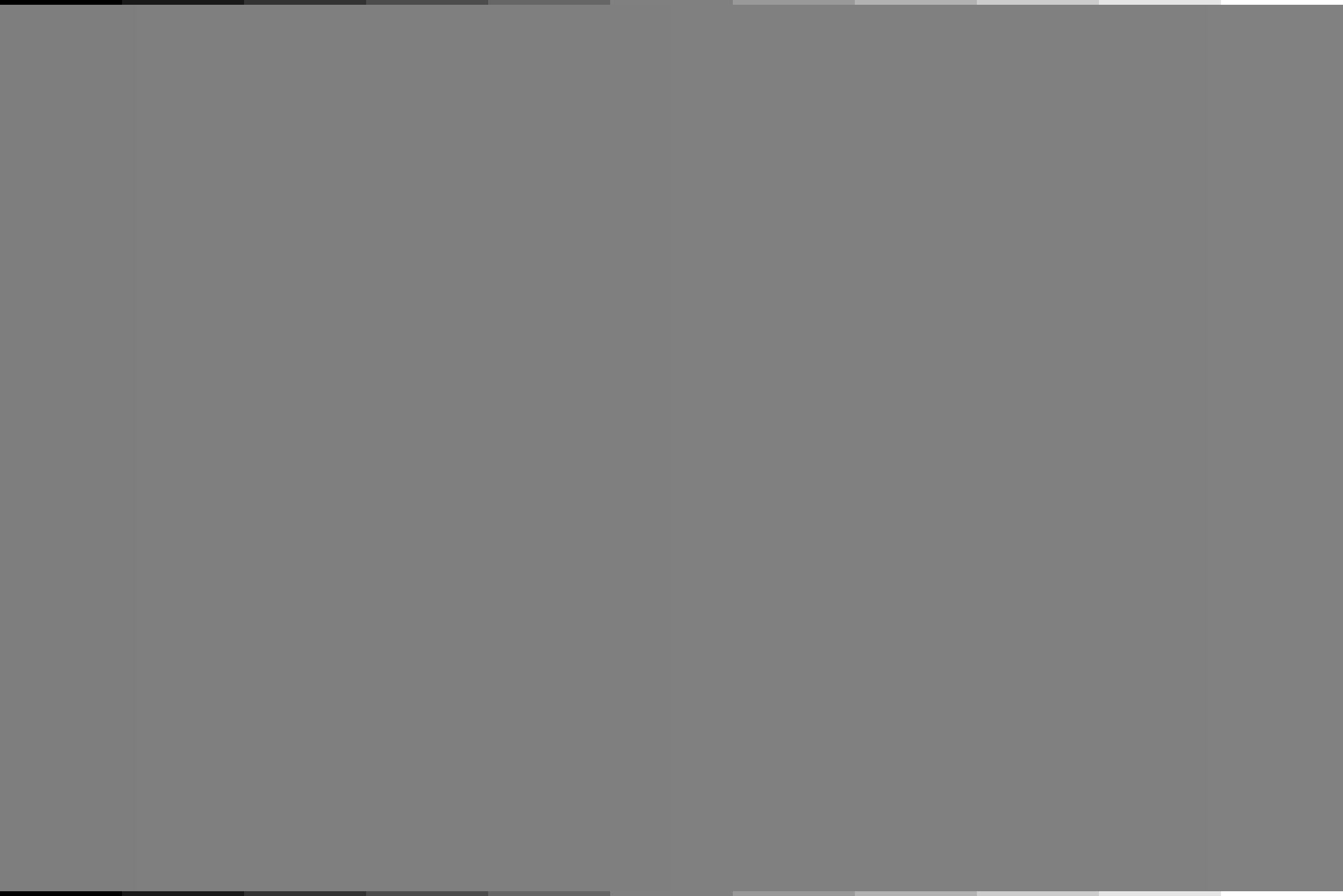}
    \caption{Our variable color band covers 99\% of the area in front of a graded color that consists of 99 bands. }
    \label{fig_10BandBGConstantPerception/099}
 \end{subfigure}
 \hfill
      \begin{subfigure}[t]{0.32\textwidth}
\includegraphics[width=0.99\textwidth]{10BandBGConstantPerception/099}
    \caption{Our variable color band covers 99\% of the area in front of a continuously graded color. }
    \label{fig_ContinousBGBGConstantPerception/099}
 \end{subfigure}
 \hfill 
    \caption{Comparison of our method with constant color. Note that Our variable color band creates constant perception}
\label{fig_099}
\end{figure}

\bibliographystyle{unsrtnat}
\bibliography{references}

\end{document}